\documentclass[10pt]{article}
\usepackage[a4paper, tmargin=0.75in, lmargin=0.75in, rmargin=0.75in, bmargin=0.75in]{geometry}
\usepackage{xcolor}
\usepackage{csquotes}
\usepackage{algorithm}
\usepackage{algpseudocode}

\usepackage[style=authoryear,backend=bibtex,maxcitenames=2,maxbibnames=99]{biblatex}
\bibliography{biblio}
\DeclareNameAlias{sortname}{family-given}%{last-first}
\makeatletter
\AtEveryBibitem{%
  \global\undef\bbx@lasthash%
  \clearfield{}}
\makeatother

\usepackage{authblk}
\usepackage{fullpage}
\usepackage{enumitem}% itemize options
\usepackage{framed,color}

\usepackage{float}%
\floatstyle{plaintop}%
\restylefloat{table}% these three lines put caption for tables ohttps://www.overleaf.com/project/5bf3e8fe09af1d23fd2c3e8fn top

\usepackage{mathrsfs}
\usepackage{amsmath}
\usepackage{amsthm}
\newtheorem{proposition}{Proposition}
\usepackage{accents}
\usepackage{amssymb}

\usepackage{graphicx}
\graphicspath{{Images/}} % Path for images' folder
\usepackage{eso-pic} % For the background picture on the title page
\usepackage{subfig} % Numbered and caption subfigures using \subfloat
\usepackage{caption} % Coloured captions
\usepackage{transparent}
\usepackage{booktabs} % For professional looking tables

\usepackage{hyperref}%
\hypersetup{
colorlinks=true,
    linkcolor=blue,
    filecolor=magenta,      
    urlcolor=blue,
    citecolor=blue,
}
\date{}
\title{\textbf{A latent space model for multivariate count data time series analysis}}
\author[1,2]{Hardeep Kaur}
\author[1]{Riccardo Rastelli}
\affil[1]{\footnotesize School of Mathematics and Statistics, University College Dublin, Ireland;}
\affil[2]{\footnotesize Insight: Centre for Data Analytics, University College Dublin, Ireland;}

\begin{document}
	
	\maketitle
\begin{abstract}
\noindent

Motivated by a dataset of burglaries in Chicago, USA, we introduce a novel framework to analyze time series of count data combining common multivariate time series models with latent position network models. This novel methodology allows us to gain a new latent variable perspective on the crime dataset that we consider, allowing us to disentangle and explain the complex patterns exhibited by the data, while providing a natural time series framework that can be used to make future predictions. Our model is underpinned by two well known statistical approaches: a log-linear vector autoregressive model, which is prominent in the literature on multivariate count time series, and a latent projection model, which is a popular latent variable model for networks. The role of the projection model is to characterize the interaction parameters of the vector autoregressive model, thus uncovering the underlying network that is associated with the pairwise relationships between the time series. Estimation and inferential procedures are performed using an optimization algorithm and a Hamiltonian Monte Carlo procedure for efficient Bayesian inference. We also include a simulation study to illustrate the merits of our methodology in recovering consistent parameter estimates, and in making accurate future predictions for the time series. As we demonstrate in our application to the crime dataset, this new methodology can provide very meaningful model-based interpretations of the data, and it can be generalized to other time series contexts and applications.
\\

\end{abstract}

\section{Introduction} \label{Introduction}

Modeling of multivariate count time series have become prominent in various applied fields. Historically, such analyses were predominantly utilized in economics and finance literature. In recent years, and in particular over the last decade, multivariate time series analysis has been increasingly associated with network analyses. In such cases, these interrelated time series are visualized through networks, with each time series acting as a node and the connections between them represented as edges. This network representation allows for a deeper understanding of the complex interdependencies within the data. For example, it helps elucidate how the behavior or attributes of individual nodes may influence or be influenced by other nodes in the network at any given time. This approach not only enhances predictive accuracy but also provides richer insights into the dynamic interactions within the data, thereby improving decision-making across various domains such as epidemiology (\cite{Dahlhaus_2000}), finance (\cite{Hardle_wang_yu2016}, \cite{DIEBOLD2014119}, \cite{Billio_2021}), biology (\cite{Shojaie_Michailidis_2010}), and social network analysis (\cite{Zhu_Pan2020}).

The dataset that we consider in this paper describes the number of burglaries in a number of areas of south Chicago, USA \parencite{CLARK2021100493}. The data exhibit complex patterns of interdependencies which are rooted in the geographical and social layout of the city \parencite{yuan2019multivariate,mohler2011self}. Trends in the number of crimes can be associated to events and other types of social dynamics, and their frequency can exhibit self-exciting or cross-exciting patterns, which may be resembling a contagious process \parencite{johnson2008repeat,mohler2011self}. Disentangling and inferring these complex patterns of the data can be a very challenging task, since they are not directly observed and typically cannot be objectively measured.

In this paper, we integrate a type of Latent Position Model (LPM) to model the hidden network representing the relations between the time series. Our objective is to ultimately uncover the network's latent structure and then use this to give a model-based visualization of the time series data, thus obtaining a highly interpretable output that can be used for prediction or qualitative analyses. This extends the existing literature on multivariate count time-series network models by introducing a novel methodology, which can directly address the modeling challenges that are posed by complex datasets such as the crime dataset that we consider here. 

More in detail, we modify a multivariate log-linear count time series model by incorporating a network modeling framework, namely the LPM. This integration is novel within the domain of multivariate count time series with network structures, particularly because the network structure is not observed and has to be inferred from the time series data. The rationale for integrating LPM into the multivariate time series context is to facilitate a straightforward and interpretable visualization, which aids in understanding pairwise relationships and unraveling the complex network structure. Moreover, our approach not only provides a geometric interpretation of these relationships but also quantifies the strength of interdependencies between the time series.  In our study, the connections between series are not merely binary; they manifest as positive, negative, or neutral, reflecting various types of interdependences. This nuanced view of connectivity introduces additional complexity into the model, opening up a relatively new research direction for the latent variable network models. Overall, our approach considerably enhances both theoretical insight and practical application in the study of network structures associated to multivariate count time series.

\subsection{Literature review}

Multivariate time series analysis comes into play in determining appropriate functions for a system of variables, generating reliable forecasts, and capturing the relationships and interactions among different entities.
To model such data, we can consider each time series individually, using, for example, with an autoregressive model (\cite{box2015time}). However, the consequence of such an approach would be a loss in capturing cross-sectional dependencies. The Vector AutoRegressive (VAR) model (\cite{book_Luetkepohl_2005}) is widely used for modeling multivariate time series data with continuous response variables. However, as the number of time series increases, the complexity of the model also increases, leading to computationally demanding estimation and over-parametrization. Moreover, when dealing with multivariate time series for continuous variables, models such as the Vector AutoRegressive Moving Average (VARMA) (\cite{book_Luetkepohl_2005}) or multivariate Generalized AutoRegressive Conditional Heteroskedasticity (GARCH) (\cite{book_Francq}) are also commonly used for analyses. 

As regards the literature on multivariate time series analysis for count data, we note that the literature on this is significantly less advanced compared to the continuous case. Some recent contributions in this topic are, for example, in electrical system reliability (\cite{Ertekin_2015}); neuroscience  \parencite{Brown_Partha_2004,Hall_Willett_2015}; econometrics (\cite{Bermúdez_Karlis_2011}); in finance (\cite{Pedeli_Karlis_2013_11}) and in environmental science (\cite{Livsey_Lund_Kechagias_Pipiras_2018}).

In the field of multivariate time series analysis, particularly in modeling systems with count data, various methodological approaches have been proposed. According to \textcite{Konstantinos_2020} and \textcite{FOKIANOS2021}, three prevalent approaches for the inferential procedures in such models can be currently considered: INteger AutoRegressive (INAR) models, parameter-driven processes, and observation-driven processes.

INAR models were first introduced by \textcite{franke1993multivariate} and \textcite{latour_1997}. The application of these models has been discussed in the papers by \textcite{Pedeli_Karlis_2013_03}, \textcite{Pedeli_Karlis_2013_11}, \textcite{SCOTTO2014233} and \textcite{Darolles_Fol_Lu_Sun2019}. 
Estimation methods for INAR models typically involve classical least squares and likelihood-based approaches. However, the application of likelihood theory, even in the simpler context of univariate INAR models, becomes particularly complex with higher-order models. Consequently, despite their suitability for modeling certain basic data structures, these models present significant challenges in estimation and prediction, particularly when dealing with models of a higher order.

\textcite{COX1981} categorized time-series models in two different ways: parameter-driven models and observation-driven models.
In parameter-driven models, parameters vary over time based on some dynamic unobserved processes. These models are also termed as state space models. Some of the applications of these models for univariate time series modeling are discussed in \textcite{Zeger1988ARM}, \textcite{Harvey1989TimeSM}, \textcite{Durbin_Koopman_2000}, and \textcite{Schnatter_Wagner_2006}. 
In addition, \textcite{Jorgensen1996}, \textcite{Jung2008}, and \textcite{Ravishanker2014HierarchicalDM} studied multivariate state space models. The estimation procedure of these models is based on likelihood approaches or on Bayesian approaches. These models are simple but computationally demanding.

Lastly, we have a class of observation-driven models. This is a class where the current parameters are considered as deterministic functions of lagged dependent variables, along with contemporaneous and lagged exogenous variables. In this approach, the parameters change randomly over time, yet they are predictable for one step ahead using past information. Univariate observation-driven models are quite prominent and widely discussed in the statistical literature; see for example \textcite{Davis2000,Fokianos_Anders_Dag2008,Fokianos_Tjøstheim_2011,Davis_Liu_2012}. The multivariate observation-driven count time series models are, instead, a quite novel topic of research. In fact, in this domain of observation-driven models, the linear and the log-linear models are a simple and popular generalization of Poisson autoregressive models for count data. \textcite{Heinen_Erick_2007}, \textcite{Andreassen_2013}, and \textcite{Lee_Tjøstheim_2017} primarily focused on studying linear models. The properties of the linear model have been discussed in \textcite{Streett_2000}, and \textcite{Ferland_Latour_Oraichi_2006}. Another relevant contribution is by \textcite{Fokianos_Tjøstheim_2011}, who talked about several advantageous characteristics of log-linear models.
Finally, \textcite{Konstantinos_2020} discussed both multivariate linear and log-linear models and used a copula-based construction for the joint distribution of the counts.

We now review the literature that connects time series models with network models. The main objective of these models is to estimate the pairwise relationship between the entities and predict future events or outcomes. A commonly studied type of multivariate time series associated with a network structure refers to situations where we observe time series data on the nodes of a network. In a recent work, \textcite{knight2016modelling} introduced the Network AutoRegressive (Integrated) Moving Average (NARIMA) models, defining multivariate continuous time series that are coupled with a network structure. These models are designed to effectively handle dynamic variations in the network structure associated with multivariate time series.
Recently, \textcite{zhu2017network} introduced a Network vector AutoRegressive model (NAR), a variant of the traditional VAR model that integrates a network structure. This model assumes that the value of a node depends not only on its own historical values but also on the average historical values of its neighboring nodes. The NAR model further incorporates node-specific characteristics or covariates, significantly simplifying the complexity by relying on a limited set of parameters. This simplification enhances manageability and interpretability. Inference is performed using least squares in two asymptotic regimes: (a) increasing time sample size ($T \rightarrow \infty$) with a fixed number of nodes $N$, and (b) both $N$ and time frames $T$ increasing. 
This methodology was further expanded by \textcite{JSSv096i05} who put forward the Generalized Network Autoregressive model (GNAR) for continuous random variables. In this setting, they introduced the node-specific effects based on neighborhood sizes. 

A different strand of literature is related to the work by \textcite{barigozzi2019nets}, who introduces a network methodology for analyzing large panels of time series data. The authors introduced a sparse VAR model by specifically focusing on two representations: a directed graph representing predictive Granger relations and an undirected graph representing contemporaneous partial correlations. This approach provides a parsimonious synthesis of the data and offers insights into the underlying structure of the panel. \textcite{NetworkGarch_2020} introduce a network GARCH model that leverages information obtained from a well-defined network structure. Their approach reduces the number of unknown parameters and significantly simplifies the computational complexity involved in the model.
Some further extensions of these models are discussed in \textcite{zhu2019network} that captures the changing behavior of quantiles over time. \textcite{Chen2020CommunityNA} enhances the flexibility of the model by \textcite{zhu2017network} by introducing the capability to accommodate diverse network effects across various network communities within the autoregressive model. The model proposed by \textcite{ZHU2019145} takes into account the sparse structure of network effects in the Network Autoregressive Model (NAM), allowing for the possibility of heterogeneity among network nodes. \textcite{Zhu_Pan2020} proposed the grouped network vector autoregression (GNAR) model, which groups network nodes by shared characteristics, with each group defined by unique parameters to capture specific behaviors and interactions. \textcite{zhu2020multivariate} introduced a spatial autoregressive model tailored for large social networks. Additionally, \textcite{kang2017dynamic} developed a model that integrates multi-scale modeling with network-based neighborhood selection, aiming to capture temporal local structures and identify significant temporal network changes while maintaining sparsity in interactions.

A central reference for our work is \textcite{zhu2017network} which combines multivariate time series with network structures and its extensions discussed so far considers continuous response data. The literature on the discrete response variable is sparse and has received relatively less attention in research. This gap is notable despite the applicability of such data across various domains, including social media, email communication, collaboration networks, disease spread modeling, financial systems, transportation, and ecological food webs. Our work contributes to this sparse literature. It relates to the multivariate time series with underlying network structure considered in the studies mentioned earlier. The key difference is that, in our setting, the time series are observed on the nodes of the network and take the form of counting processes. Additionally, the weights on the edges quantify various types of interdependencies. 

Theoretical and methodological advancements in modeling networks for multivariate count time series are currently active areas of ongoing research. Some more recent work by \textcite{armillotta2023} has made a significant contribution by demonstrating the suitability of multivariate observation-driven count time series models for modeling time-varying network data. Their extension of the NAR model to count data led to the formulation of the Poisson Network Autoregression (PNAR) model. This model uses linear and log-linear approaches and incorporates copulas to capture interdependencies among count variables. It also allows for alternative distributions like the negative binomial. Their work extensively analyzes two asymptotic regimes discussed in \textcite{zhu2017network} and advances a theoretical framework for asymptotic inference. The authors employ Quasi Maximum Likelihood Estimation (QMLE) for parameter estimation. 
 
\textcite{bracher2022endemic} also discusses an application of a model akin to the Network Autoregressive (NAR) model for multivariate count data. \textcite{amillotta2022generalized} in their work, discuss a statistical framework that consolidates the findings from 
previous studies by \textcite{zhu2017network} and \textcite{armillotta2023}. The authors present a comprehensive framework that accommodates the case of both continuous and count responses measured over time for each node of a known network, demonstrating its practical potential.  For high-dimensional discrete data, \textcite{mark2017network} constructs a mathematical model that includes a statistical learning guarantee for estimating parameters in complex multivariate point process models. \textcite{Pandit_2020} presents a p-lag multivariate discrete valued autoregressive model. The main challenge in this model is the presence of a large number of unknown parameters that need to be estimated. The analysis is done using a $l1$-regularized M-estimator when the loss functions are strongly convex. 

With regards to the networks literature, our work pivots on LPMs, which were introduced and popularized as a statistical model for social networks by \textcite{Hoff20021090}. The highly influential work of \textcite{Hoff20021090} lead to a significant increase in the literature on LPMs. However, despite this increase in research, we are not aware of any work integrating LPMs with multivariate time series, and, specifically, count time series data. This conceptual link represents one of our original contributions to both the networks and time series research literature. One paper in this area is by \textcite{ahelegbey2017bayesian}, which presents a Bayesian hierarchical model combining a Covariance Graphical Model (CGM), and a VAR model. However, their work focused on financial time series, whose values are continuous and not counts. This is a major distinction from our paper, as their contribution adds to the literature on multivariate continuous time series using specifically a sparse network structure. Another significant difference is that \textcite{ahelegbey2017bayesian} model the network using a CGM. In contrast, in our paper, we use a LPM framework both for modeling and inference purposes. 

Another relevant contribution in this context is that of \textcite{tafakori2022measuring}, where a latent space approach is used to model default probabilities of financial institutions. In a similar framework, latent variable models have also been employed by \textcite{hledik2023dynamic} to study systemic risk and financial stability. To our knowledge, no work has specifically applied LPM to the study of multivariate count time series data, highlighting a clear gap in the research literature.

The remainder of this paper is structured as follows: Section \ref{models} describes the Poisson Log-linear model, LPM, and the proposed Time series Latent Position Model (TSLPM). Section \ref{Inferential Procedure} outlines the inferential process. Section \ref{Simulation study} contains simulation studies exploring the performance of the TSLPM in terms of parameter estimation and prediction. Section \ref{real datasets} includes the application of the proposed model using the Chicago burglary dataset. Section \ref{discussion} concludes the paper with a conclusion and discussion on the possible extensions.

\section{Time Series Latent Position Model (TSLPM)}
\label{models}

\subsection{Poisson log-linear models} \label{Poisson Log-linear model}
Let $\mathbf{y}_{t}=\{y_{it},\ i=1,2, \dots ,N$\}, for $t=1,2, \dots, T$, denote a multivariate count time-series made of $N$ time series and $T$ time points. Assume a Poisson distribution on $\mathbf{y}_{t}$ and let $ \boldsymbol{\lambda}_{t} =\{\lambda_{it},\ i=1, \dots ,N$\}, for $t=1,\dots, T$, be the corresponding intensity process. Define $\mathscr{F}_{t}$ be the $\sigma$-field generated by $\mathbf{y}_{t-1}$. According to the model specification, we make the assumption that $\boldsymbol{\lambda}_t = E(\mathbf{y}_{t}|\mathscr{F}_{t-1})$.

Considerable emphasis has been placed on observation-driven models, particularly the Poisson log-linear model, for the estimation and inference of multivariate time series for count data \parencite{Doukhan2017MultivariateCA,FOKIANOS2021,Davis_Fokianos_Holan_Joe_Livsey_Lund_Pipiras_Ravishanker_2021}. A Poisson log-linear model of order $1$ introduced by \cite{Konstantinos_2020}, is formalized as follows:
\begin{gather}
   \mathbf{y}_{t}| \mathscr{F}_{t-1}^{\mathbf{Y},\boldsymbol{\lambda}} \sim \text{Poisson}(\boldsymbol{\lambda}_{t}) \notag\\
   \log(\boldsymbol{\lambda}_{t}) = \alpha + \mathbf{B} \log(\mathbf{y}_{t-1} + \mathbf{1}_{N})
   \label{log-linear VAR model}
\end{gather}
where $\log(\boldsymbol{\lambda}_{t})$ is the log link applied on all elements of $\boldsymbol{\lambda}_{t}$ and $\mathbf{1}_{N}$ is a $N$-dimensional unit vector. The value of $\log(\boldsymbol{\lambda}_{t})$ is determined by the lagged values of the actual time series via $\log(\mathbf{y}_{t-1} + \mathbf{1}_{N})$ in Eq.~\ref{log-linear VAR model}. In the given model, $\alpha$ may either be a constant or a vector of dimension $N$, and $\mathbf{B}$ is a matrix of size $N \times N$. These are the parameters that must be estimated.

By contrast, the linear model variant for multivariate Poisson VAR(1) discussed by \cite{Konstantinos_2020} is defined for each $i= 1,2, \dots , N$, as follows:
\begin{gather}\label{linear VAR model}
    \mathbf{y}_{t}| \mathscr{F}_{t-1}^{\mathbf{Y},\boldsymbol{\lambda}} \sim \text{Poisson}(\boldsymbol{\lambda}_{t}) \notag\\
    \boldsymbol{\lambda}_{t}= \boldsymbol{\alpha} + \mathbf{B}\mathbf{y}_{t-1}
\end{gather}
 In this case, the entries of $\boldsymbol{\alpha}$ and $\textbf{B}$ are all assumed to be non-negative. This constraint is imposed on the model parameters to ensure that the intensity vector is always non-negative.

 Despite its convenient formulation, the linear model has several limitations that result in the log-linear model typically being a more suitable modeling choice; see \cite{Fokianos_Tjøstheim_2011} for univariate and \cite{FOKIANOS2021} for multivariate time series modeling and related discussions. The rationale behind favoring a log-linear autoregressive models over linear autoregressive models is that the former do not impose any constraints on the parameters regarding positivity, facilitating the analysis of datasets exhibiting both positive and negative correlations. Additionally, the log-linear model may be expanded through the inclusion of covariates thus making it an optimal choice for a more comprehensive statistical analysis of multivariate count time series. For a detailed discussion on these models, see \textcite{FOKIANOS2021}.

\subsection{Latent position models}

LPMs \parencite{Hoff20021090} have gained prominence for their ability to offer a readily interpretable visualization of a network via a latent social space. The extensive body of literature on LPMs has found widespread application across various disciplines, particularly within the domain of social network analysis. For network data, we often represent the relationships between nodes using an $N \times N$ symmetric adjacency matrix, which we can generically denote as $\mathbf{A}$. In this context, $N$ refers to the number of observed nodes, and the entries $a_{ij}$ in the matrix indicate the relationship between node $i$ and node $j$. Typically, an adjacency matrix is a binary matrix of $1$s and $0$s which describe the presence or absence of an edge, respectively. In this section we review this binary framework, although we note that later in this paper we will be more interested in a weighted network framework, where $a_{ij}$ can take any value in $\mathbb{R}$, again, to represent the type of relation that the two corresponding nodes have. 

Let $\mathbf{Z}$ be a $N \times d$ matrix representing the coordinates of the $N$ nodes in a $d$-dimensional social space. Using the notation $\mathbf{z}_{i}$, we refer to the $i$-th row of the matrix $\mathbf{Z}$, which indicates the $d$ coordinates of the $i$-th node in the latent space. We highlight that inference aimed at estimating the number of latent dimensions $d$ from the data is becoming a prominent aspect of the research on LPMs; see \cite{Handcock2007301}, \cite{sewell2017latent}, \cite{Friel20166629}, \cite{D_Angelo2020324}, and \cite{Gwee_2023} for their contribution. 
However, in this paper, we only focus on the case $d=2$. This choice facilitates a clear visualization of the network structure and it simplifies inference. This choice is indeed very common in the literature and the simplification allows us to focus on and develop other aspects of our methodology. We emphasize that the choice of the number latent dimensions is an important future direction of our work.

We adopt a modeling approach based on conditional independence, assuming that each edge is determined solely by the positions of the nodes, without direct influence from other edges in the network. Consequently, the probability of the presence of an edge existing between nodes $i$ and $j$ is expressed as a function that captures the relationship between the latent positions $\mathbf{z}_{i}$ and $\mathbf{z}_{j}$ associated with these nodes, given by Eq.~\ref{eq:lpm_hoff_1}. 
\begin{equation}\label{eq:lpm_hoff_1}
\mathbb{P}(\boldsymbol{Y}|\boldsymbol{Z},\alpha) = \prod_{i < j} \mathbb{P}(y_{ij}|\mathbf{z}_{i},\mathbf{z}_{j},\boldsymbol{\theta})
\end{equation} 
here, the product notation is intended over all pairs of nodes $(i,j)$ such that $i<j$, for $i=1,2,\dots ,N$ and $i=1,2,\dots ,N$, whereas $\boldsymbol{\theta}$ is the collection of parameters that do not refer to edges or nodes.

In these models, the nodes with similar characteristics (i.e. positions) tend to possess a higher probability of forming an edge between them. The definition of similarity in latent characteristics can be approached in various ways, and \textcite{Hoff20021090} has introduced two distinct approaches known as the distance model and the projection model. For a comprehensive review of both models, see \parencite{Salter-Townshend2012243,LPNM_2023}.

The distance model by \textcite{Hoff20021090} posits that as the proximity between two nodes increases, the probability of a connection between them also increases. Conversely, as the distance between nodes grows larger, the probability of a connection decreases. This framework provides an easily interpretable visualization of geometric social space. The log odds that node $i$ and $j$ will form a tie are usually given as: 
\begin{equation}\label{eq:lpm_hoff_2}
\begin{aligned}
	\zeta_{ij} = \rm
 logodds(y_{ij}=1|\mathbf{z}_{i},\mathbf{z}_{j},\alpha) = \alpha - |\mathbf{z}_{i} - \mathbf{z}_{j}| 
\end{aligned}
\end{equation}
where $\alpha\in \mathcal{R}$ is called the intercept and $|\cdot|$ represents the Euclidean norm.
The Euclidean distance is frequently employed for this purpose, although alternative distance functions can be explored. For a few related contributions to the latent distance model, see \textcite{Handcock2007301} for clustering of highly connected nodes in the network via a latent mixture mode, \textcite{Krivitsky2008} which uses actor-specific random effects along with model-based clustering of nodes,  \textcite{angelo2019900} and \textcite{D_Angelo2020324} for introducing a Euclidean distance LPM for multi-view networks to model. Further related contributions have appeared over the last years, including \parencite{hoff2021additive, Gormley200790, Gollini2016246, aliverti2019spatial, rastelli2023continuous}. 

An alternative latent position model, also proposed by \textcite{Hoff20021090} is called the projection model. This model proposes that two nodes are highly likely to form a tie if their respective latent positions are pointing in the same direction ($\mathbf{z}_{i}^{\top} \mathbf{z}_{j} >> 0$), or conversely, they are less likely to connect if they point in opposite directions ($\mathbf{z}_{i}^{\top} \mathbf{z}_{j} << 0$). In other words, the probability that the two nodes will connect is higher when the angle between them is small, and the probability is lower when the angle gets closer to $180$ degrees. Here, $(.)^{\top}$ represents the transpose of a vector or matrix. The log odds for this model are defined as:
\begin{equation}\label{eq:lpm_hoff_3}
\begin{aligned}
	\zeta_{ij} = \rm logodds(y_{ij}=1|\mathbf{z}_{i},\mathbf{z}_{j},\alpha) = \alpha  + \mathbf{z}_{i}^{\top}\mathbf{z}_{j}
\end{aligned}
\end{equation}
Here, the magnitude of the dot product between the latent positions $\mathbf{z}_{i}^{\top} \mathbf{z}_{j}$ of node $i$ and node $j$ quantifies the degree of similarity or dissimilarity between the nodes. In the latent projection model, it is more straightforward to visualize the latent space using polar coordinates, where the positions of nodes can be understood as a combination of direction and magnitude (i.e. distance from the center of the space). This polar coordinate representation allows for a clearer understanding of the relationship between the nodes, with the direction indicating their orientation and the magnitude representing the strength of their latent characteristics. The statistical literature centered around this model includes \parencite{hoff2005bilinear, hoff2007modeling, hoff2011hierarchical, hoff2021additive,young2007random,nickel2008random}. Some applications of this model have been explored by \textcite{Durante20171547} and \textcite{Durante201829}, where the model is used to characterize and test for differences in brain connectivity networks.

Both the latent distance model and the latent projection model provide meaningful insights through geometric latent space representations. The latent distance models are popular due to their clarity and intuitive interpretations. However, latent projection models offer advantages over distance models, particularly in their ability to accommodate both positive and negative relationships between entities. Since our primary aim is to effectively visualize and quantify the pairwise relationships among time series, which may be involve either positive or negative interactions, we focus our attention on the latent projection model. Ultimately, this model can facilitate a model-based visualization of the data using a latent embedding of the series.

\subsection{The TSLPM}

\subsubsection{Model definition}

We introduce a hierarchical model consisting of a multivariate log-linear autoregressive time series model combined with latent projection network model. 
We call our model the Time Series Latent Position Model (TSLPM), of order 1, and define it as follows:
\begin{gather} 
y_{it}|\mathscr{F}_{t-1} \sim Pois(\lambda_{it}) \notag \\
\log(\lambda_{it}) = \alpha + \sum_{j=1}^N \gamma_{ij} \log(y_{j(t-1)}+1)+\sum_{k=1}^K  \delta_{k} x_{ik} \notag \\
 \gamma_{ij}=   \begin{cases} 
      \beta_{i}  &  i=j \\
      \mathbf{z}_{i}^{\top}\mathbf{z}_{j} & i \neq j \\
   \end{cases}
   \label{eq:TSLPM_model}
\end{gather}
for all $i = 1, \dots, N$ and $t = 1, \dots, T$. The model parameters are $\alpha \in \mathbb{R}$, $\boldsymbol{\beta} = \{\beta_1, \dots, \beta_N\} \in \mathbb{R}^N$, $\textbf{Z} = \{\mathbf{z}_i \in \mathbb{R}^2| i = 1, \dots, N\}$ and $\boldsymbol{\delta} = \{\delta_1, \dots, \delta_K\} \in \mathbb{R}^K$.

The model states that the Poisson log rate of the marginal conditional mean of any series $i$ is determined by two main components. The first component is the log rate of the previous count of series $i$ itself, associated to the autoregressive coefficient $\beta_i$. The second component consists of the log rates of the counts of all the other series, each weighted by a suitable coefficient $\gamma_{ij}$, for all $j$s. The parameter $\alpha$, which may be consistent across all series or vary among them, allows for the possibility of a non-zero mean process and significantly affects the long-term forecasts generated by the model. Another aspect of this model is the presence of covariates, indicated in Eq.~\ref{eq:TSLPM_model} by the term $x_{ik}$, for $K$ observed covariates. These additional variables are included in the log-linear model using the coefficients $\delta_1, \dots, \delta_K$. We point out that, similarly to the $\alpha$ parameter, the parameters $\delta_{k}$ may be equal across all series or differ among them. In this version of the model, we consider the same $\alpha$ and the same $\delta$s for all series.

The parameter $\gamma_{ij}$ plays a significant role in informing the network structure that identifies the interrelations between the time series. In the TSLPM, the $(i,j)$th element of matrix $\mathbf{B}$, represented by $\gamma_{ij}$, corresponds to the autoregressive coefficients in the Poisson log-linear model specified by the Eq.~\ref{log-linear VAR model}. We modify the structure of the log-linear model by incorporating the projection modeling approach on the elements of matrix $\mathbf{B}$. Consequently, each non-diagonal entry of the matrix $\mathbf{B}$ is the dot product of the latent positions associated to two different series, series $i$, and series $j$, thus measuring the dependence of the series $\mathbf{y}_{i}$ and $\mathbf{y}_{j}$. As regards interpretation of these parameters, the relationships between series are significantly influenced by the angular proximity of their latent position vectors. Vectors that are closely aligned, demonstrated by smaller angles between them, are associated with a positive association between the two corresponding series. Conversely, vectors that form obtuse angles, are indicative of negative associations. Orthogonal vectors indicate the absence of a meaningful relationship, i.e. the two series in this case are conditionally independent given the model parameters.

\subsubsection{Conditions to impose stationarity} \label{Stationarity Conditions}
In this section, we discuss on stationarity and ergodicity of the family of time series that we consider in this paper, building on the discussions presented by \textcite{Doukhan2017MultivariateCA} and \textcite{Konstantinos_2020}.

 We define the $\ell^{1}$-norm of a matrix $\mathbf{B}$ as $\lVert{\mathbf{B}} \rVert_1 = \underset{1 \leq j \leq N}{max} \sum_{i=1}^{N} |b_{ij}| $, which corresponds to the maximum absolute column sum of the matrix. Furthermore, the spectral norm of a matrix $\mathbf{B}$ is defined as $\lVert{\mathbf{B}} \rVert_2 = \sqrt{\rho (\mathbf{B}^{\top} \mathbf{B})} = \sigma_{max}(\mathbf{\mathbf{B}})$, where $\rho(.)$ denotes the spectral radius of a matrix and $\sigma_{max}$ is the largest singular value of $\mathbf{B}$. In the case where $\mathbf{B}$ is a symmetric matrix in $\mathcal{R}^{N \times N}$ (i.e., $\mathbf{B}^{\top}= \mathbf{B}$) then $\lVert{\mathbf{B}} \rVert_2 = \rho (\mathbf{B}) =  \underset{1 \leq i \leq N}{max} |\Omega_i|$, where, $\Omega_i$ denotes the $i$th eigenvalue of matrix $\mathbf{B}$.

The TSLPM model is a variant of the Poisson VAR(1) model. It is perhaps worth pointing out the stability condition of the VAR model. The condition is as follows:   
 \begin{proposition}[\cite{book_Luetkepohl_2005}] \label{Condition_1} Assume $\mathbf{y}_{t}=(y_{1t},y_{2t},.....,y_{Nt})^{\top}$ denotes an $N \times 1$ vector of time-series. The basic VAR of order 1 has the form:
$$
\mathbf{y}_{t}= \alpha + \mathbf{B} \mathbf{y}_{t-1}+ \boldsymbol{\epsilon}_{t}, ~~~ t = 1, \dots, T
$$
where $\mathbf{B}$ is a $N \times N$ matrix with autoregressive coefficients and $\mathit{\epsilon}_{t}$  is an $N \times 1$ zero mean white noise vector process.

The VAR(1) is stable if the roots of
$det (\mathbf{I}_{N}- \mathbf{B} z) = 0$ lie outside the complex unit circle (i.e., have a modulus greater than one), or, equivalently, if all the eigenvalues of the matrix $\mathbf{B}$ have absolute value less than one. Assuming the process has been initialized in the infinite past, a stable VAR(1) process is stationary and ergodic with time-invariant means, variances, and autocovariances. 
\end{proposition}

Next, we discuss the conditions that are critical for achieving stationarity and ergodicity of our TSLPM, borrowing from the literature on related models. The main goal is to obtain stationarity and ergodicity of the joint process $(\mathbf{y}_{t},\boldsymbol{\lambda}_{t})$ due to the dependency structure between the count data 
$\mathbf{y}_{t}$ and the intensity process $\boldsymbol{\lambda}_{t}$.

\begin{proposition}[\cite{Doukhan2017MultivariateCA,Konstantinos_2020}]  \label{Prop. Ergodicity of log-linear model}
Consider the model of Eq.~\eqref{log-linear VAR model}. Suppose that $\lVert{\mathbf{B}} \rVert_1 < 1$. Then there exists a unique causal solution \{$(\mathbf{y}_{t},\log(\boldsymbol{\lambda}_{t})$\} to the model of Eq.~\eqref{log-linear VAR model} which is stationary, ergodic and satisfies $\text{E} \lVert \log(\mathbf{y}_{t} + \mathbf{1}_N) \rVert_{r}^{r}  < \infty$ and $\text{E} \lVert \log(\boldsymbol{\lambda}_{t}) \rVert_{r}^{r} < \infty$ and $\text{E}[\exp(r \lVert \log(\boldsymbol{\lambda}_{t}) \rVert_{1})]< \infty$ for any $r \in \mathcal{N}$.    
\end{proposition}
The above proposition also holds true whenever $\lVert{\mathbf{B}} \rVert_2 < 1$, thus this condition is sufficient to guarantee stationarity and ergodicity of the joint process $(\mathbf{y}_{t},\boldsymbol{\lambda}_{t})$. For proof and detailed discussion, see \textcite{Konstantinos_2020}. Leveraging the above results, we have obtained an alternative condition for stationarity and ergodicity, i.e. the condition $ \underset{1 \leq i \leq N}{max} |\Omega_i(\textbf{B})| < 1$.

We now discuss further on conditions that are sufficient to achieve stationarity, extending Proposition~\ref{Prop. Ergodicity of log-linear model}. We rely here on the Gershgorin Circle Theorem (GCT) which has been a cornerstone for introducing bounds on eigenvalues of a square matrix.
\begin{proposition}[\cite{varga2011geršgorin}] \label{Condition_2} % 
 Given a real symmetric $N\times N$ matrix $\mathbf{B}$ with generic element $b_{ij}$, define the Gershgorin disc for the $i$-th row as the closed disc $D(b_{ii},r_{i})$ centered at $b_{ii}$ with radius $r_{i}$, which is given by:
\begin{equation*}
r_{i} = \sum_{i \neq j} |b_{ij}|
\end{equation*}
where $r_{i}$ is the sum of the absolute values of the non-diagonal entries in the $i$-th row. According to the Gershgorin Circle Theorem (GCT), every eigenvalue of matrix $\mathbf{B}$ lies within at least one of the Gershgorin discs ${\displaystyle D(b_{ii},r_{i}).}$
Consequently, the boundaries of the union of all discs establish the potential range of the eigenvalues. Thus, $\min_{i}(b_{ii} - r_{i})$ and $ \max_{i}(b_{ii} +r_{i})$ establish the lower and upper bounds on the eigenvalues, respectively:
\begin{equation*}
\min_{i}(b_{ii} - r_{i}) < \Omega_i(\mathbf{B}) < \max_{i}(b_{ii} +r_{i})
\end{equation*}
\end{proposition}
Hence, the Proposition~\ref{Condition_2} supplements these criteria by providing explicit bounds on the eigenvalues, derived from the Gershgorin disc analysis. This introduces a rigorous set of necessary conditions to guarantee stationarity of a TSLPM:
\begin{equation} \label{Final condition from propositions}
-1<\min_{i}(b_{ii} - r_{i}) < |\Omega_{max}(\mathbf{B})| < \max_{i}(b_{ii} +r_{i}) < 1
\end{equation} 
Satisfying the above inequality ensures the stationarity and ergodicity of the joint process $(\mathbf{y}_{t}, \boldsymbol{\lambda}_{t})$ ensuring that the process remains well-behaved over time. Critically, we note that these conditions are sufficient, but not necessary for stationarity and ergodicity of the joint process.

\paragraph{Example.} For instance, let us assume that we model a multivariate time series $\{Y_1, \dots Y_5\}$ using a TSLPM of order $1$. Let us consider the following interaction matrix $\mathbf{B}$ to analyze and determine the stationarity and ergodicity of the associated multivariate time series process.\\

\begin{center}
\begin{equation*}
\centering
    \mathbf{B} = \begin{bmatrix}
        0.12 & 0.02 & 0.08 & 0.08 & -0.14 \\
        0.02 & 0.20 & -0.02 & -0.02 & 0.10 \\
        0.08 & -0.02 & 0.07 & 0.06 & -0.13 \\
        0.08 & -0.02 & 0.06 & 0.06 & -0.12 \\
        -0.14 & 0.10 & -0.13 & -0.12 & 0.27
    \end{bmatrix}
\end{equation*}
\end{center}

 By applying the Gershgorin Circle Theorem (GCT) as stated in Proposition
\ref{Condition_2}, we obtain 5 discs given by $D(0.12, 0.32),\ D(0.2, 0.16),\ D(0.07, 0.29),\ D(0.06, 0.28)$ and $D(0.27, 0.49)$. These discs suggest that the eigenvalues of $\mathbf{B}$ are bounded as follows:
\begin{equation*}
-0.22 < \Omega_{i}(\mathbf{B}) < 0.76
\end{equation*}
Given the actual eigenvalues of $\mathbf{B}$, which are $0.5,\ 0.21,\ 0.01,\ 0.003,\ -0.0009$, we confirm that the largest eigenvalue is $0.5$. This observation is consistent with the established bounds and also satisfies the condition from Eq.~\ref{Final condition from propositions}. 

\subsubsection{Prior specification}
We apply a Bayesian estimation framework to estimate the model parameters. Specifically, we incorporate prior information for the parameters $\alpha$, $\boldsymbol{\beta}$, $\mathbf{Z}$, and $\boldsymbol{\delta}$ by assigning independent and noninformative priors to each of them. Each parameter, namely $\alpha$, $\boldsymbol{\beta}$, $\mathbf{Z}$, and $\boldsymbol{\delta}$ is assigned an independent normal prior with a mean of zero and a standard deviation of 100.
This choice of a high standard deviation is deliberate, aiming to minimize the influence of prior assumptions on the posterior distributions. By employing broad, noninformative priors, we ensure that the analysis remains less biased by prior beliefs, allowing the observed data to play a more significant role in shaping the posterior densities. Furthermore, the specifications allow $\alpha$, $\boldsymbol{\beta}$, $\mathbf{Z}$, and $\boldsymbol{\delta}$ to take both positive and negative values, reflecting their potential impacts in the model.

\subsubsection{Parameter interpretation}
In this section we discuss more in detail how the TSLPM parameters can be interpreted.
To facilitate this, we also express our TSLPM model using an alternative form which is equivalent to the form given in Eq.~\ref{eq:TSLPM_model}:
\begin{gather} \label{TSLPM_alternateform}
 \lambda_{it} = \exp(\alpha + \sum_{K}  \delta_{k}    {x}_{ik})  (y_{i(t-1)} + 1)^{\beta_{i}} \prod_{j}  (y_{j(t-1)}+1)^{\gamma_{ij}}
\end{gather}
The exponential function in the model translates the log-linear relationship into a multiplicative form for $\boldsymbol{\lambda}$. 
This approach provides a different perspective on the interpretation of parameters, focusing on proportional changes rather than additive changes. In other words, this emphasis is particularly meaningful for predicting the effect of individual parameters on the count data generated from TSLPM.

To begin with, the intercept $\alpha$ and covariates effects $\boldsymbol{\delta}$ determine a  baseline for the count data. The effect of $\alpha$ on the Poisson rate is common across all series and is multiplicative through an exponential function, so a positive $\alpha$ will tend to increase the time series values, and a negative $\alpha$ will tend to decrease them. Similarly, the coefficients $\boldsymbol{\delta}$ can add more nuances by differentiating the baselines for the series using the covariates information.

The series-specific autoregressive parameter $\boldsymbol{\beta}$ determines instead the magnitude of the multiplicative effect of the preceding period on current counts. A value equal to zero means that the preceding time series value has no effect on the current one. A higher $\boldsymbol{\beta}$ value gives a typical autoregressive structure with positive autocorrelations and sustained trends. By contrast, a negative value suggests an alternation of positive and negative autocorrelations and a more erratic behavior.

As regards $\gamma_{ij}$, each of these parameters determines the strength of a multiplicative effect given by $y_{j(t-1)}+1$. Similarly to the $\boldsymbol{\beta}$ parameters, also the $\gamma_{ij}$ are autoregressive terms that capture the interactions between the series. A value of zero indicates no associations between series $i$ and $j$. A value greater than zero gives positive cross autocorrelations and sustained trends. A negative value suggests a more erratic behavior through negative cross autocorrelations.

\section{Inferential procedure} \label{Inferential Procedure}
\subsection{Estimation}
In this paper, we explore two main approaches for the inference of the TSLPM. The first approach involves optimizing the likelihood function using the Limited-memory Broyden-Fletcher-Goldfarb-Shanno (LBFGS) algorithm. This algorithm is particularly suited for large-scale optimization problems, where the computing cost increases at least with the square of the number of series. The LBFGS algorithm is efficient in this optimization task as it utilizes limited memory to store information from previous iterations, thereby circumventing the need for full Hessian matrix computation. The algorithm iteratively updates the current candidate solution by efficiently combining past gradients and curvature differences. The steps of the LBFGS algorithm as discussed by \textcite{NoceWrig06} are outlined in the Algorithm \ref{alg:LBFGS}.
\begin{algorithm} [htbp]
\caption{The LBFGS algorithm}
\label{alg:LBFGS}
\begin{algorithmic}
\Require
1. Choose an initial guess for all the model parameters $\alpha, \boldsymbol{\beta}$, and $\mathbf{Z}$. Set $r=0$. \\
\vspace{2pt} % Adjust the space as needed
2. Compute the gradient of the objective function at the current point.\\
\vspace{2pt} % Adjust the space as needed
3. Update the search direction $p_r$ by approximating the inverse Hessian matrix $\mathbf{H}_{r}^{-1}$ using information from the previous iterations.\\
\vspace{2pt} % Adjust the space as needed
4. Select a step size %$alpha_k$
that sufficiently reduces the objective function.\\
\vspace{2pt} % Adjust the space as needed
5. Update the current point by taking the step determined in the line search.\\
\vspace{2pt} % Adjust the space as needed
6. Store the updated estimate of the inverse Hessian matrix by using the information from the current iteration and the previous iterations.\\
\vspace{2pt} % Adjust the space as needed
7. Check for convergence criteria. If the criteria are met, stop. Otherwise, increment $r$ and repeat steps 2-6.\\
\end{algorithmic}
\end{algorithm}
The algorithm is not guaranteed to converge to a global maximum of the objective function, however, it will converge to a local optimum and it generally shows good performance on our problem using very little computation demand.

The second approach which we have opted for is Hamiltonian Monte Carlo (HMC). HMC is a Markov chain Monte Carlo (MCMC) algorithm that efficiently navigates high-dimensional spaces by using derivatives of the density function to inform sampling decisions from the posterior distribution  \parencite{Neal2011MCMCUH,betancourt2013hamiltonian}. Unlike simpler sampling methods such as the random-walk Metropolis or Gibbs sampling \parencite{1987PhLB..195..216D, Neal2011MCMCUH}, HMC leverages the gradient information of the log-density function, facilitating a more effective exploration of the parameter space.

Hamiltonian dynamics are utilized in the Hamiltonian Monte Carlo (HMC) method to simulate two distinct components. The first component is a $d$-dimensional parameter space, denoted as  \textbf{$\boldsymbol{\theta}$} = $(\theta_1, \theta_2, \dots \theta_d)$. The second component comprises $d$-dimensional auxiliary momentum variables, $\mathbf{m}= (m_1, m_2, \dots , m_d)$. These momentum variables typically follow a multivariate normal distribution. This distribution has a mean of zero and is characterized by a covariance matrix ($\boldsymbol{\Sigma}$). The purpose of this setup is to help rotate and scale the target distribution, as described by \textcite{betancourt2011geometry} in their work on the geometry of Hamiltonian Monte Carlo. 
 
The joint distribution from which HMC samples can be expressed reads as follows:
\begin{equation*}
 \text{p}(\boldsymbol{\theta}, \mathbf{m}) \propto \exp({\mathcal{L}(\boldsymbol{\theta}) - \frac{1}{2} \mathbf{m}^{\top} \mathbf{m})}
\end{equation*}
where $\mathcal{L}(\boldsymbol{\theta}) $ denotes the log-probability for $\boldsymbol{\theta}$. This augmented non-normalized model acts as a fictitious Hamiltonian system reflecting the total negative energy of the particle $\log(\text{p}(\boldsymbol{\theta}, \mathbf{m}))$ for some particle's negative potential energy  $\mathcal{L}(\boldsymbol{\theta})$ and the kinetic energy $\frac{1}{2} \mathbf{m}^{\top} \mathbf{m}$. The HMC algorithm procedures are detailed in Algorithm \ref{alg:HMC}.

The algorithm  (\cite{Hoffman2019NeuTralizingBG}) employs the leapfrog integrator to numerically simulate the Hamiltonian dynamics over time. This integrator preserves volume and is reversible, conserving the Hamiltonian to an approximation. Given the step size $\epsilon$ and time point $t$, the leapfrog integrator updates the states as follows:
\begin{gather}
     \mathbf{m}_{t+\frac{\epsilon}{2}} = \mathbf{m}_{t} + \frac{\epsilon}{2} \nabla_{\boldsymbol{\theta}} \mathcal{L}(\boldsymbol{\theta}_{t}) \notag \\ 
      \boldsymbol{\theta}_{t+\epsilon} = \boldsymbol{\theta}_{t} + \epsilon~\mathbf{m}_{t+\frac{\epsilon}{2}} \notag \\ 
     \mathbf{m}_{t+\epsilon} = \mathbf{m}_{t+\frac{\epsilon}{2}} + \frac{\epsilon}{2} \nabla_{\boldsymbol{\theta}} \mathcal{L}(\boldsymbol{\theta}_{t+\epsilon}) \notag \\
     \label{HMC_USING_LEAPFROG}
\end{gather}

Following $L$ repetitions of these steps, the proposed state ($\hat{\boldsymbol{\theta}}, \hat{\mathbf{m}}$) is evaluated using a Metropolis acceptance step in Eq.~\ref{metropolis_acceptance_step} to decide on its inclusion in the Markov chain. If the proposal is not accepted, the previous parameter value is returned for the next draw and used to initialize the next iteration.
\begin{algorithm} [htb]
\caption{The HMC algorithm}\label{alg:HMC}
\begin{algorithmic}
\Require
1. Choose the initial state $s_{t-1}= (\boldsymbol{\theta}, \mathbf{m})$ from the parameter state of interest.\\
\vspace{2pt} % Adjust the space as needed
2. Sample the new momentum variable $ \mathbf{m}^\star \sim N(\textbf{0}, \textbf{1})$.\\
\vspace{2pt} % Adjust the space as needed
3. Update position and momentum ($\boldsymbol{\theta}, \mathbf{m}^\star$) by applying $L$ leapfrog steps starting from state $s_{t-1}^* = (\boldsymbol{\theta}, \mathbf{m}^\star)$. \\
\vspace{2pt} % Adjust the space as needed
4. The new state $\hat{s_{t}}    =$ ($\hat{\boldsymbol{\theta}}, \hat{\mathbf{m}}$) is proposed.\\
\vspace{2pt} % Adjust the space as needed
5. The Metropolis algorithm is applied to calculate the acceptance probability of the proposal ($\hat{\boldsymbol{\theta}}, \hat{\mathbf{m}}$), which has been generated by transitioning from the state $s_{t-1}^* = (\boldsymbol{\theta}, \mathbf{m}^\star)$ and is given as:
\begin{equation}
    \text{p}_{accept}(s_{t-1}^*) = min \left\{ 1, \frac{p(\hat{\boldsymbol{\theta}}, \hat{\mathbf{m}})}{p(\boldsymbol{\theta}, \mathbf{m}^*)} \right\}
    \label{metropolis_acceptance_step}
\end{equation}
\vspace{2pt} % Adjust the space as needed
6. Accept the move from $s_{t-1}^*$ to $s_{t}$ with probability $p_{accept}$, otherwise the current state is preserved within the sample and used to initialize the next iteration. \\
\end{algorithmic}
\end{algorithm}

The leapfrog integrator is accurate up to $\mathcal{O}(\epsilon^{2})$. To maintain a high target acceptance rate (commonly set at $0.8$), the algorithm adapts to use smaller step sizes $\epsilon$. This can improve sampling efficiency (effective samples per iteration) at the cost of increased $L$ (number of iterations) to keep the total distance traveled roughly constant. This increase is costly due to the higher number of gradient computations per iteration as indicated in the leapfrog steps in Eq.~\ref{HMC_USING_LEAPFROG}. Thus, the goal is to balance the acceptance rate to optimize both efficiency and computational expense.

\subsection{Procrustes analysis of nodal positions} \label{Procrustes}
Parameter identification poses a very intricate challenge in the estimation procedure of latent variable models. In this study, we have adopted the projection model to estimate the positions of nodes within the latent space. This method brings to light a critical issue: the model depends on the latent positions only through the pairwise distances between them. As a result, the parameters become non-identifiable when subjected to certain transformations of the point process. Differently from other types of latent variable network models (e.g. the distance model), these transformations are limited to rotations and reflections, since the projection model lacks translation invariance. 

Focusing on a single parameter configuration, as it would be the case in an optimization setting, removes to some extent the issues related to non-identifiability, since these transformations would not alter our interpretations of the latent space. However, the situation becomes more complex in a Bayesian framework. Here, we generate posterior samples and attempt to summarize these by calculating estimators, such as the posterior mean. Applying transformations like rotations or reflections to any configuration of the sample will affect the posterior summaries. One way to explain this is by noting that it is impossible for us to know if the latent space was rotated or reflected during the sampling process. Thus, in a Bayesian framework, the posterior sample of the TSLPM lacks identifiability, creating an obstacle in providing meaningful insights or summaries.

To address this issue, we employ a popular methodology known as Procrustes matching, also referred to as Procrustes analysis or Procrustes superimposition. This technique seeks to find the best alignment or correspondence between two or more sets of points, typically represented as coordinate matrices in a multidimensional space, denoted here as $\mathbf{X}$ and $\textbf{Y}$. In Procrustes matching, $\mathbf{X}$ undergoes a rigid transformation, consisting of rotations or reflections (pertinent to the projection model), to get as near to the reference $\mathbf{Y}$ as possible. The optimal match minimizes the sum of squared distances between corresponding points across the matrices, defined as:
$$ R^2 =\sum_{i=1}^{N}\sum_{k=1}^{d} (y_{i,k}-x_{i,k})^2 $$
where $\mathbf{x}_{i}$ and $\mathbf{y}_{i}$ are the coordinates of point $i$ in configuration $\mathbf{X}$ and $\mathbf{Y}$, respectively.

Considering that the projection model lacks translation invariance, we formulate an optimization problem by introducing just rotations and reflections onto the matrix $\mathbf{X}$. For a more detailed discussion, see \textcite{Hoff20021090}.
The transformation applied to $\mathbf{X}$ is represented as:
$$ \textbf{x}_{i}^{\top} = \textbf{O}^{\top} \textbf{x}_i  $$
where $\textbf{O}$ is an orthogonal matrix that facilitates a rotation and/or reflection transformation.
Therefore, the modified sum of squared distances between points is expressed as:
\begin{equation} \label{R_sq_for_projection}
     R^2 =\sum_{i=1}^{n} (\textbf{y}_{i}-\textbf{O}^\top \textbf{x}_i)^\top  (\textbf{y}_{i}-\textbf{O}^\top \textbf{x}_i).  
\end{equation}
By minimizing the above equation, we can estimate the optimal $\textbf{O}$ which can then be applied to the original $\textbf{X}$ configuration so it is best matched with $\textbf{Y}$. The measure of the "match" between the two configurations is the minimized value of $R^{2}$ – known as the Procrustes sum of squares.

We adopt this procedure on all posterior samples that we obtain using HMC. As regards the reference configuration, we typically choose the maximum-a-posteriori solution, so that all other points configurations are rotated to match this one. As an exception, in the simulation studies that follow in the next section, we use the true positions as a reference whenever these are available. We emphasise again that this operation does not change the qualitative interpretation of each individual solution, in that the posterior value associated to any given sample remains unchanged before and after the Procrustes matching.

\section{Simulation study} \label{Simulation study}
We implement a simulation study to assess the performance of the TSLPM model in estimating all the model parameters. For this goal, we run a simulation study on a variable number of time series with different sample sizes, so that a multitude of settings is examined.

Since our parameters' priors are noninformative, we devise a data-generating process that is capable of generating realistic time series data, as follows:
\begin{gather*}
    \alpha \sim Uniform(0,3) \notag \\
     \boldsymbol{\beta} \sim Uniform(-1,1) \notag \\
     \text{\textbf{Z}} \overset{\mathrm{iid}}{\sim} Normal(0, \sigma^2) \notag \\
\end{gather*}
The intercept parameter $\alpha$ is generated from a uniform distribution on a narrow interval, specifically chosen to prevent the series from experiencing exponential growth. Similarly, the parameter $\boldsymbol{\beta}$ is simulated from a uniform distribution, with values ranging from $-1$ to $1$. The rationale behind this choice is also to avoid degeneracy, and the specific choice of values is motivated by the results in Proposition-\ref{Prop. Ergodicity of log-linear model} in Section \ref{Stationarity Conditions} of the paper.

In the proposed model, a two-dimensional latent space is employed to facilitate the visualization of network structure. The latent positions, denoted as $\mathbf{z}_{1}, \dots, \mathbf{z}_{N}$, are generated using independent and identical distributions within this space. Each component $z_{ik}$ of a latent position $\mathbf{z}_i$ follows a normal distribution with a mean of zero and variance $\sigma^2 > 0$. 
Initially, the variance $\sigma^2$ for these normal distributions is set to a small value, specifically $0.01$.
This small value makes it very likely that the stability conditions set in Section \ref{Stationarity Conditions} are satisfied. So, the initial latent space that we generate gives stable series, but also it exhibits an almost-independence of the series, due to the interaction terms being almost zero.
Thus, before proceeding with the time series generation, we transform the latent space to make it more interesting, and closer to a borderline setting. We do this by iteratively adjusting the latent space, by expanding the latent positions by a small factor. This iterative adjustment is continued until the derived interaction matrix fail to meet the stability conditions outlined in Section \ref{Stationarity Conditions}. With this expansion procedure, we are still guaranteed to satisfy the stability conditions for the time series, while making the datasets more interesting since the latent space would actually play a stronger role. This contrasts with a setting where latent positions are all close to the center of the space, the latent space effect remains weak, and the generated series are independent. 

We use the STAN programming language (\textcite{STAN_SOFTWARE}) to formulate our statistical models and to perform parameter estimation.
For this purpose, we utilized two distinct approaches: an optimization procedure, notably the LBFGS algorithm and a Bayesian method, specifically Hamiltonian Monte Carlo (HMC). STAN implements a framework to use efficiently both procedures.

In the subsections below, we go through some synthetic data experiments that we conducted to explore the performance of the TSLPM in parameter estimation under varying network sizes and time series sample sizes, and in prediction under varying network sizes.

\subsection{Estimation performance using LBFGS}  \label{EST_performance_sim_study}
\begin{figure}[htbp]
    \centering
     \includegraphics[width = 0.496 \textwidth]{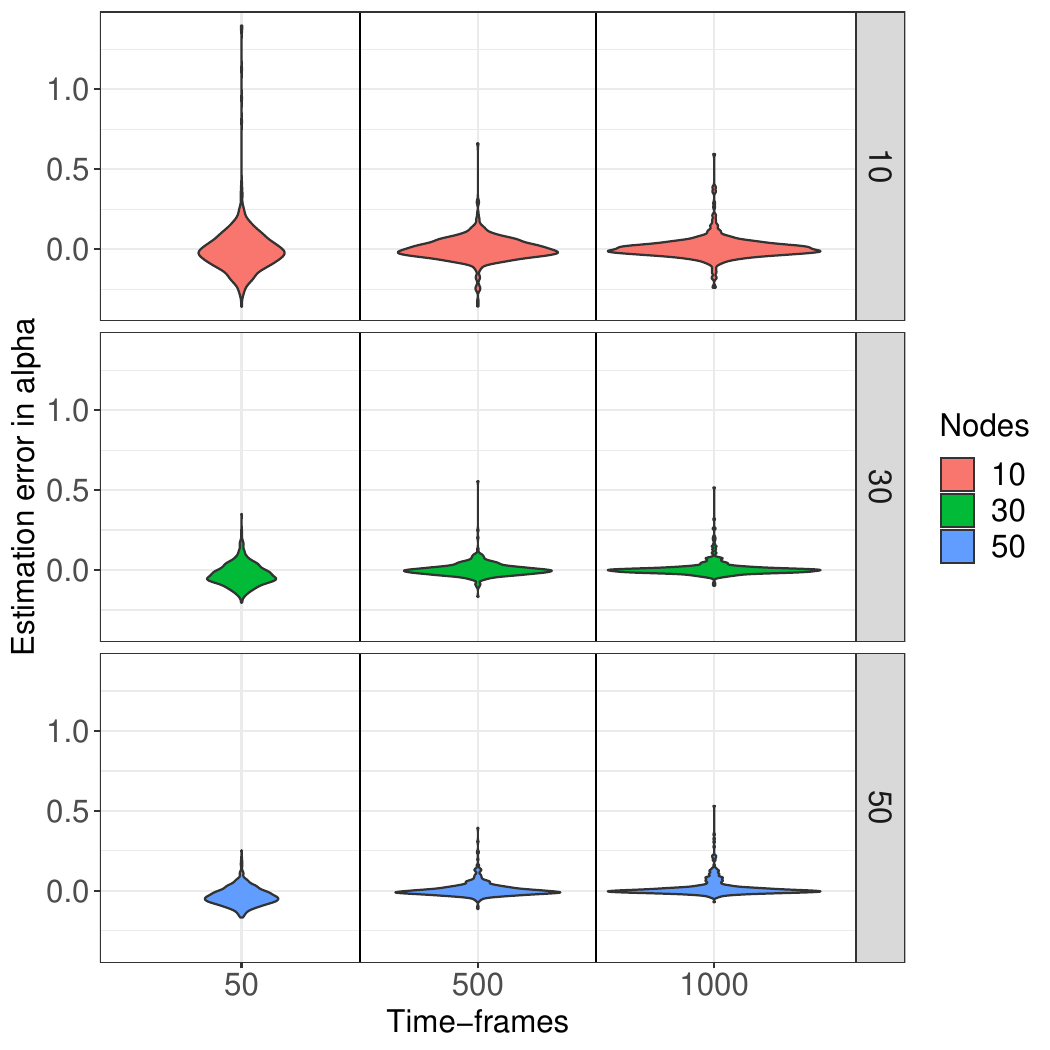}
     \includegraphics[width = 0.496 \textwidth]{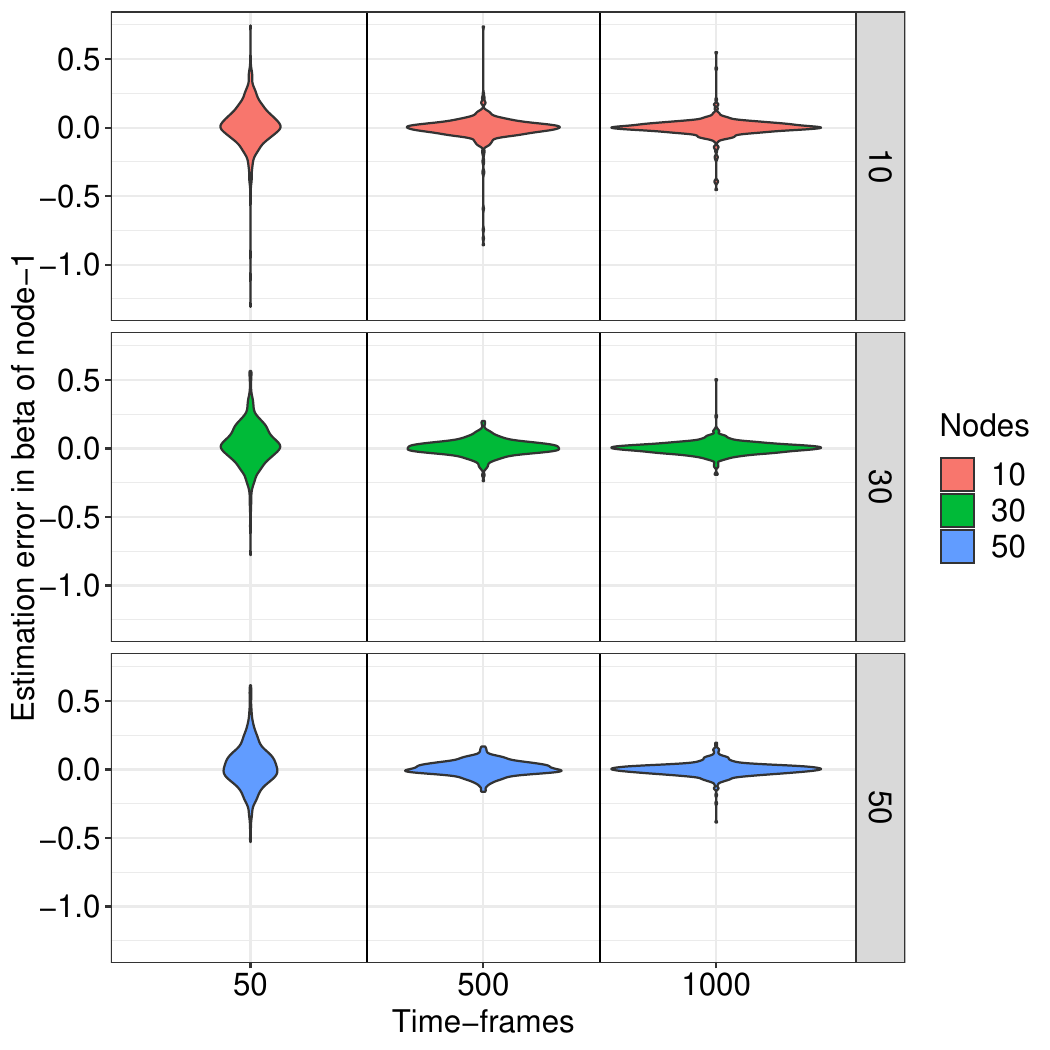}
    %\vspace*{-3mm}
    \caption{Estimation errors across 500 different network datasets: Left panel shows the error for the parameter $\alpha$, and the right panel shows the error for the parameter $\boldsymbol{\beta}$, with node counts $N =\{10, 30, 50\}$ and time points $T = \{50, 500, 1000\}$.}
    \label{fig: estimation_error_alpha_casestudy1}
\end{figure}

\begin{figure}[htbp]
    \centering    
    \includegraphics[width = 0.496 \textwidth]{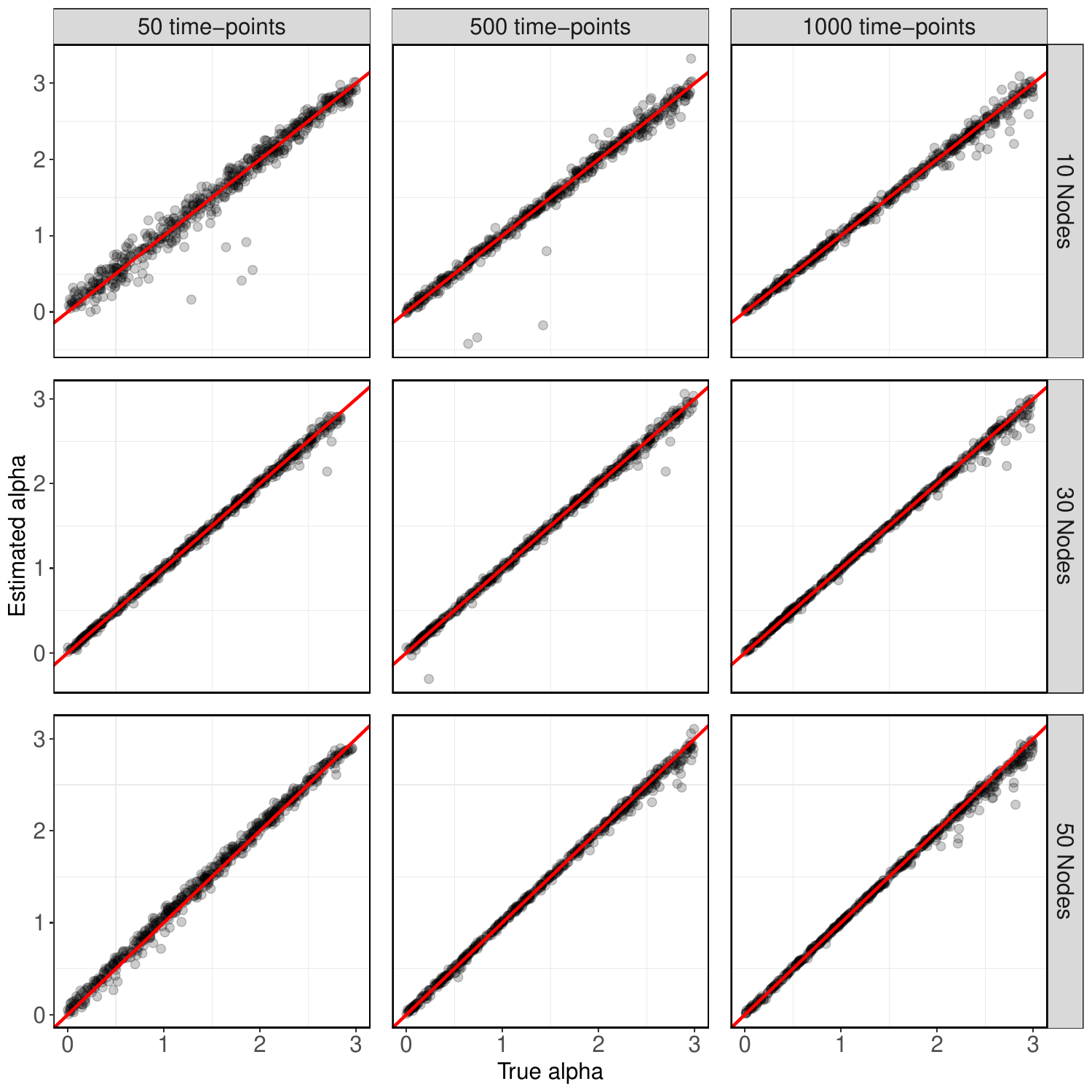}
    \includegraphics[width = 0.496 \textwidth]{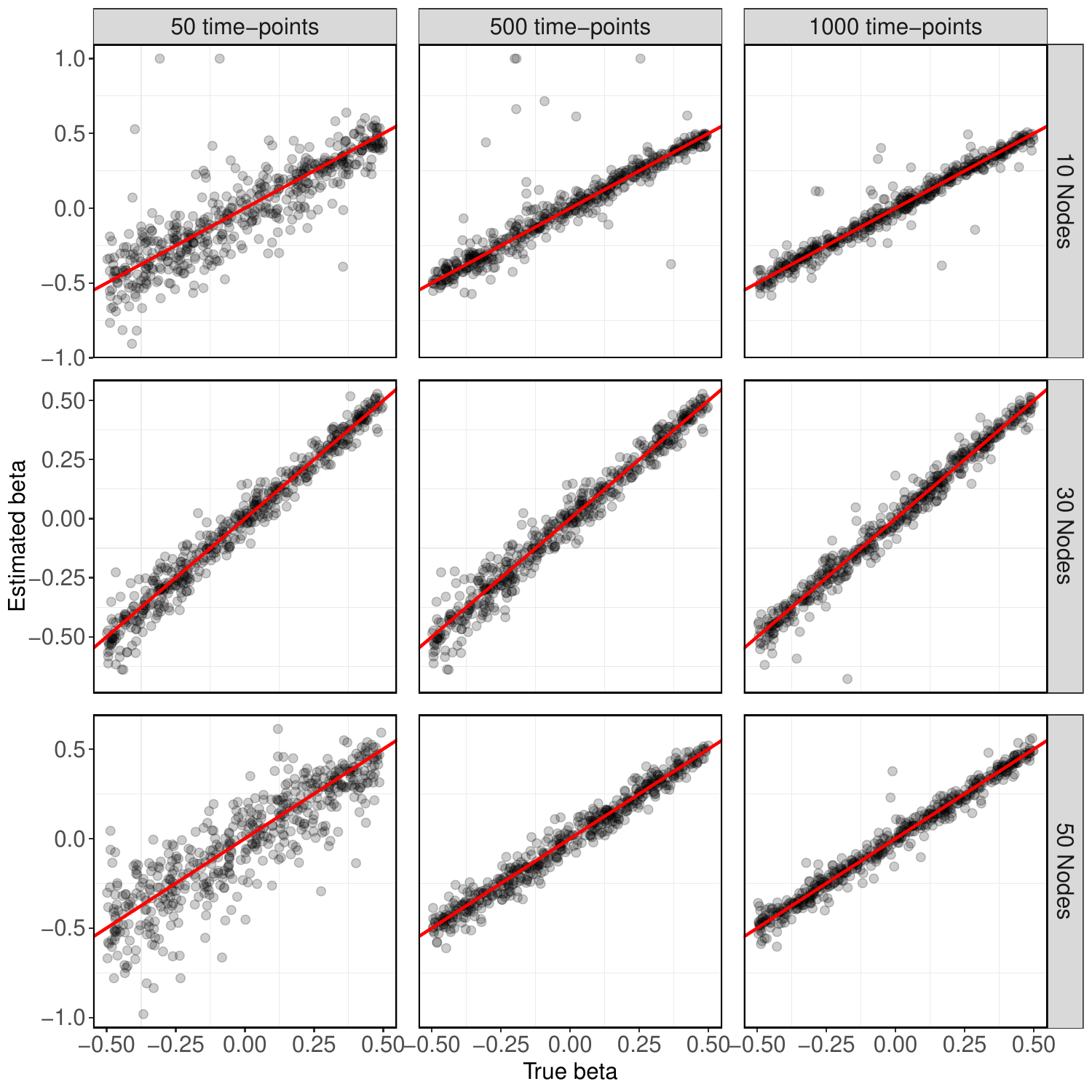}
    \caption{True values (x-axis) versus estimated values (y-axis) for the parameters across 500 different network datasets with nodes $N =\{10, 30, 50\}$ and time points $T = \{50, 500, 1000\}$. Left: $\alpha$, Right: $\beta$}

    \label{fig: truevsest_alpha}
\end{figure}

We consider datasets with $N =\{10, 30, 50\}$ and time frames $T = \{50, 500, 1000\}$ and we simulate multivariate time series datasets with associated network structures using the TSLPM. In the simulation procedure, we generate $500$ different realizations of a multivariate time series with different random seeds, for each choice of $N$ and $T$. In this study, we employ the LBFGS optimization procedure to obtain point estimates of the parameters by maximizing the joint posterior from the model. We illustrate the estimation error for the parameters for $\alpha$ and $\boldsymbol{\beta}$ using violin plots in Figure \ref{fig: estimation_error_alpha_casestudy1}, across the $500$ generated datasets. 
Similarly, we report a comparison between true and estimated values in Figure~\ref{fig: truevsest_alpha}.
Based on the simulated results, we can draw the following conclusions. The violin plots in Figure \ref{fig: estimation_error_alpha_casestudy1} show some trends as $N$ and $T$ change. These plots illustrate a notable reduction in estimation errors as the sample points increase for both parameters $\alpha$ and $\boldsymbol{\beta}$. Additionally, with the expansion of network sizes, the range of estimation errors also demonstrates a decreasing pattern. Analogous conclusions can be drawn from the analysis of scatter plots in Figure \ref{fig: truevsest_alpha}. In datasets with limited sample sizes, the estimated values exhibit a relatively higher level of dispersion, whereas they become more concentrated and approach the true value as the sample size increases. This observation aligns with the expectation that larger sample sizes enhance the accuracy and precision of parameter estimation. These findings highlight the effectiveness of the LBFGS optimizer in achieving satisfactory parameter estimation results, even in the presence of just limited data. 

Another noteworthy observation emerges from the examination of scatterplots comparing the true and estimated values of $\alpha$ in Figure \ref{fig: truevsest_alpha}. In datasets with larger sample sizes, a larger error is evident for higher values of $\alpha$. %This phenomenon can be attributed to overdispersion, a characteristic often encountered in count data scenarios where the variance of the counts surpasses their mean. 
The higher value of $\alpha$ leads to a systematic deviation in the estimated values towards higher magnitudes, which introduces challenges in accurately estimating the parameter $\alpha$. This reaffirms the rationale behind the imposed bounds on $\alpha$ during the data generation process. 

\begin{figure}[htbp]
    \centering
    \includegraphics[width = 0.6 \textwidth]{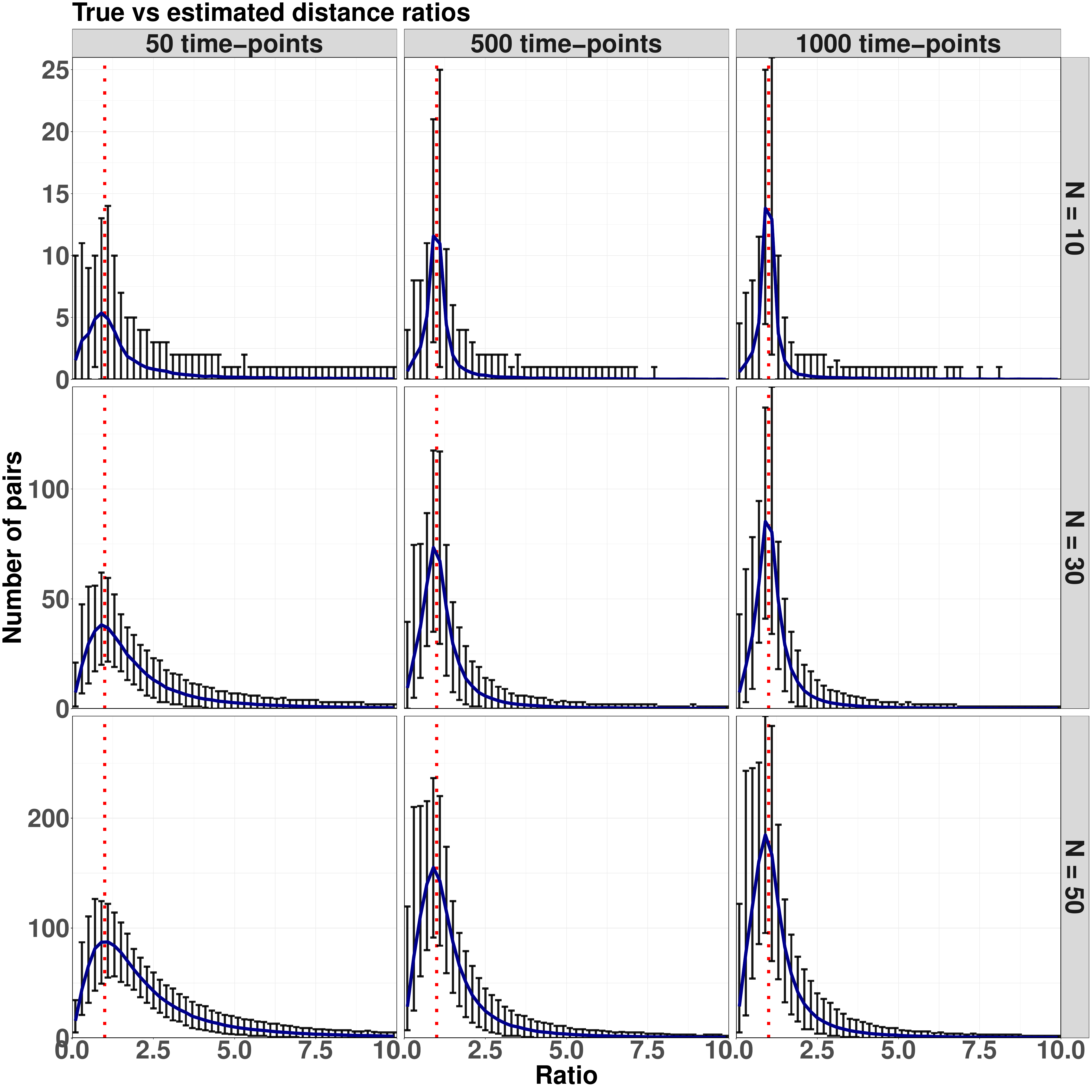}
    \caption{Distribution of pairwise distance ratios, comparing estimated latent positions with true latent positions among $500$ different network datasets with nodes $N =\{10, 30, 50\}$ and time points $T = \{50, 500, 1000\}$. }
    \label{fig: ratio_plot_lp}
\end{figure}

The plot in Figure \ref{fig: ratio_plot_lp} displays the average curve for the distribution of ratios of pairwise distances, similarly to the approach used by \textcite{Sewell20151646, Sewell2016105}. 
The distance ratios of $N$ nodes are calculated by taking the ratio of the pairwise distances between the estimated latent positions and the pairwise distances between the true latent positions. So for each dataset we look at $ \frac{\lVert \hat{\mathbf{z}_{i}} - \hat{\mathbf{z}_{j}} \rVert} { \lVert \mathbf{z}_{i} - \mathbf{z}_{j} \rVert }$, giving us the distribution curve consists of $\frac{N(N-1)}{2}$ such ratios and a total of $500$ curves for $500$ different realizations. A narrow resultant plot centered at $1$ conveys that the obtained latent space is quite similar to the true latent space.  

The bars on the graph of Figure \ref{fig: ratio_plot_lp} represent the $95\%$ quantile, indicating the number of pairs that fall within a specific ratio range. It is notable that these average distributions are consistently centered around $1$, as indicated by a red dotted line in the plot. Moreover, as the sample points increase, these distributions exhibit a remarkable narrowing around the central value of $1$. This trend holds across various network sizes, underlining the robustness of the estimation results for the latent space.

\subsection{Forecasting} 
In this section, we employ forecasting as an evaluation tool to test the predictive performance of TSLPM, comparing it against two established models - the Vector AutoRegression (VAR) \parencite{sims_1980,Lutkepohl_1999} and the Poisson Network AutoRegression (PNAR) (\textcite{armillotta2023}), across various network sizes. 
Our comparative analysis involves out-of-sample testing at each node using distinct datasets. It is important to note that the VAR model does not account for network structure explicitly, whereas the PNAR model incorporates and leverages a \textit{known} underlying patterns or mechanisms that govern node connections. We emphasize that our TSLPM is similar to the PNAR approach in that it uses a network structure to determine the interactions, however, differently from PNAR, our method assumes that the network is not observed.

\subsubsection{Predictions for a small-sized network}
We consider here an initial setting where we generate small-sized $5$-node networks, with a time series observed on each node for $1000$ observations each. A total of $300$ datasets are generated with this format. For each simulated network, we use different seeds and parameter specifications, employing the same data-generating process as discussed in Section \ref{Simulation study}. We fit each dataset using our TSLPM, but also two similar other methods, i.e. PNAR (\textcite{armillotta2023}) and VAR \parencite{sims_1980,Lutkepohl_1999}. 

All the predictions that we make are intended as $h$ steps ahead, meaning that once we reach the last available observation, we predict the following $h$ values of each series, for some $h\geq 1$. We use two distinct techniques to make the $h$ steps ahead forecast of the time series. The first method uses one-step forecasting steps, where predictions are generated one step at a time for future periods, each time using the real observed time series. This setup mimics a live prediction system where more data becomes available for each prediction. In contrast, the second approach, which we refer to as multi-step forecasting, entails generating forecasts for multiple future time steps simultaneously within a single computation cycle. Since in this simulation study we possess the true generated data, we proceed by splitting the data into training set and test set. The first $80\%$ of the observations are used as training, whereas the remaining $20\%$ is used as test.

\begin{figure}[htbp]
    \centering
    \includegraphics[width = 0.9 \textwidth]{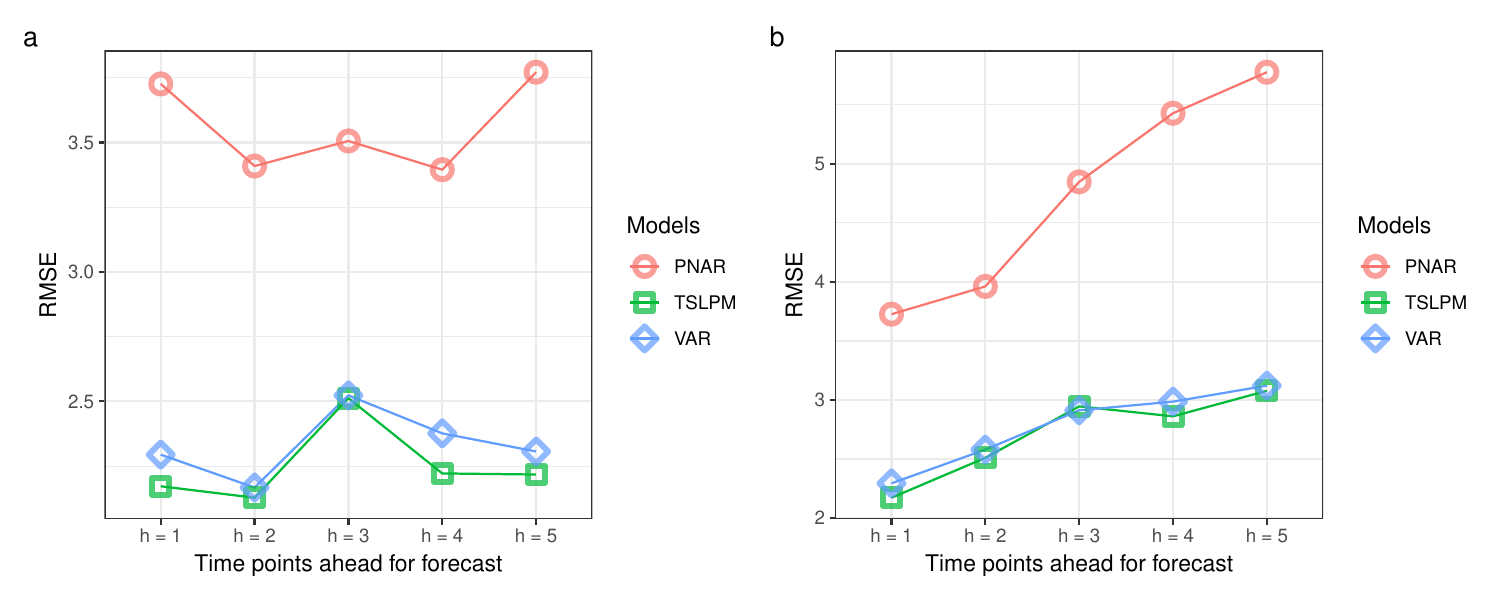}
    \caption{Forecasting, case $N = 5$. RMSE of three models namely TSLPM, VAR, and PNAR in small-sized networks at each h-step ahead in the future using two different forecasting algorithms. a) One-step forecasting algorithm, and b) Multi-step forecasting algorithm.}
    \label{MAE_ERRORBARPLOT_N5}
\end{figure}

For all models and for all prediction tasks, we calculate the conditional means to obtain point estimates, i.e. $E(\mathbf{y}_{t+h|t}) = \hat{\mathbf{y}}_{t+h}$.
Then, to measure the accuracy of our predictions, we consider the Root Mean Squared Error (RMSE). We evaluate the RMSE for $h=1, \dots, 5$, for all three models using both one-step (left panel) and multi-step forecasting algorithms (right panel) in Figure \ref{MAE_ERRORBARPLOT_N5}. 
The evaluation and comparison of each model's performance involve estimating these errors at each future time point. In both the left and right plots of Figure \ref{MAE_ERRORBARPLOT_N5}, the mean of $300$ different simulated time series observed on Node-1 of the networks is depicted. These plots consistently show that the PNAR models yield the highest RMSE at each future time point. This is quite unexpected since, in this application, PNAR uses the true underlying network structure as input. By contrast, slight variations are observed between the VAR and TSLPM models, with TSLPM demonstrating marginally better performance at certain time points in both one-step and multi-step forecasts. The similarity of results between TSLPM and VAR is not particularly suprising because, ultimately, our TSLPM is a variant of VAR for count data where we add more structure on the interaction terms.

\begin{figure}[htbp]
    \centering
    \includegraphics[width = 0.9 \textwidth]{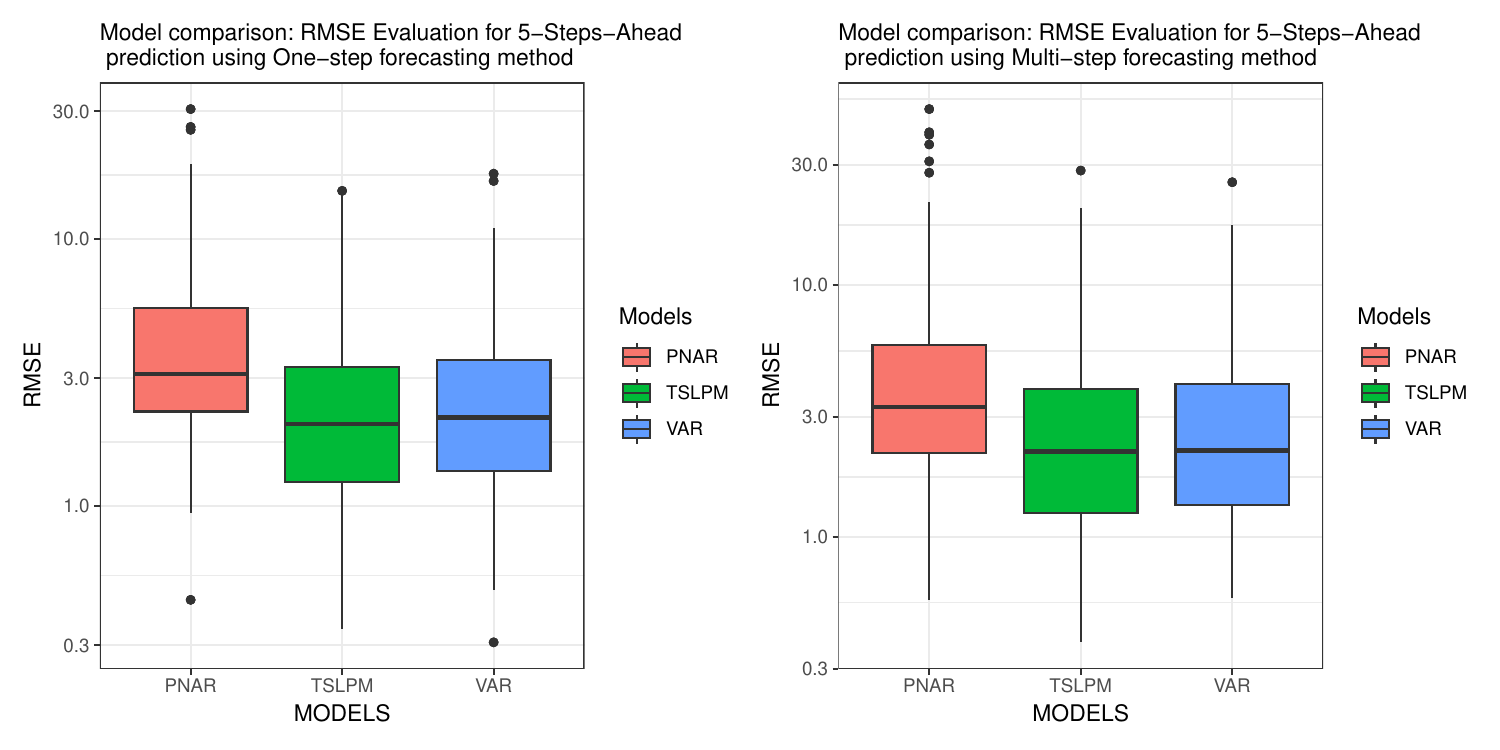}
    \caption{Forecasting, case $N = 5$. RMSE evaluation in small-sized networks using TSLPM, VAR, and PNAR models across two future time points using two different forecasting algorithms. Left: $5$-steps ahead prediction using a five $1$-step ahead forecast (aggregated error). Right: $5$-steps ahead prediction using a multi-step approach (aggregated error).}
    \label{RMSE_BOXPLOT_M5}
\end{figure}

In Figure \ref{RMSE_BOXPLOT_M5} we plot instead the distribution of the RMSE as a boxplot for each model. 
In this case, the panels show only the $5$-step ahead forecast that is either obtained by cumulating five $1$-step ahead predictions or via a joint $5$-steps ahead prediction. In no scenario the PNAR model outperforms the other two models, with similar distributions of errors for the VAR and TSLPM, echoing our findings from Figure \ref{MAE_ERRORBARPLOT_N5}.

\subsubsection{Moderately sized network}
\begin{figure}[htbp]
    \centering
    \includegraphics[width = 0.8 \textwidth]{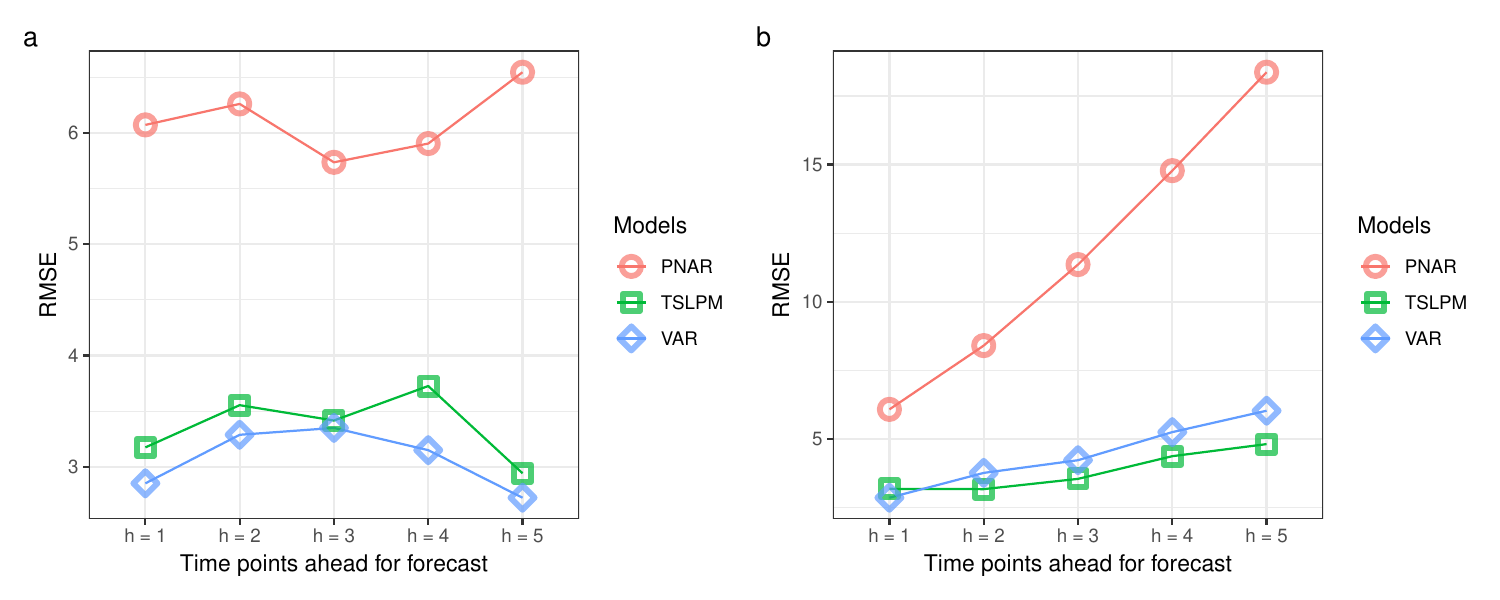}
    \caption{Forecasting, case $N = 30$. RMSE of three models namely TSLPM, VAR, and PNAR in small-sized networks at each h-step ahead in the future using two different forecasting algorithms. a) One-step forecasting algorithm, and b) Multi-step forecasting algorithm.}
    \label{MAE_ERRORPLOT_N30}
\end{figure}

\begin{figure}[htbp]
    \centering
    \includegraphics[width = 0.9 \textwidth]{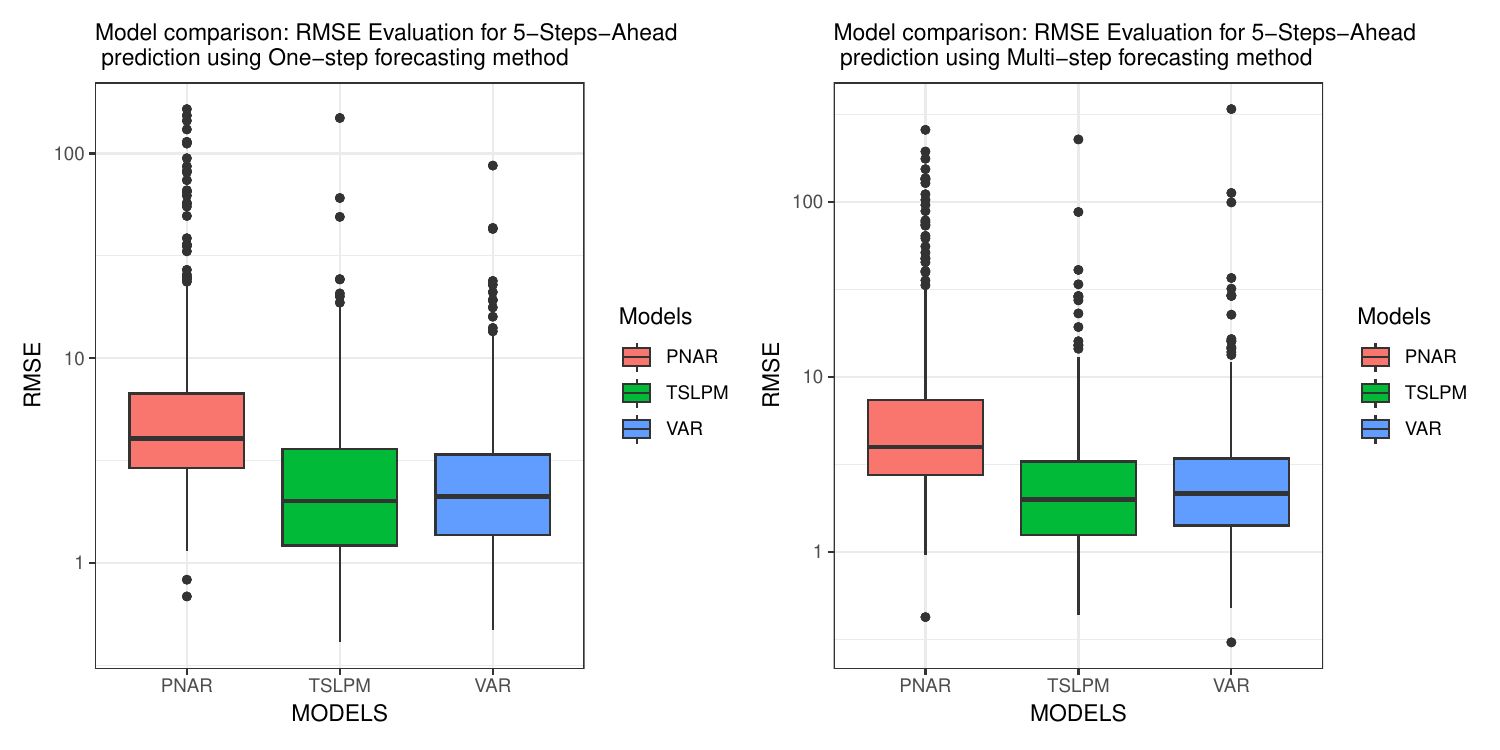}
    \caption{Forecasting, case $N = 30$. RMSE evaluation in small-sized networks using TSLPM, VAR, and PNAR models across two future time points using two different forecasting algorithms. Left: $5$-steps ahead prediction using a five $1$-step ahead forecast (aggregated error). Right: $5$-steps ahead prediction using a multi-step approach (aggregated error).}
    \label{RMSE_BOXPLOT_N30}
\end{figure}

We consider another setting involving a moderately sized network comprising of $30$ nodes, each with a time series observed over a length of $T=1000$. The simulation regime used in generating the time series on the nodes of the networks, as well as the approach used to generate the forecast, remains unchanged from that used for small-sized networks. We replicate the same plots to compare the performance of the models.
In Figure \ref{MAE_ERRORPLOT_N30}, on average, the RMSE is smaller in TSLPM compared to both the other models using a multi-step forecasting algorithm. 
The distributions of RMSE in Figure \ref{RMSE_BOXPLOT_N30} convey similar findings regarding the models as observed in the small-sized networks. The PNAR model continues to perform poorly, while the distribution of errors appears similar for VAR and TSLPM. 

\subsubsection{Large network}
\begin{figure}[htbp]
    \centering
    \includegraphics[width = 0.8 \textwidth]{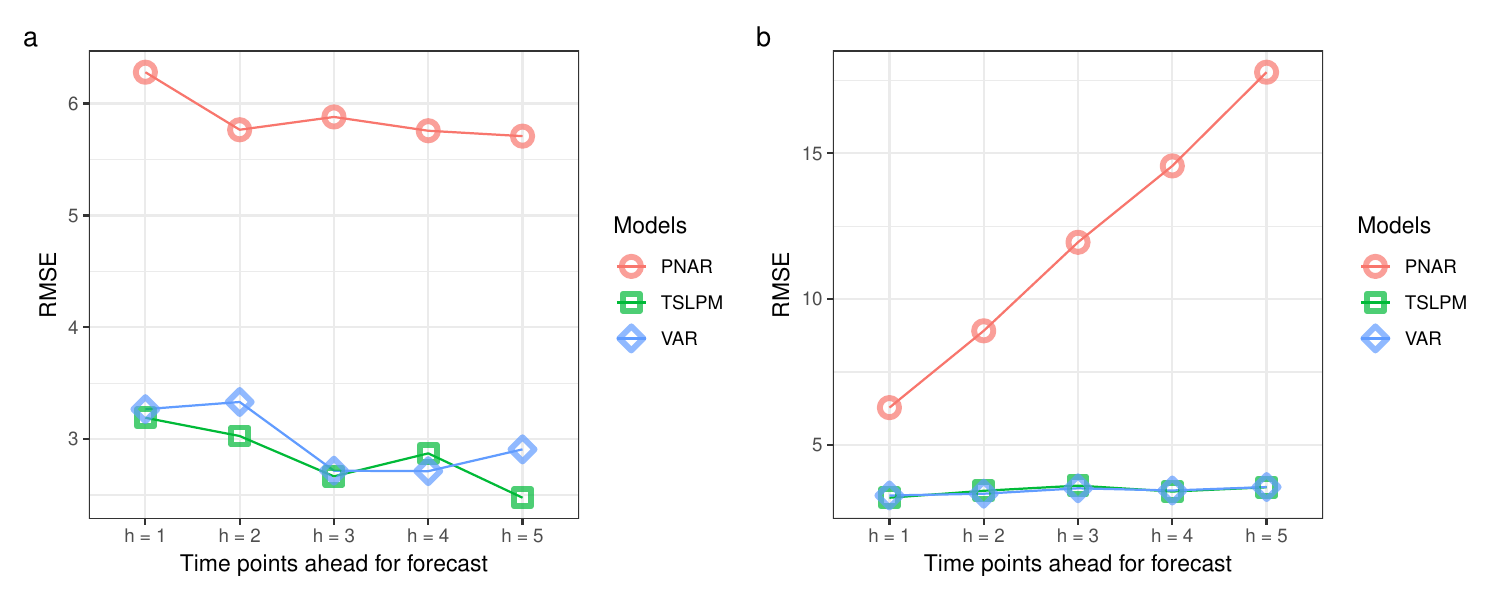}
    \caption{Forecasting, case $N = 50$. RMSE of three models namely TSLPM, VAR, and PNAR in small-sized networks at each h-step ahead in the future using two different forecasting algorithms. a) One-step forecasting algorithm, and b) Multi-step forecasting algorithm.}
    \label{MAE_LINEERRORPLOT_N50}
\end{figure}

\begin{figure}[htbp]
    \centering
    \includegraphics[width = 0.9 \textwidth]{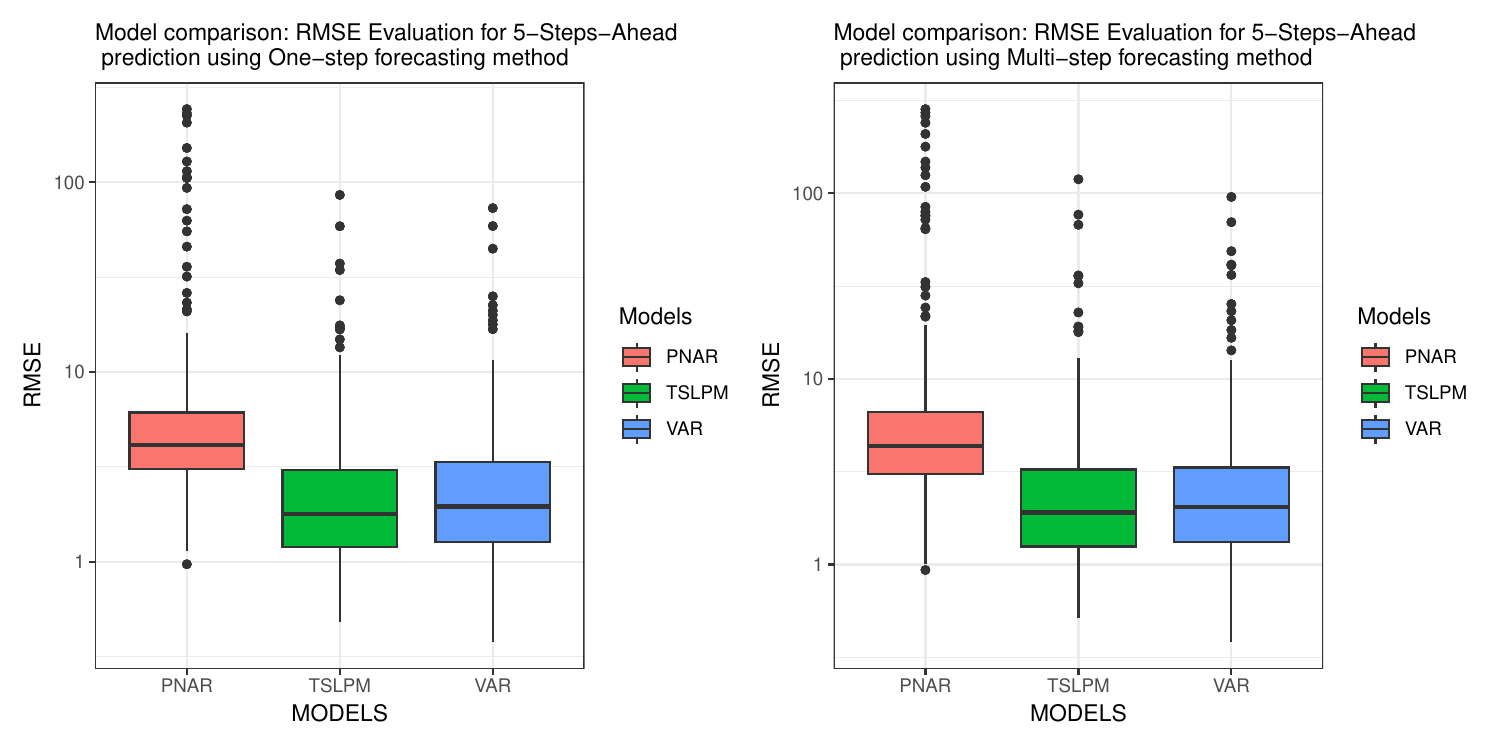}
    \caption{Forecasting, case $N = 50$. RMSE evaluation in small-sized networks using TSLPM, VAR, and PNAR models across two future time points using two different forecasting algorithms. Left: $5$-steps ahead prediction using a five $1$-step ahead forecast (aggregated error). Right: $5$-steps ahead prediction using a multi-step approach (aggregated error).}
    \label{RMSE_BOXPLOT_N50}
\end{figure}

We now assess the predictive performance of the model to examine networks comprising of $50$ nodes, each with a time series observed over a length of $T=1000$. The simulation regime and forecasting approach employed remains consistent with those used for smaller networks. We reproduce the same type of plots to facilitate a comparative analysis of the model's performance.
The Figures \ref{MAE_LINEERRORPLOT_N50} and \ref{RMSE_BOXPLOT_N50} depict whether the errors are at individual time points in the future or aggregated up to five steps ahead. The PNAR model consistently exhibits poorer performance, while the VAR and TSLPM demonstrate similar predictive accuracy.

\section{Chicago crime data} \label{real datasets}
We consider now the analysis of burglary data within the south city of Chicago, Illinois, spanning from 2010 to 2015, utilizing crime data sourced from \textcite{CLARK2021100493}\footnote{\url{ https://github.com/nick3703/Chicago-Data}}. The dataset comprises counts of burglaries across $N = 552$ census block groups for $72$ months. For each census block, we observe the monthly time series representing the number of burglaries recorded. In addition, we have geographical data from which we can create a network of neighbor census blocks: two blocks are linked by an undirected edge if they share a border.

The south side of Chicago is characterized by a relative racial and socio-economic homogeneity which is distinct from the rest of America. Although restricting the dataset to south Chicago reduces some socio-economic variability, residual differences persist. In order to understand these variations, we incorporate covariates into our model to reveal unique socio-economic and demographic characteristics for each region, while exploring potential relationships among census block groups to uncover underlying shared characteristics. Specifically, we consider population size, the number of young males (15-20 years old), and the mean percentage of unemployed individuals. The maximum number of burglaries within a given month in a census block is $17$. Additionally, there is a clear seasonality trend in the data, as already documented in \textcite{CLARK2021100493, armillotta2023}.

\subsection{Methodology } \label{crime_methodology}
First, we construct a similarity matrix using the geographical location of the census block groups, resulting in a network adjacency matrix. This matrix describes which census blocks are geographically adjacent (i.e. they share a border). Through this process, we construct a $552 \times 552$ binary adjacency matrix, effectively mapping the direct neighborhood connections among the block groups.
Since our method cannot handle $552$ nodes due to computational limitations, we employ a network clustering algorithm to simplify the network, enabling us to examine the connections between a few macro-areas of Chicago that are comprised of clusters of block groups. For the sake of clarity in the subsequent discussion, these clusters of block groups will henceforth be referred to as ``regions''.

\begin{figure}[htbp]
\centering
\includegraphics[width = 0.49 \textwidth]{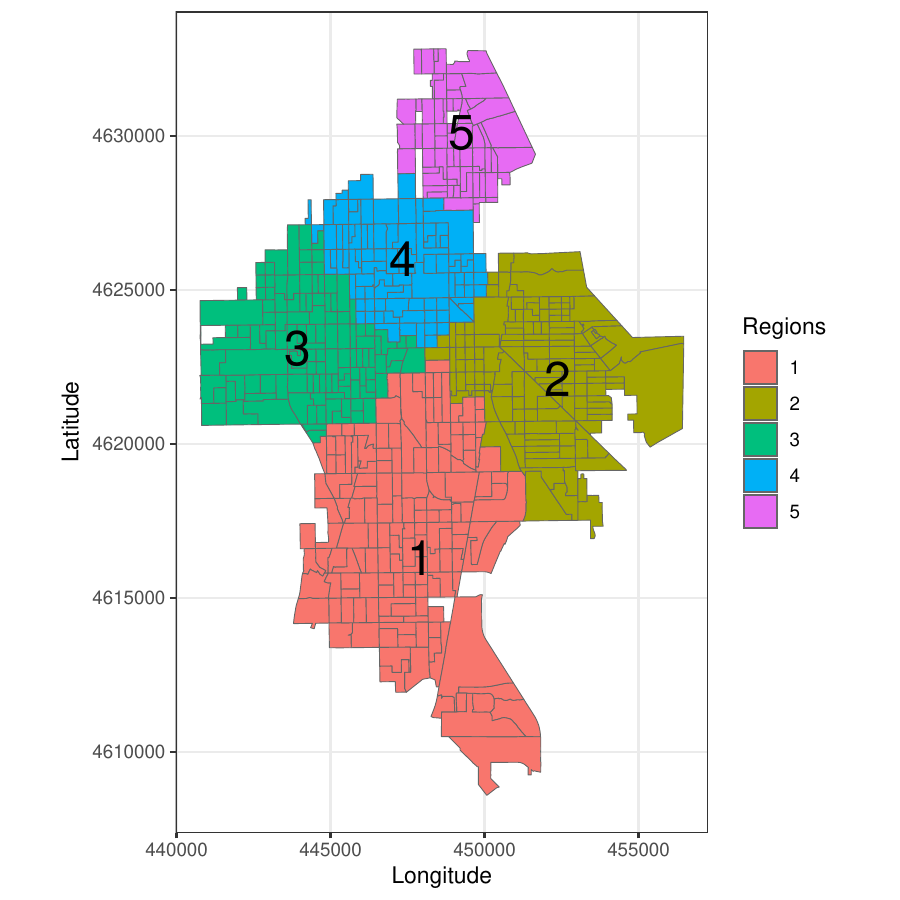}
\includegraphics[width = 0.49 \textwidth]{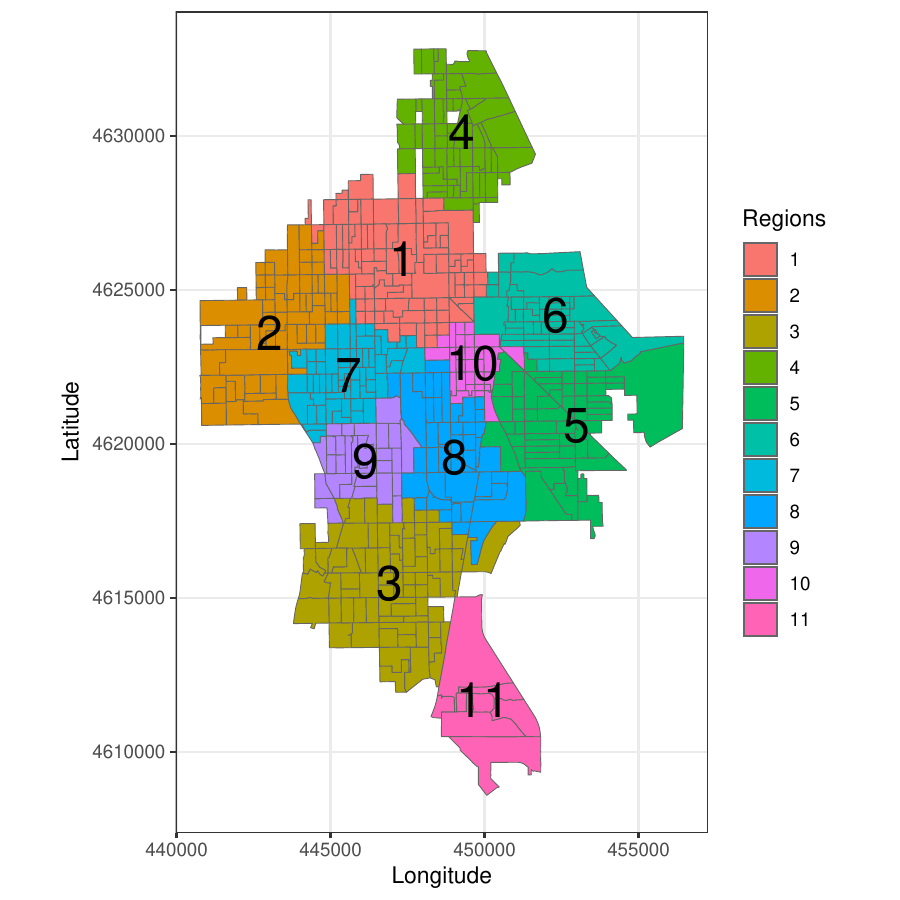}
\caption{Clustering results on the network of $552$ census block groups in South Chicago, with each color representing a different aggregated region. Left: aggregation into $5$ regions. Right: aggregation into $11$ regions.}
\label{Network_gna}
\end{figure}

This aggregation reduces the computational problems and it facilitates the interpretation of the results. Among the various clustering algorithms available, we choose the Girvan-Newman algorithm. Since this algorithm relies on edge-betweenness to capture and determine the importance of edges, we believe that the algorithm's mechanism is coherent with the network topology that is induced by geographical information. While we choose to employ the Girvan-Newman algorithm, it is important to note that any clustering algorithm could be used in this context. The specific choice of algorithm is not crucial to our goals; our primary objective is not to identify the most effective clustering but rather to demonstrate a methodology for quantifying and interpreting the relationships between the obtained regions. The clustering approach partitions the network into either five or eleven regions, as shown in Figure \ref{Network_gna}, using the modularity of the network as a measure to determine the optimal number of groups. Eleven regions achieve the highest modularity (0.78), but with only a slight increase from the modularity of five regions (0.72), making the latter a perfectly viable (and perhaps more parsimonious) alternative. Remarkably, the clusters found show excellent cohesion from a geographical point of view, in that none of the clusters are split into disconnected areas. The two partitions are shown in Figure \ref{Network_gna}.

\begin{figure}[htbp]
    \centering
    \includegraphics[width = 0.7 \textwidth]{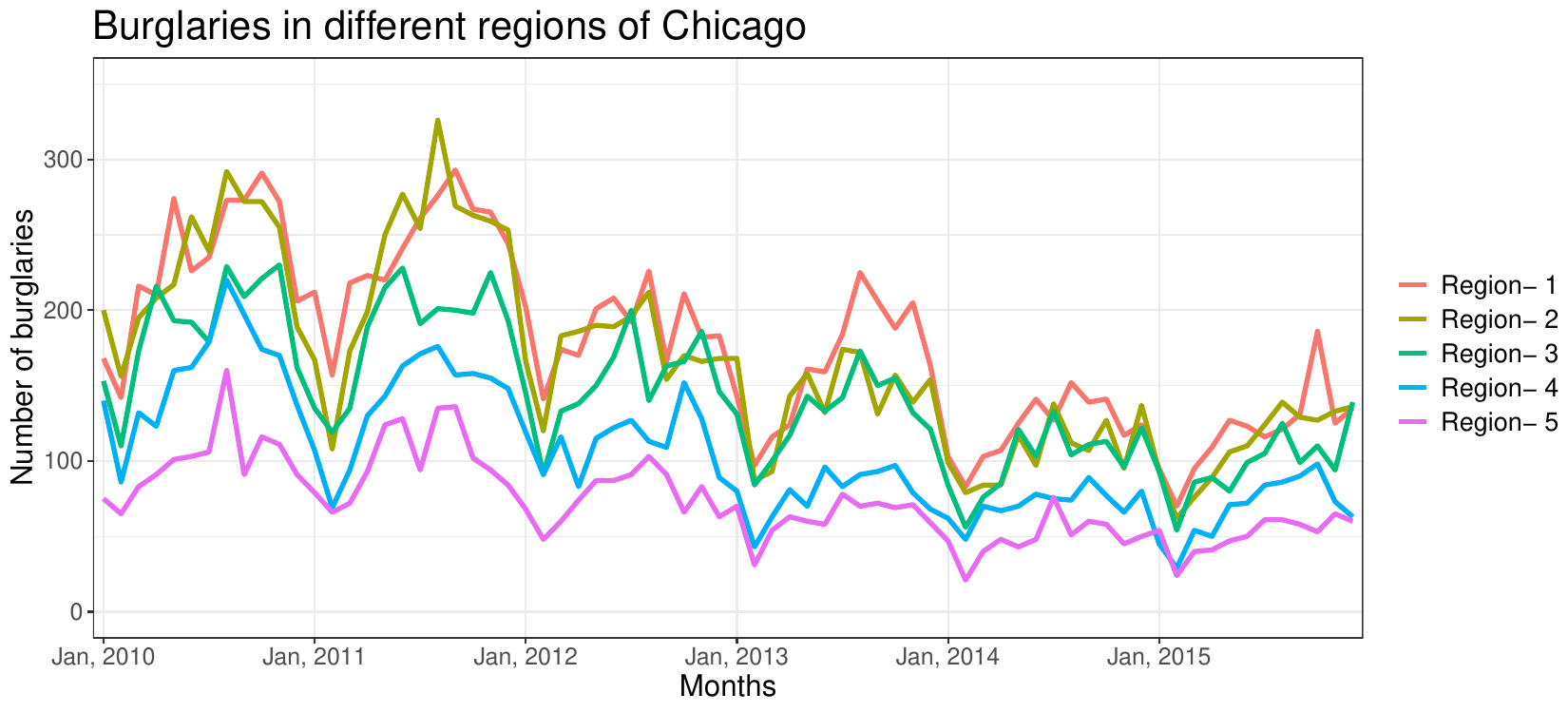}
    \includegraphics[width = 0.7 \textwidth]{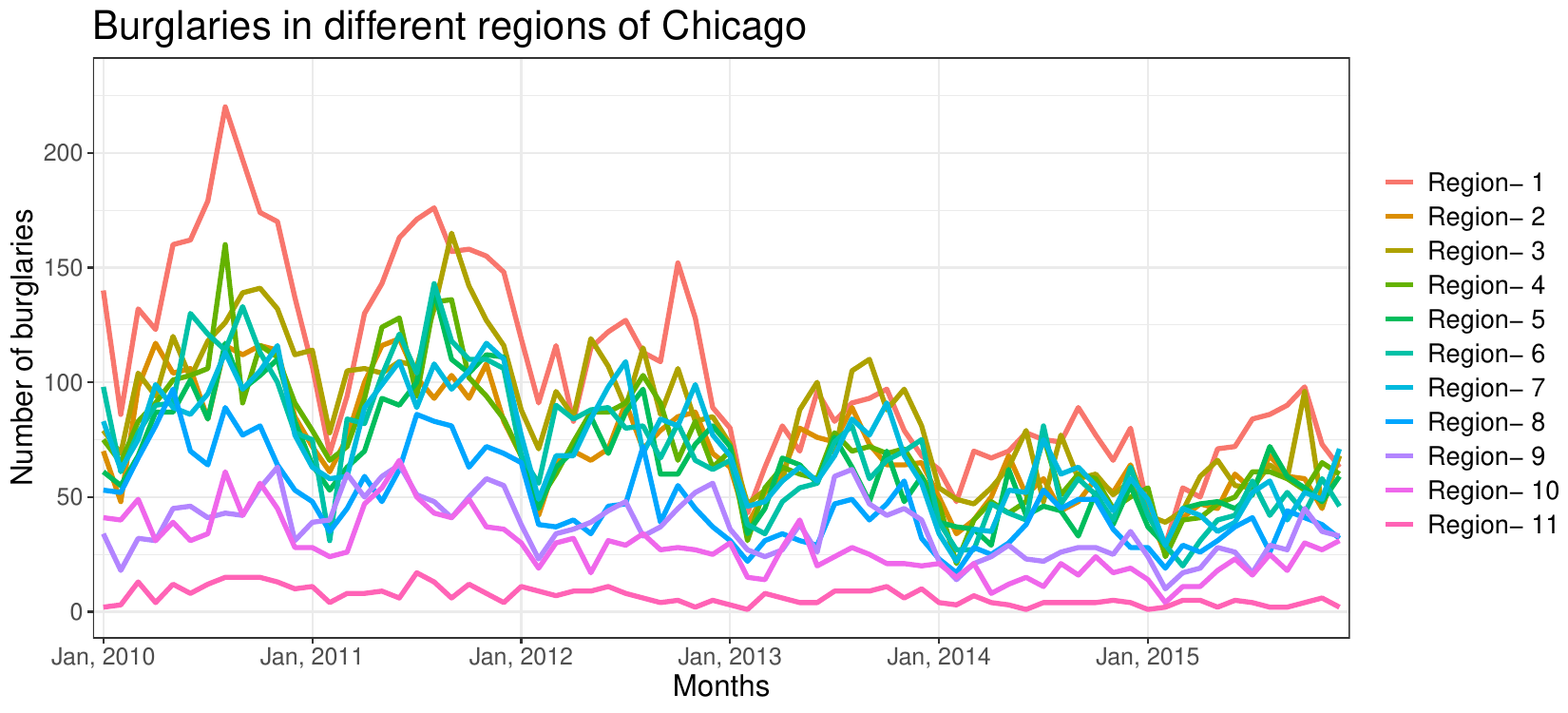}
    \caption{Monthly number of cases of burglaries in the south side of Chicago (2010-2015). Left: aggregated over $5$ regions. Right: aggregated over $11$ regions.}
    \label{time_Series_plot_c5_11}
\end{figure}

We proceed by aggregating the crimes for each block within a region to create an aggregated time series. These aggregated time series reveal a clear annual pattern in crime rates, as illustrated in Figure \ref{time_Series_plot_c5_11}.

\subsection{TSLPM for the crime dataset}

As a template model for the crime data, we extend the model of Eq.~\ref{eq:TSLPM_model} as follows:
\begin{gather}
    y_{it} \sim \text{Pois}(\lambda_{it}) \notag\\
\log(\lambda_{it}) = \alpha_{i} + \beta_{i} \log(y_{i(t-1)} + 1) + \sum_{j} \gamma_{ij} \log(y_{j(t-1)}+1) + \eta_{i} \log(y_{i(t-12)} + 1)+\sum_{k}  \delta_{k} {x}_{ik}
\label{crime_dta_model_withcov_and_seasonality}
\end{gather}
There are some differences with respect to the model of Eq.~\ref{eq:TSLPM_model}, namely: the intercept $\boldsymbol{\alpha}$ may depend on the node, and we introduce a seasonality effect (with a lag of $12$ months) with associated coefficients $\boldsymbol{\eta}$. The parameters $\boldsymbol{\delta}$ can obviously be different for each covariate, but they are the same parameters that are shared across different nodes. This is to simplify the interpretation of the results and to avoid potential non-identifiability issues.

This model essentially postulates that the average rate of change in the number of burglaries is determined by four main components. The first is the number of burglaries that happened last month measured through the parameter $\boldsymbol{\beta}$, the second is the number of burglaries that happened $12$ months before determined by coefficient $\boldsymbol{\eta}$, the third is the number burglaries that happened in the other areas during the previous month quantified through the latent space interactions $\boldsymbol{\gamma}$, and, finally, the fourth is the effect given by the node-specific covariates, captured by the coefficients $\boldsymbol{\delta}$. 

To enhance the predictive accuracy and interpretability of TSLPM, we have incorporated specific covariates that may highlight similarities across the regions. These covariates are the size of the population (hereafter,  \texttt{pop}), the count of young males aged 15-20 years (\texttt{ym}), and the mean unemployment rate (\texttt{unemp}) within each region. These variables were chosen due to their potential impact on crime rate, and, in Eq.~\ref{crime_dta_model_withcov_and_seasonality}, they are indicated with $\mathbf{X} = \{x_{ik}: i=1,\dots, N$ and $k=1,2,3\}$, respectively. 

As regards seasonality, to effectively account for the annual seasonal effect, our analysis starts from Jan 2011 whereas the available data from January 2010 to December 2010 are employed just to calculate the seasonality effects from Jan 2011 onwards. This methodological approach allows us to utilize the initial year's data as a seed for the first observable year in our analysis, ensuring that the model adequately captures the annual seasonal trend from the outset. This strategy not only mitigates the issue of missing initial seasonal data but also enhances the predictive accuracy and relevance of the model by accurately reflecting the cyclical nature of crime rates.

All the computations were implemented in STAN and R. We used HMC for fitting TSLPM to obtain Markov chains for $10,000$ iterations with a burn-in period of $5,000$ iterations, and a thinning rate of $5$, thus obtaining $1000$ samples to represent the posterior distributions of the model parameters. For the model parameters $ \boldsymbol{\alpha, \beta, \eta}, \mathbf{Z}$, and $\boldsymbol{\delta}$, we employ non-informative prior, specifying independent normal priors with a mean of zero and a standard deviation of $100$ for each. This choice of high standard deviation for the priors is deliberate, aiming to minimize the influence of prior choices on the posterior distributions. This approach ensures that the analysis remains less biased by prior assumptions, allowing the data to play a more significant role in shaping the posterior densities.

We used the $\hat{R}$ statistic (also referred to as the potential scale reduction factor) and visual examination of the chains to check for their convergence (\cite{BDA_Gelman2013}). To assess the quality and reliability of the samples of parameter estimates, the Effective posterior Sample Size (ESS) was checked. Divergence transitions for the Markov chains were also checked and eliminated. Sampling efficiency in STAN was managed by setting the tree depth parameter, which controls the simulation steps per iteration in the No-U-Turn Sampler (i.e. the sampler used by STAN). Adjusting this parameter may increase the computational load, so, we have carefully maintained a balance between allowing comprehensive exploration and managing computational resources. Our results all showed good convergence and reliable results for all models fitted.

\subsection{Model selection} \label{model selection}
We proceed our analysis by fitting, on both datasets with $5$ and $11$ clusters, the model in Eq.~\ref{crime_dta_model_withcov_and_seasonality} and a number of nested simpler models.
In order to choose the best model overall, we use the Deviance Information Criterion (DIC), which stands out as particularly suitable for Bayesian multivariate time series models, offering a balanced approach to model selection. We also mention some viable alternatives such as the Widely Applicable Information Criterion (WAIC) proposed by \textcite{watanabe2010asymptotic} and leave-one-out cross-validation (LOO-CV) by \textcite{vehtari2017practical}, although their applicability to models with time dependence is limited due to the potential for including future information in the analysis of past predictions. An alternative, Leave-Future-Out Cross-Validation (LFO-CV), specifically addresses this issue for univariate time series models, see \textcite{burkner2020approximate} for more detailed discussion. However, its use becomes computationally demanding with more intricate models, making DIC a more pragmatic option for many Bayesian time series applications, especially when considering the balance between accuracy and computational feasibility. 

For the TSLPM selection process of Chicago's regional burglary network, we consider a set of $2^{3} = 8$ alternative multivariate time series network models based on a different combination of the set of parameters $\{\boldsymbol{\alpha}, \boldsymbol{\beta}, \boldsymbol{\eta} \}$. Additionally, we integrate three covariates into these models in various configurations individually, in pairs, or all together—yielding a total of $50$ alternative models. This approach allows us to examine how different parameters and covariate combinations influence our understanding of the patterns observed across the regions.

We initiated our analysis by applying the methodology outlined in Section \ref{crime_methodology} for each of the proposed model configurations, generating $1,000$ posterior samples of parameters for each model. Following the assessment of parameter estimates for quality and reliability, we examined potential collinearity to address any non-identifiability issues among variables, utilizing pairs scatter plots for this analysis. This examination revealed the presence of collinearity among some of the covariates, across all model configurations. In particular, the covariates \texttt{pop} and \texttt{ym} exhibited pronounced collinearity, compromising their simultaneous inclusion and thereby rendering the affected models unsuitable for fitting. Moreover, in models where the coefficients $\boldsymbol{\alpha}$ varied across different time series, collinearity among these coefficients was evident using pairs plot, signalling potential non-identifiability issues.

Given these challenges, the initial set of models was refined to a subset of $20$ viable candidates, by making $\alpha$ constant across different series, and by removing one of the covariates. From there, we proceeded to calculate the DIC for each of the $20$ models, for both the $5$ clusters and $11$ clusters datasets.

\subsection{Crime data \texorpdfstring{$5$}{5} clusters} \label{Model_fit_5_regions}
We consider the analysis of the burglaries in five non-overlapping regions of Chicago from year 2011 to 2015. As regards model selection, the analysis of the DIC values from Table \ref{tab:model_comparison_5regions} reveals that the best model is the one where the set of parameters  $\{\boldsymbol{\beta,\eta} \}$ varies across nodes and $\alpha$ is constant. 
\begin{table}[htbp]
\centering
\caption{DIC values for model comparisons for the $5$ regions dataset.}
\label{tab:model_comparison_5regions}
\begin{tabular}{@{}lcc@{}}
\toprule
Model   & DIC for 5 regions \\ \midrule
\{$\beta_{i}, \eta_i, \alpha, \delta_{\texttt{pop}}$\}               & -46138.50   \\
\{$\beta_{i}, \eta_i, \alpha, \delta_{\texttt{ym}}$\}                   & -45297.54      \\
\{$\beta_{i}, \eta_i, \alpha, \delta_{\texttt{unemp}}$\}              & -47787.59  \\
\{$\beta_{i}, \eta, \alpha, \delta_{\texttt{pop}}$\}               & -43940.26  \\
\{$\beta_{i}, \eta, \alpha, \delta_{\texttt{ym}}$\}                  & -43104.12   \\
\{$\beta_{i}, \eta, \alpha, \delta_{\texttt{unemp}}$\}               & -47255.93 \\
\{$\beta, \eta_i, \alpha, \delta_{\texttt{pop}}$\}                 & -43492.49 \\
\{$\beta, \eta_i, \alpha, \delta_{\texttt{ym}}$\}                  & -44010.15 \\
\{$\beta, \eta_i, \alpha, \delta_{\texttt{unemp}}$\}              & -45590.13\\
\{$\beta, \eta, \alpha, \delta_{\texttt{pop}}$\}                   & -41448.41  \\
\{$\beta, \eta, \alpha, \delta_{\texttt{ym}}$\}                       & -41420.87  \\
\{$\beta, \eta, \alpha, \delta_{\texttt{unemp}}$\}                  & -45075.88  \\
\{$\beta_{i}, \eta_i, \alpha, \delta_{\texttt{pop}}, \delta_{\texttt{unemp}} $\}          & \textcolor{red}{-48507.85}  \\
\{$\beta_{i}, \eta_i, \alpha, \delta_{\texttt{ym}}, \delta_{\texttt{unemp}} $\}             & -48397.05    \\
\{$\beta_{i}, \eta, \alpha, \delta_{\texttt{pop}}, \delta_{\texttt{unemp}} $\}             & -45772.26      \\
\{$\beta_{i}, \eta, \alpha, \delta_{\texttt{ym}}, \delta_{\texttt{unemp}} $\}             & -47136.06  \\
\{$\beta, \eta_i, \alpha, \delta_{\texttt{pop}}, \delta_{\texttt{unemp}} $\}           & -46106.72   \\
\{$\beta, \eta_i, \alpha, \delta_{\texttt{ym}}, \delta_{\texttt{unemp}} $\}              & -45903.04  \\
\{$\beta, \eta, \alpha, \delta_{\texttt{pop}}, \delta_{\texttt{unemp}} $\}            & -44293.39   \\
\{$\beta, \eta, \alpha, \delta_{\texttt{ym}}, \delta_{\texttt{unemp}} $\}               & -43942.64 \\ \bottomrule
\end{tabular}
\end{table}
The covariates to include are \{\texttt{pop}, \texttt{unemp}\}. The model formulation takes the form discussed in Eq.~\ref{crime_dta_model_withcov_and_seasonality} with the sole distinction being that the parameter $\alpha$ is held constant across all regions.

% In this study, we utilize an alternative form of the TSLPM model to assess how model parameters affect burglary counts, which takes the form:

% \begin{gather} 
%     y_i^{t} \sim \text{Pois}(\lambda^{t}_{i}) \notag\\
% \lambda^{t}_{i} = \exp(\alpha)  (y^{t-1}_{i} + 1)^{\beta_{i}} \prod_{j}  (y^{t-1}_{j}+1)^{\gamma_{ij}} (y^{t-12}_{i} + 1)^{\eta_{i}} \exp(\sum_{k}  \delta_{k}    {x}_{ik})
% \label{TSLPM_crimedta_alternateform_R5}
% \end{gather}

% The exponential function in the model translates the log-linear relationship into a multiplicative impact in the original scale of $\boldsymbol{\lambda}$. 
% This approach provides a different perspective on the interpretation of parameters, focusing on proportional changes rather than additive changes. This emphasis is particularly meaningful for models predicting rates or counts, like TSLPM.

\begin{figure}[htbp]
    \centering
    \includegraphics[width = 0.45 \textwidth]{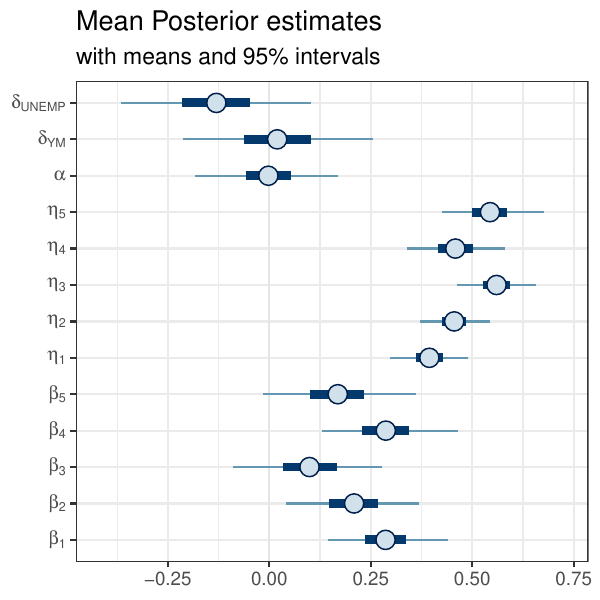}
    \includegraphics[width = 0.45 \textwidth]{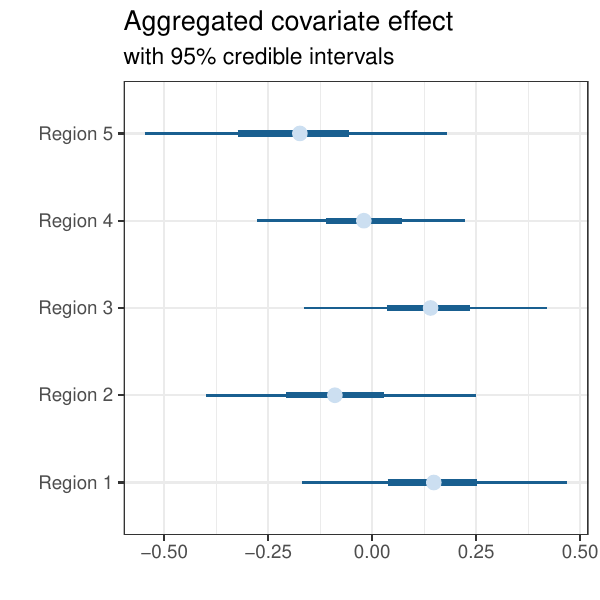}
    \caption{Left: posterior averages with 95\% credible intervals for the model parameters in the $5$ clusters data, incorporating covariates. Right: aggregated mean effects of covariates (population and unemployment) on each region, shown with 95\% credible intervals.}
    \label{EST_95CI_C5_POP_UNEMP_ym_credibleinterval_covariate_effect_C5}
\end{figure}

Figure \ref{EST_95CI_C5_POP_UNEMP_ym_credibleinterval_covariate_effect_C5} shows the posterior summaries for all model parameters except the latent positions. 
These parameters characterise the individual behavior of nodes without taking into account the dependencies between nodes. We can see from these results that all $\boldsymbol{\beta}$ and $\boldsymbol{\eta}$ parameters tend to be positive, which indicates positive autocorrelations and a strong association due to seasonality. This can translate into a noticeable persistence of burglaries over time within the same region.
The seasonality effect seems to be strongest for node $3$, corresponding to the west region, and weakest (although still different from zero) for the south region (node $1$).

The role of external factors, such as population and unemployment, are explored through the expression $ \exp(\sum_{k} \delta_{k} x_{ik})$. This setup of the linear combination of terms allows for the analysis of both collective and individual impacts of these factors on burglary rates. The model predicts that an increase in these covariates, assuming a simultaneous rise by their standard deviations, leads to proportional changes in burglary counts, either enhancing or reducing them based on the direction of change shown in the right panel of Figure \ref{EST_95CI_C5_POP_UNEMP_ym_credibleinterval_covariate_effect_C5}. 

\begin{figure}[htbp]
    \centering 
     \includegraphics[width = 0.45 \textwidth]{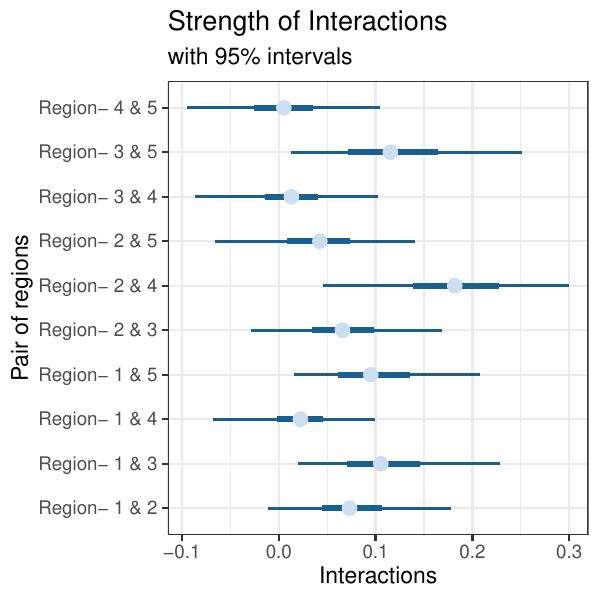}
    \caption{Mean estimates with 95\% credible intervals for pairwise interactions among the five different regions of Chicago.}
    \label{fig: pairwise_interactions_c5}
\end{figure}

We now delve into the detailed analysis of pairwise interactions between the regions, a primary aspect of this study. The integration of the interaction term $\gamma_{ij}$ into our model facilitates this exploration of pairwise interactions among different regions, leveraging the spatial positioning $\mathbf{Z}$ of these regions as a fundamental element for interpreting their interactions. Through this methodology, our analysis offers both quantitative insights and visual representations to discern specific patterns of burglaries on the south side of Chicago, facilitating a deeper understanding of the geographical connections in the incidence of burglaries. 
First, we analyze the posterior distribution of the $\boldsymbol{\gamma}$ parameters, in Figure \ref{fig: pairwise_interactions_c5}.
The credible intervals illustrate the posterior distribution of connections between these regions, where positive values indicate a positive relationship in burglary rates and vice versa. 

As regards the latent positions, the model fit resulted in 1,000 posterior samples, representing the latent positions of each region within a latent space. To align these sampled coordinates for analysis, we applied a Procrustes transformation, as discussed in Section \ref{Procrustes}, using the MAP as a reference for orientation. 
 
\begin{figure}[htbp]
    \centering
        \includegraphics[width = 0.65 \textwidth]{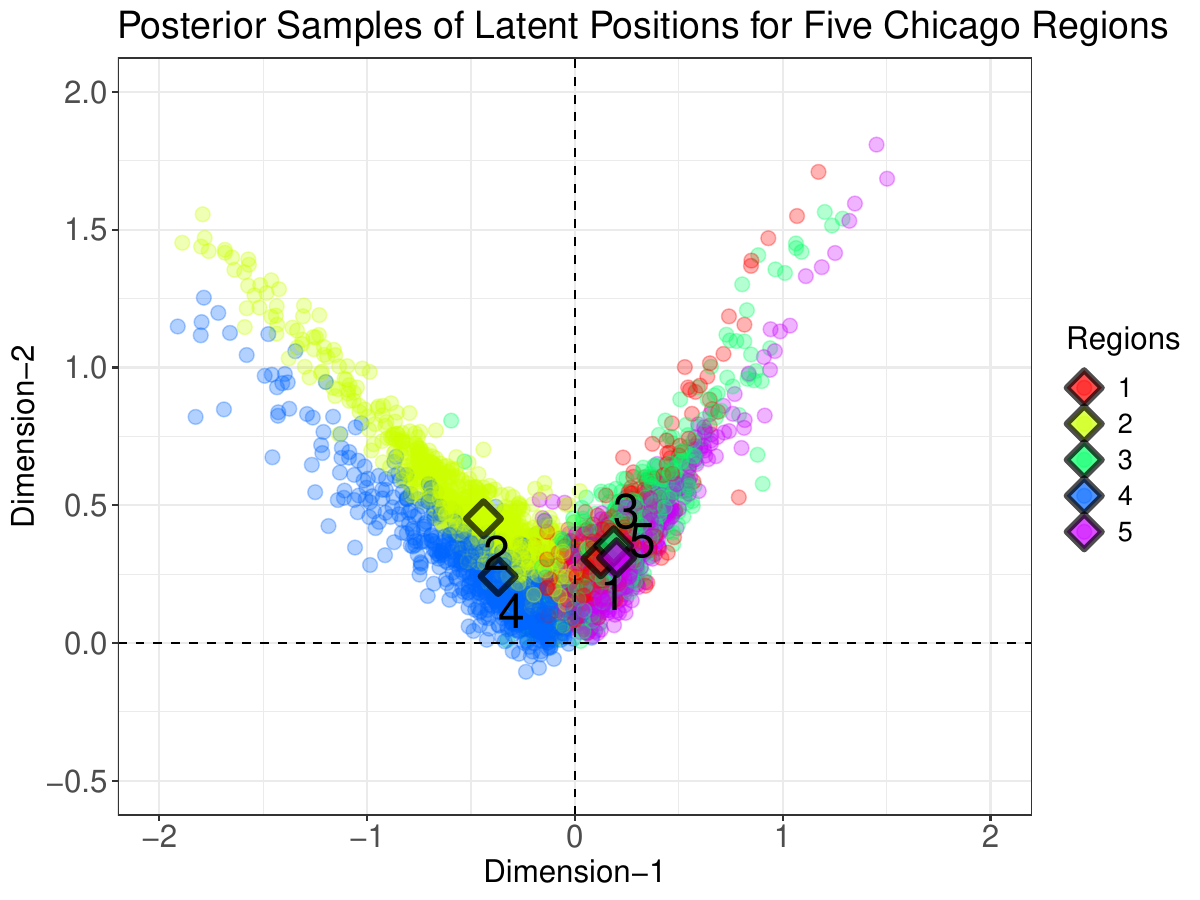}
        \caption{Posterior samples of the latent positions for $5$ different regions of Chicago. the posterior means are also shown in black and labeled in accordance to the region.}
        \label{fig:latent_positions_c5}
    \end{figure}

We show in Figure \ref{fig:latent_positions_c5} the posterior samples made of 1,000 points from the posterior distribution of the estimated latent positions of each region post-Procrustes transformation. 
Here, larger, solid points denote the posterior means, providing a clear visual representation of the estimated final locations of the regions.

\begin{figure}[htbp]
    \centering 
    \includegraphics[width = 0.95 \textwidth]{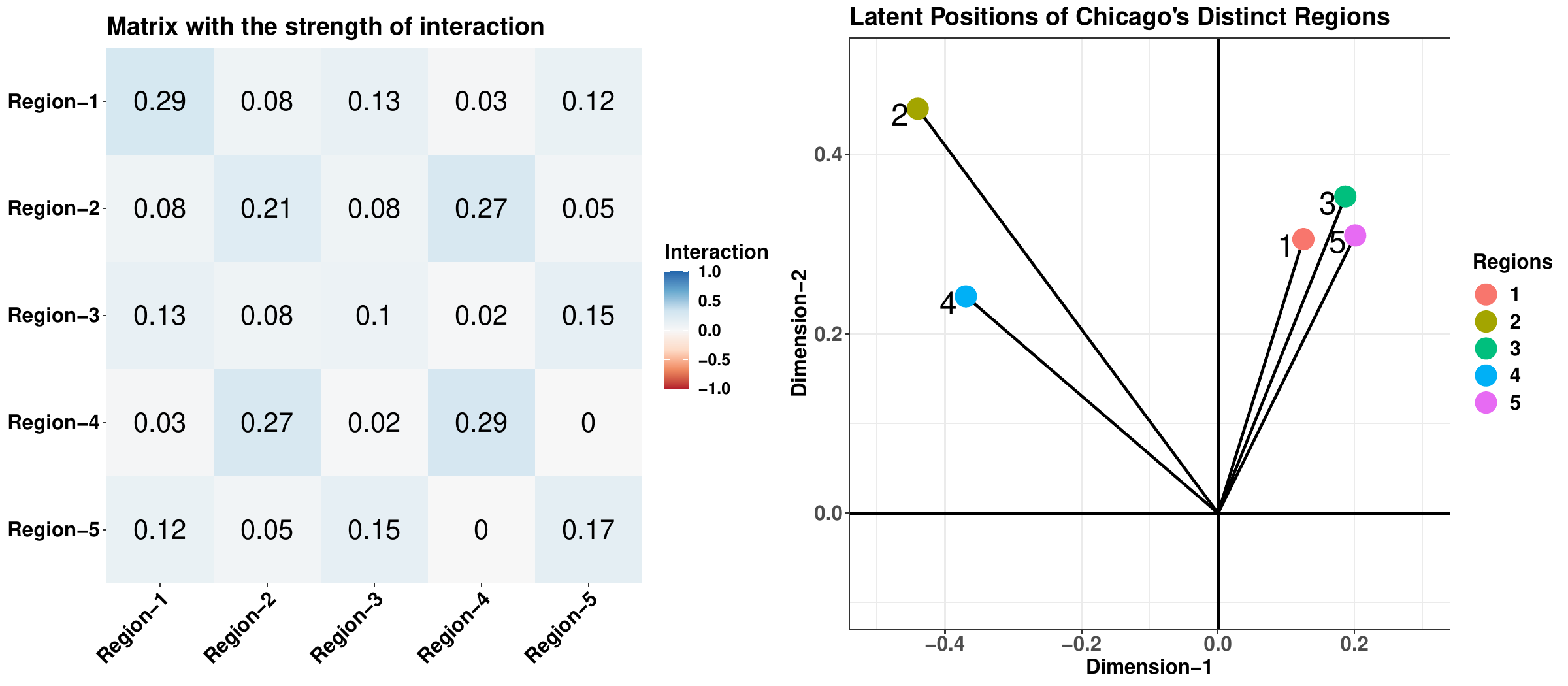}
    \caption{Left: A matrix of connections between the five regions of Chicago. Right: latent space representation showing the posterior mean of locations for each region in Chicago. Each color represents a different region. }
    \label{fig: Latent_pos_and_interactions_C5}
\end{figure}

Additionally, we show on the left panel of Figure \ref{fig: Latent_pos_and_interactions_C5} the posterior average of the interactions matrix with pairwise connections ($\boldsymbol{\gamma}$s) in the non-diagonal entries and autoregressive effects ($\boldsymbol{\beta}$s) on the diagonal. 
On the right panel of the same figure, instead, we show again the latent space representation using the posterior means and highlighting the directions they point at. This two plots allow for an appreciation of how the interdependencies between the series are captured via the latent space.

The left panel of Figure \ref{fig: Latent_pos_and_interactions_C5} reveals distinct patterns of burglary persistence over time, with the south and north-west regions of Chicago (regions $1$ \& $4$) exhibiting the strongest autoregressive effects. In contrast, the north and eastern sides of Chicago (regions $2$ \& $5$) show moderately weaker effects, with the weakest observed in the western region (region $3$). This reflects the average varying degrees of historical burglary rates across regions, as reflected in the diagonal elements of the matrix of interactions indicating the strength of these autoregressive effects.
 
From the non-diagonal elements, and from the latent positions on the right panel, we can see that regions $2$ and $4$ appear aligned, and so are the regions $1$, $3$, and $5$. This alignment, alongside the quantified interactions (ranging from $1$ for strong positive to $-1$ for strong negative) reveals the nuanced connections between regions. Although interactions are generally modest, the adjacent regions $2$ and $4$ share the most pronounced connection. Further analysis of the regions $1$, $3$, and $5$ reveals nuanced differences in their interconnections, with the strongest linkage between regions $3$ (west) and $5$ (northernmost), and a slightly lesser connection with region $1$ (south). The interaction between regions $1$ and $5$, the geographically most distant pairs, is found to be the weakest. These findings point to noteworthy connections in burglary occurrences across Chicago's varied geographic landscapes.

\begin{figure}[htbp]
    \centering
    \includegraphics[width = 0.99 \textwidth]{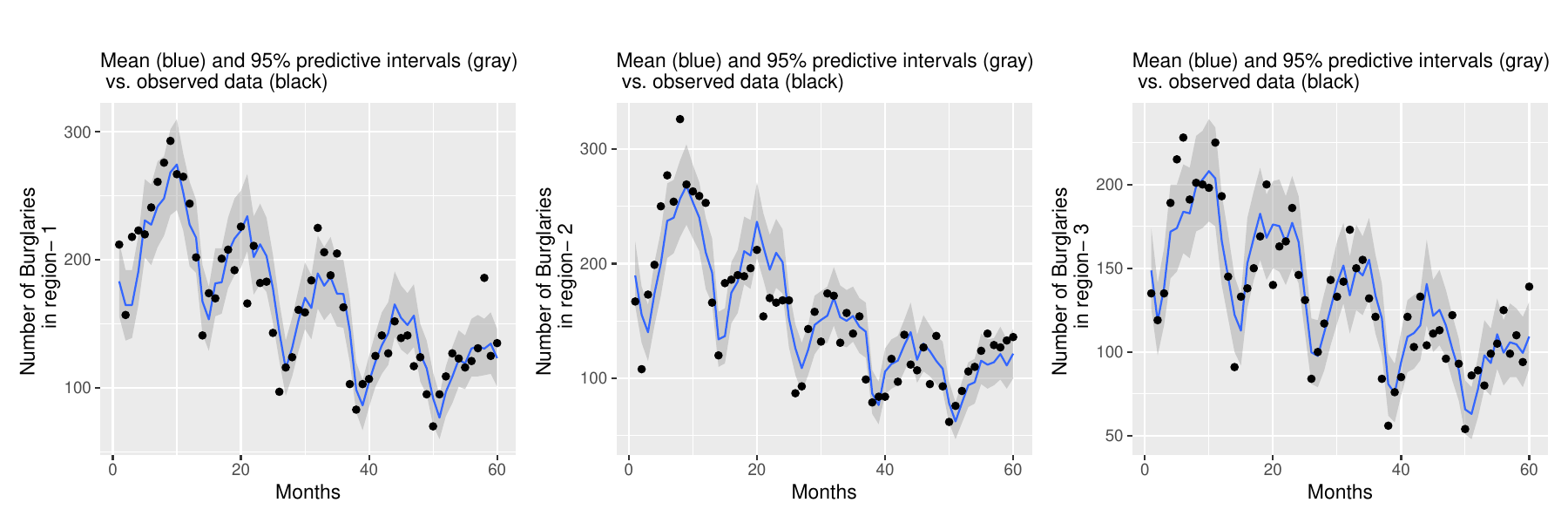}
    \caption{Number of burglaries in the first three of five distinct regions of Chicago (2011-2015). The points are the observed number of burglaries. The blue line represents the prediction of the model with 95\% prediction intervals shown in gray.}
    \label{True_vs_simulated_series_c5}
\end{figure}

Finally, we employed posterior predictive checks to evaluate our TSLPM on these data, focusing on one-step-ahead predictions. This method generates new predictions from the posterior predictive distribution based on the fitted model and then compares these values against the observed data to identify any significant discrepancies. As illustrated in Figure \ref{True_vs_simulated_series_c5}, the observed time series data for burglaries are juxtaposed with the predictions, falling within the $95\%$ confidence interval the vast majority of times. 
The findings detailed in Table  \ref{tab:percentage_coverage_C5_different_cov} demonstrate the coverage of observed burglary counts by the posterior 95\% credible intervals across the five regions. Overall, the model fit reveals minimal variance in the coverage of observed data demonstrating its effectiveness.
\begin{table}[htbp] 
\centering
\caption{Prediction coverage of the observed data using a 95\% posterior credible interval for the $5$ regions data.}
\label{tab:percentage_coverage_C5_different_cov}
\begin{tabular}{@{}lcc@{}}
\toprule
Region                           & Percentage coverage \{\texttt{pop}, \texttt{unemp}\}  \\ \midrule
Region-1    & 80                 \\
Region-2    & 73                   \\
Region-3    & 78                  \\
Region-4    & 77                   \\
Region-5    & 85                   \\ \bottomrule
\end{tabular}
\end{table}

\subsection{Chicago's eleven-regional network data} \label{Model_fit_11_regions}
We extend the analysis to the $11$ regions dataset following the model selection strategy defined in Section \ref{model selection}. 
We select the final model by employing the DIC to identify the optimal model. Table \ref{tab:model_comparison_11regions} reveals that the model exhibiting the lowest DIC has a variable $\boldsymbol{\beta}$ for each region, with $\alpha$ and $\eta$ held constant across series and employing {(\texttt{pop}, \texttt{unemp})} as a pair of covariates. 
\begin{table}[htbp]
\centering
\caption{DIC values for model comparisons for the $11$ regions dataset.}
\label{tab:model_comparison_11regions}
\begin{tabular}{@{}lcc@{}}
\toprule
Model   & DIC for 11 regions \\ \midrule
\{$\beta_{i}, \eta_i, \alpha, \delta_{\texttt{pop}}$\}                 & -38711.20            \\
\{$\beta_{i}, \eta_i, \alpha, \delta_{\texttt{ym}}$\}                   & -36858.17           \\
\{$\beta_{i}, \eta_i, \alpha, \delta_{\texttt{unemp}}$\}               & -35506.23           \\
\{$\beta_{i}, \eta, \alpha, \delta_{\texttt{pop}}$\}                   & -40506.05           \\
\{$\beta_{i}, \eta, \alpha, \delta_{\texttt{ym}}$\}                    & -40367.67           \\
\{$\beta_{i}, \eta, \alpha, \delta_{\texttt{unemp}}$\}                & -40592.83           \\
\{$\beta, \eta_i, \alpha, \delta_{\texttt{pop}}$\}                    & -33705.27           \\
\{$\beta, \eta_i, \alpha, \delta_{\texttt{ym}}$\}                     & -32448.32           \\
\{$\beta, \eta_i, \alpha, \delta_{\texttt{unemp}}$\}                 & -30790.28           \\
\{$\beta, \eta, \alpha, \delta_{\texttt{pop}}$\}                     & -35812.89           \\
\{$\beta, \eta, \alpha, \delta_{\texttt{ym}}$\}                     & -35700.33           \\
\{$\beta, \eta, \alpha, \delta_{\texttt{unemp}}$\}                 & -35713.67           \\
\{$\beta_{i}, \eta_i, \alpha, \delta_{\texttt{pop}}, \delta_{\texttt{unemp}} $\}         & -39529.08           \\
\{$\beta_{i}, \eta_i, \alpha, \delta_{\texttt{ym}}, \delta_{\texttt{unemp}} $\}                   & -39115.18           \\
\{$\beta_{i}, \eta, \alpha, \delta_{\texttt{pop}}, \delta_{\texttt{unemp}} $\}                  & \textcolor{red}{-42335.00}              \\
\{$\beta_{i}, \eta, \alpha, \delta_{\texttt{ym}}, \delta_{\texttt{unemp}} $\}                     & -42242.19           \\
\{$\beta, \eta_i, \alpha, \delta_{\texttt{pop}}, \delta_{\texttt{unemp}} $\}                  & -33243.90            \\
\{$\beta, \eta_i, \alpha, \delta_{\texttt{ym}}, \delta_{\texttt{unemp}} $\}                    & -33520.29           \\
\{$\beta, \eta, \alpha, \delta_{\texttt{pop}}, \delta_{\texttt{unemp}} $\}                 &  -36016.88                   \\
\{$\beta, \eta, \alpha, \delta_{\texttt{ym}}, \delta_{\texttt{unemp}} $\}                    & -36390.81           \\ \bottomrule
\end{tabular}
\end{table}
The model adopts a the same form as the model of Eq. \ref{crime_dta_model_withcov_and_seasonality}, with the only difference being that the parameters $\alpha$ and $\eta$ are constant across all regions. Thus, a key difference with previous models is that, in this optimal model, the parameter $\eta$ is constant across all regions. 

\begin{figure}[htbp]
    \centering
    \includegraphics[width = 0.45 \textwidth]{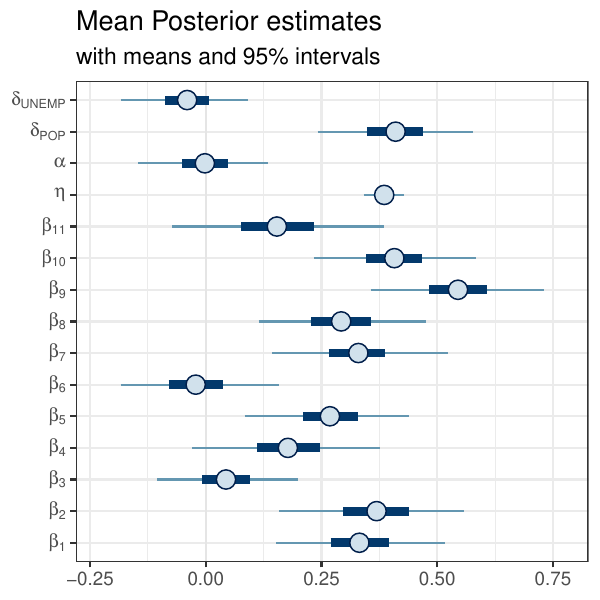}
    \includegraphics[width = 0.45 \textwidth]{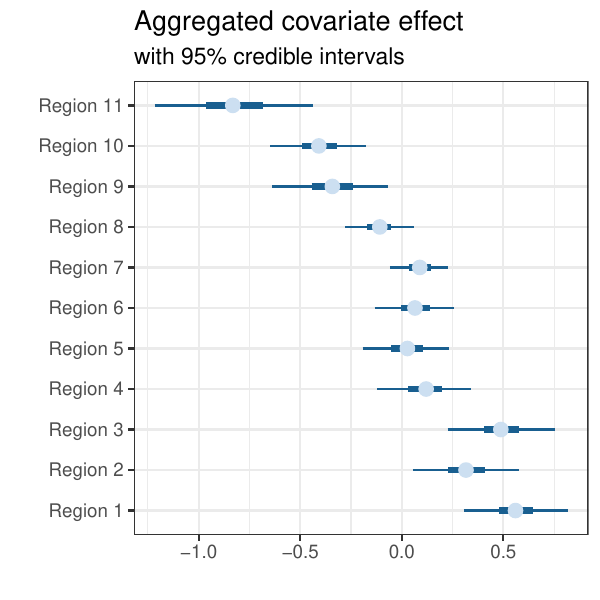}
    \caption{Left: mean posterior estimates with 95\% credible intervals for the model fitted to data across $11$ regions, incorporating covariates. Right: aggregated mean effects of covariates (population and unemployment) on each region, shown with 95\% credible intervals.}
    \label{EST_95CI_C11_POP_UNEMP_ym_credibleinterval_covariate_effect_C11}
\end{figure}

We can observe from the left panel of Figure \ref{EST_95CI_C11_POP_UNEMP_ym_credibleinterval_covariate_effect_C11}, that the parameters $\boldsymbol{\beta}$ generally take positive values, although we notice more diverse values compared to the $5$ clusters solution.
In particular, some regions exhibit small negative $\boldsymbol{\beta}$ values within their 95\% credible intervals, suggesting that negative autocorrelations may also be possible. 
The $\boldsymbol{\beta}$ values show a diminished persistence of burglaries over time in the north-most and south-most regions of Chicago (regions $3$, $11$, and $4$). The western, eastern, and central regions (regions $1$, $2$, $7$, $8$, $10$ and $5$) have comparatively strong autoregressive effects. Notably, the strongest effect is observed in the southwestern region (region $9$) and the weakest in the northeast region (region $6$). The parameter $\eta$ takes a small positive value, suggesting that seasonal burglary patterns are likely to persist from one year to the next across all regions.

The model explores the impact of external factors like population and unemployment on burglary rates, with results shown in the right panel Figure \ref{EST_95CI_C11_POP_UNEMP_ym_credibleinterval_covariate_effect_C11}. An increase in these covariates leads to corresponding changes in burglary counts, with regions $1$, $2$, and $3$ notably experiencing an increase in burglary cases due to these combined effects.
Additionally, if we look at the coefficients of the covariates, depicted in the left panel of Figure \ref{EST_95CI_C11_POP_UNEMP_ym_credibleinterval_covariate_effect_C11}, we note a small coefficient for unemployment and a fairly large coefficient for population. These findings are consistent with previous studies on the same Chicago burglary data, such as those reported by \textcite{CLARK2021100493} and \textcite{armillotta2023}. 

\begin{figure}[htbp]
 \centering
 \includegraphics[width = 0.8 \textwidth]
{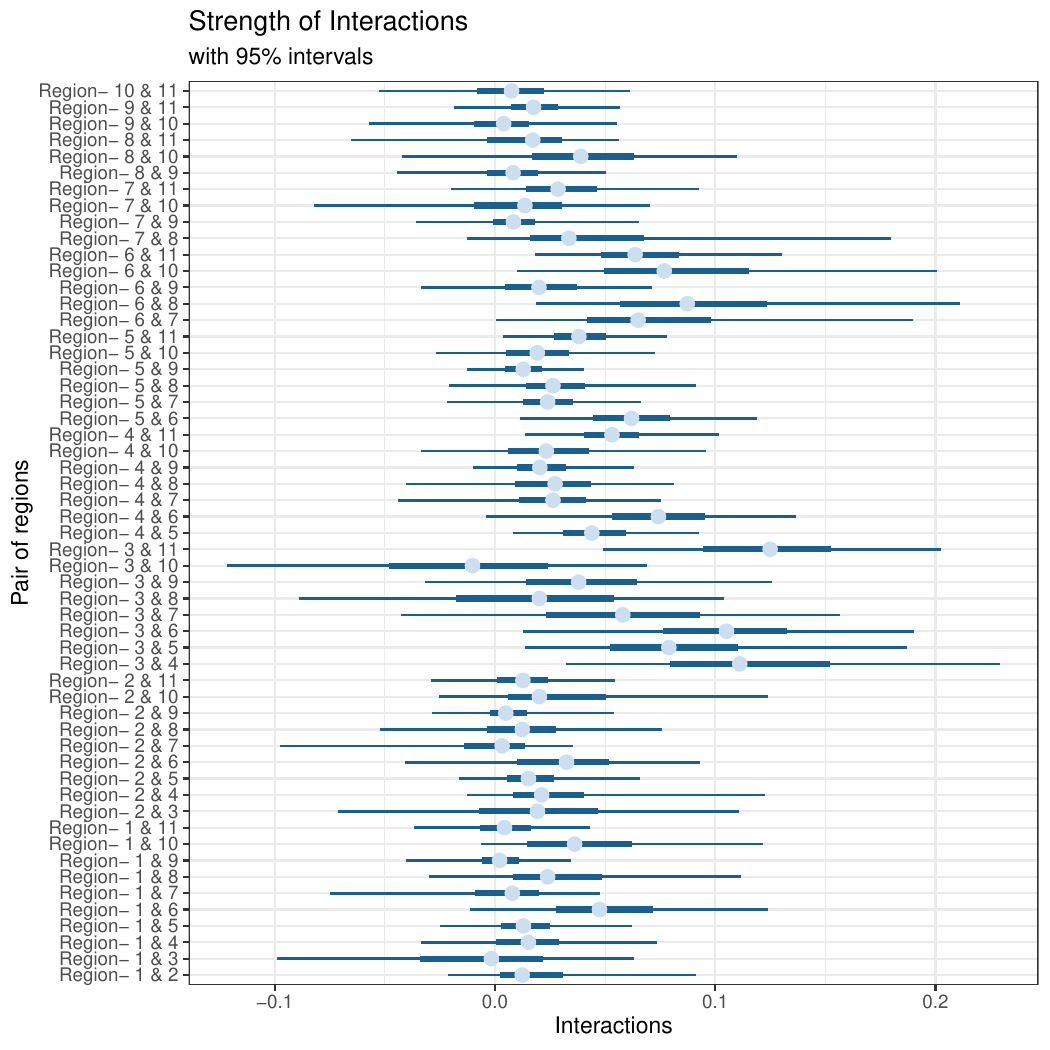}
\caption{Mean estimates with 95\% credible interval of pairwise interactions between $11$ regions of Chicago.}
    \label{fig:pairwise_interactions_c11}
\end{figure}

As regards the interaction parameters $\boldsymbol{\gamma}$, the posterior distribution of connections between all regions are depicted in Figure \ref{fig:pairwise_interactions_c11}.   
We note that some interactions are fairly strong so this suggests that a model capable of capturing these dependencies can be useful for these data.

\begin{figure}[htbp]
\centering
       \includegraphics[width = 0.8 \textwidth]{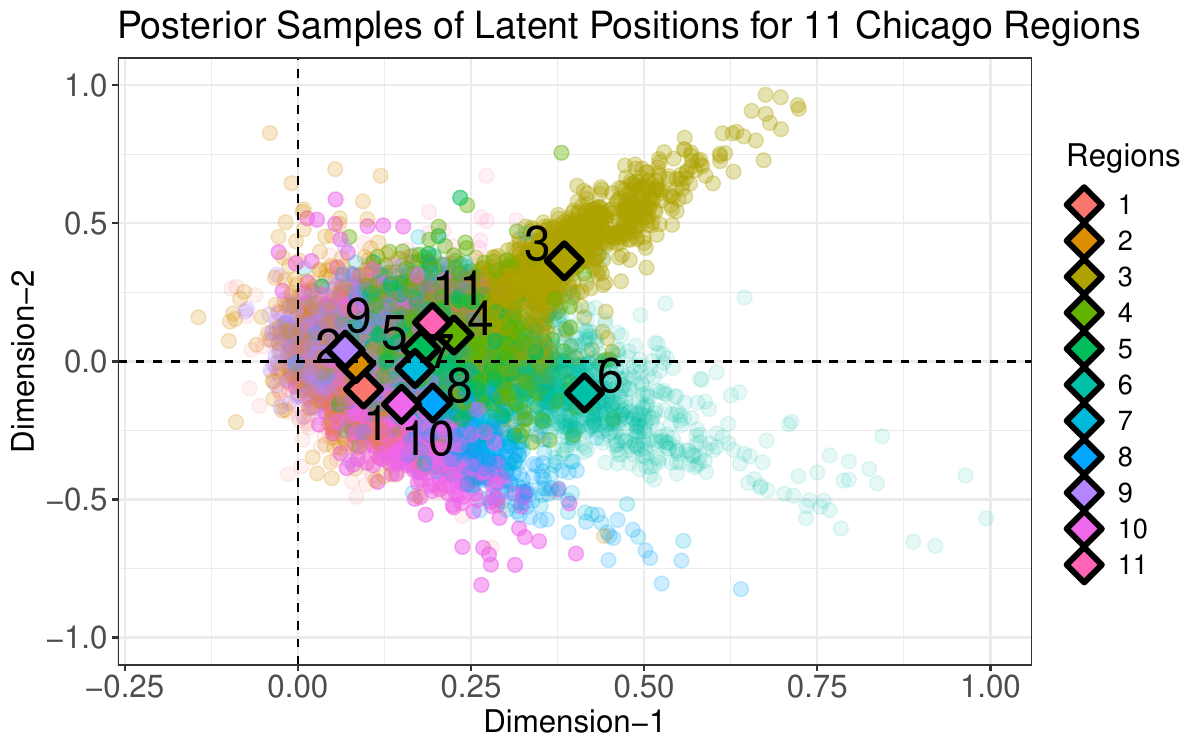}
    \caption{Posterior samples of Latent positions for $11$ different regions of Chicago. The large squared points represent the posterior means of the regions.}
    \label{sub:latent_positions_C11}
\end{figure}

\begin{figure}[htbp]
\centering 
\includegraphics[width = 1 \textwidth]{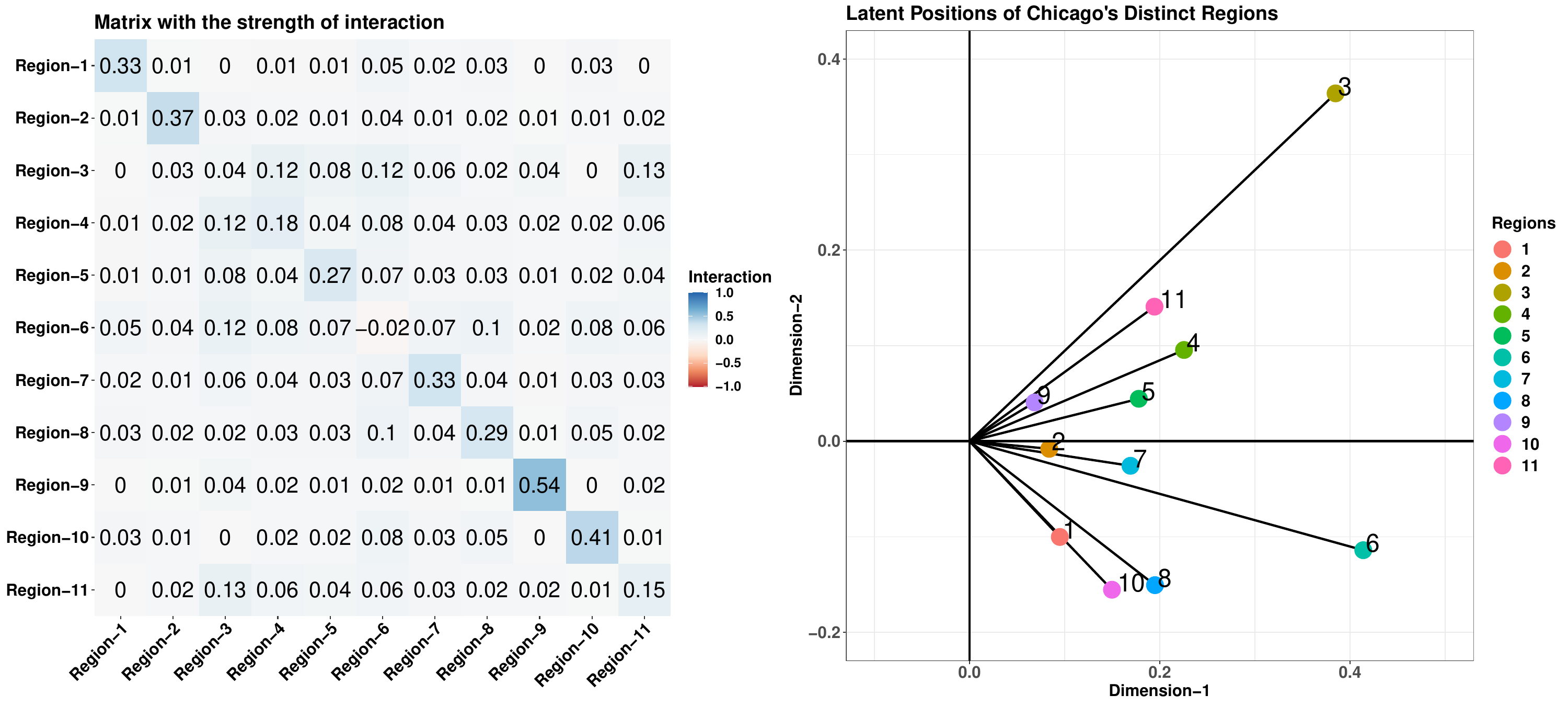}
\caption{Left: A matrix of connections between all $11$ regions of Chicago. Right: latent space representation showing the posterior mean of locations for each region.}
\label{fig: Latent_pos_and_interactions_C11}
\end{figure}

Now, we show a latent space with all posterior samples for each of the $11$ regions, post-Procrustes transformation, in Figure \ref{sub:latent_positions_C11}.
Here, the posterior means are denoted by larger, squared points, offering a clear visual representation of the estimated final locations of the regions.
We also show, in Figure \ref{fig: Latent_pos_and_interactions_C11}, the posterior averages of the latent positions besides the corresponding interactions and $\boldsymbol{\beta}$ parameters, to facilitate the interpretation of the latent space.
The latent space plot reveals a pattern of connections among the regions positioned in the northernmost, southernmost, south-eastern, and western areas (regions $3$, $4$, $5$, $9$, and, $11$) due to their similar directions in the latent space. Similarly, regions in the eastern, western, and central regions of Chicago (regions $1$, $2$, $6$, $7$, $8$, and, $10$) also exhibit similar latent directions. Although overall interactions are modest, the strongest interactions are between regions $3$ and $11$ in the south, with significant connections also observed between regions $3$ and $4$ (north), and between region $6$ in the northeast and region $8$ in the center. The interaction matrix highlights another inter-group relation between northeastern region $6$ with region $3$ in the south. Overall, there are notable similarities observed among the northern, southern, and eastern regions, while the western regions appear to remain distinct. 

\begin{figure}[htbp]
    \centering
    \includegraphics[width = 1 \textwidth]{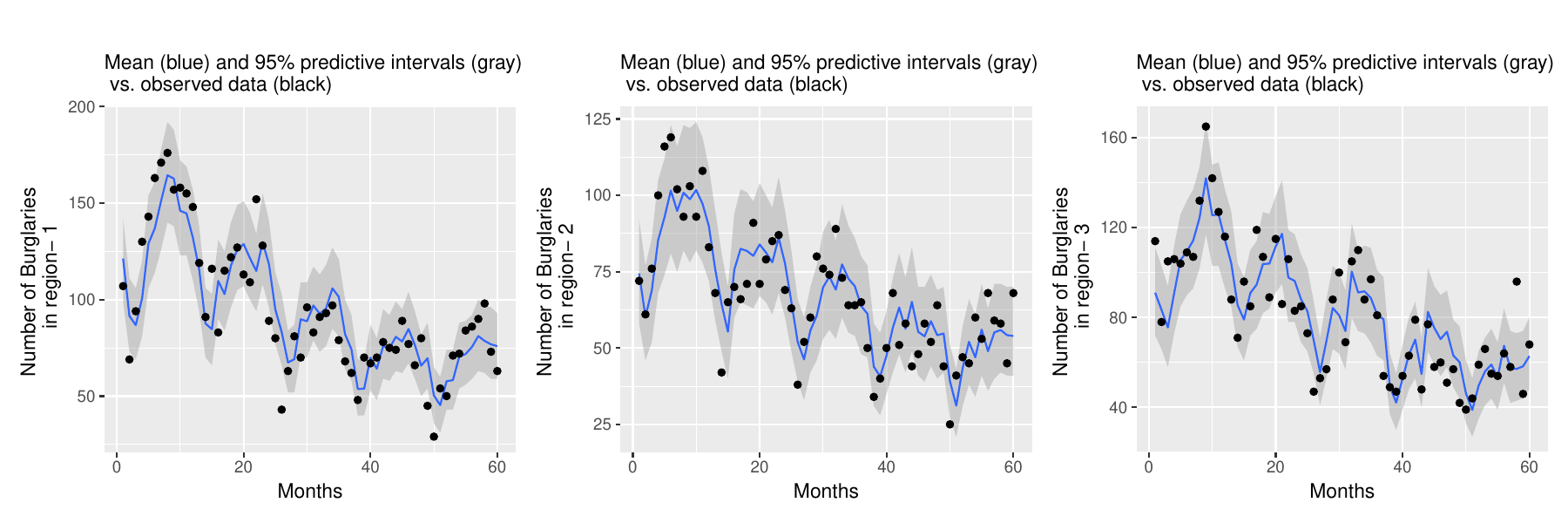}
    \caption{Number of burglaries in the first three of eleven regions of Chicago. The points are the observed number of burglaries. The blue line represents the mean predictions of the model with 95\% prediction intervals shown in gray.}
    \label{True_vs_simulated_series_c11}
\end{figure}

Finally, Figure \ref{True_vs_simulated_series_c11} presents the observed time series against predictions for the first three regions, including 95\% credible intervals. 

These show good agreement and overall a very good model fit. This is also confirmed by Table \ref{tab:percentage_coverage_C11_different_cov}, showing the extent to which the observed burglary counts are included in the posterior 95\% credible intervals for each region.
\begin{table}[htbp]
\centering
\caption{Prediction coverage of the observed data using a 95\% posterior credible interval for the $11$ regions data.}
\label{tab:percentage_coverage_C11_different_cov}
\begin{tabular}{@{}lcc@{}}
\toprule
Region                           & Percentage coverage \{\texttt{pop}, \texttt{unemp}\}  \\ \midrule
Region-1    & 77                    \\
Region-2    & 87                    \\
Region-3    & 83                    \\
Region-4    & 87                    \\
Region-5    & 77                     \\
Region-6    & 85                    \\
Region-7    & 85                    \\
Region-8    & 90                    \\
Region-9    & 85                    \\
Region-10    & 85                  \\
Region-11    & 95                  \\ \bottomrule
\end{tabular}
\end{table}

\section{Conclusion and discussion} \label{discussion}
In this paper, we have introduced the TSLPM model, which provides a novel framework to analyse multivariate count time series data using a latent variable network approach. The TSLPM combines the flexibility of VAR models for time series, with the interpretable and useful representations that are obtained with latent variable network models. In the paper, we discussed model identifiability, stationarity, parameter interpretation, inference under a Bayesian setting, hence highlighting the main features of the model and how to use it. Through a series of simulation experiments, we have validated the TSLPM performance for parameter estimation and for prediction. The results indicate that the geometry of the estimated latent space closely matches with that of the true latent space, thus providing accurate modeling of the time series dependencies. When compared to other viable models in terms of forecasting accuracy, the TSLPM consistently performs as a very competitive approach, thus making it a suitable model for the analysis of multivariate count time series.

We showed an application of our new methodology on a dataset of crimes in the south city of Chicago. The model revealed connections in burglary occurrences across various geographic regions of Chicago. This reflects the idea that the frequency of crimes can follow rather complex patterns of interdependencies, which go beyond the simple geographical layout of the areas. Our results show a meaningful interaction between various areas of the city, with a some tendency to form clusters in the 5 regions case. The resulting inferred latent space is highly interpretable and offers a new model-based way to visualize the connections between the areas of the city. This latent space may serve as a starting point for further analysis, for example aiming at quantifying endogenous vs exogenous risk for each area, or for making future predictions regarding risk. Our crime data application very strongly suggests that the TSLPM methodology can be used in other contexts using an equivalent formulation: a natural application of these methods could be in finance and in the analysis of multivariate financial time series. In fact, extensions to multivariate time series data with continuous responses would be fairly straightforward. In the recent work by \textcite{zhu2017network}, a network vector autoregressive model (NAR) is proposed for continuous data. The model incorporates the information from the known network structure. An extension to this model is when the network structure is not known which is one of the assumptions in TSLPM. These two model structures can be exploited to introduce a network model for multivariate time series with continuous response.

Another potential extension is to consider a TSLPM of order greater than one. While we have partially done this using the seasonality effect in our real data application, a more formal and comprehensive approach may be considered to incorporate autoregressive coefficients of higher order. This inclusion serves to capture enduring dependencies and trends within the data, thus enhancing the precision and reliability of our forecasts. As regards the model definition, another avenue for expansion is the integration of edge covariates. In numerous real network applications, data related to nodes and edges are readily accessible. By introducing these covariates into the model, we have the potential to significantly boost predictive performance.

Finally, in relation to the networks literature, a number of possible extensions can definitely be considered, including latent distance models and dynamic models. While the literature on dynamic network models has been fast growing, the domain of models addressing multivariate time series with underlying network structures remains relatively underrepresented. To our knowledge, the work by \textcite{kang2017dynamic} is one of the few papers in this sparse literature. Consequently, this area constitutes an open research question, awaiting further exploration.

\section{Acknowledgement}
This publication has emanated from research supported in part by a grant from the Insight Centre for Data Analytics which is supported by Science Foundation Ireland under Grant number $12/RC/2289\_P2$. We also extend our gratitude to Nicholas John Clark for providing the crime dataset analyzed in this paper.

\section{Appendix}

\subsection{Supplementary tables}
The results from fitting the TSLPM model to Chicago data, encompassing 5 regions, are presented in Table \ref{tab:Posterior_estimates_C5}, and for 11 regions, in Table \ref{tab:Posterior_estimates_C11}. These are the supplementary tables represent the quantified results of parameter estimates as given in the left panel of Figure \ref{EST_95CI_C5_POP_UNEMP_ym_credibleinterval_covariate_effect_C5}
 and \ref{EST_95CI_C11_POP_UNEMP_ym_credibleinterval_covariate_effect_C11}.
\begin{table}[htbp]
\centering
\begin{minipage}[b]{0.45\linewidth}
\centering
\begin{tabular}{lrr}
  \toprule
Parameter & Mean & 95\% Credible Interval \\ 
  \midrule
  $\beta_1$ & 0.29 & (0.14,0.44) \\ 
  $\beta_2$ & 0.21 & (0.04,0.37) \\ 
  $\beta_3$ & 0.10 & (-0.09,0.28) \\ 
  $\beta_4$ & 0.29 & (0.13,0.47) \\ 
  $\beta_5$ & 0.17 & (-0.02,0.36) \\ 
  $\eta_{1}$ & 0.39 & (0.30,0.49) \\ 
  $\eta_2$ & 0.46 & (0.37,0.54) \\ 
  $\eta_3$ & 0.56 & (0.46,0.66) \\ 
  $\eta_4$ & 0.46 & (0.34,0.58) \\ 
  $\eta_5$ & 0.54 & (0.42,0.68) \\ 
  $\alpha$ & 0.00 & (-0.18,0.17) \\ 
  $\delta_{\texttt{pop}}$ & 0.02 & (-0.21,0.26) \\ 
  $\delta_{\texttt{unemp}}$ & -0.13 & (-0.37,0.10) \\ 
   \bottomrule
\end{tabular}
\end{minipage}
\caption {Mean estimates with 95\% credible interval for the model fitted to data across 5 regions with covariates.}
\label{tab:Posterior_estimates_C5}
\end{table}

\begin{table}[htbp]
\centering
\begin{minipage}[b]{0.45\linewidth}
\centering
\begin{tabular}{lrr}
  \toprule
Parameter & mean & 95\% Credible Interval \\ 
  \midrule
  $\beta_{1}$ & 0.33 & (0.15,0.52) \\ 
  $\beta_{2}$ & 0.37 & (0.16,0.56) \\ 
  $\beta_{3}$ & 0.04 & (-0.10,0.20) \\ 
  $\beta_{4}$ & 0.18 & (-0.03,0.38) \\ 
  $\beta_{5}$ & 0.27 & (0.08,0.44) \\ 
  $\beta_{6}$ & -0.02 & (-0.18,0.16) \\ 
  $\beta_{7}$ & 0.33 & (0.14,0.52) \\ 
  $\beta_{8}$ & 0.29 & (0.11,0.48) \\ 
  $\beta_{9}$ & 0.54 & (0.36,0.73) \\ 
  $\beta_{10}$ & 0.41 & (0.23,0.58) \\ 
  $\beta_{11}$ & 0.15 & (-0.07,0.39) \\ 
  $\eta$ & 0.39 & (0.34,0.43) \\ 
  $\alpha$ & -0.00 & (-0.15,0.13) \\ 
  $\delta_{\texttt{pop}}$ & 0.41 & (0.24,0.58) \\ 
  $\delta_{\texttt{unemp}}$ & -0.04 & (-0.18,0.09) \\ 
   \bottomrule
\end{tabular}
\end{minipage}
\caption{Mean estimates with 95\% credible interval for the model fitted to data across 11 regions with covariates.}
\label{tab:Posterior_estimates_C11}
\end{table}

\printbibliography

@article{nickel2008random,
author = {Nickel, C.L.M.},
title = {Random dot product graphs: a model for social networks},
school={Johns Hopkins University},
year = {2007},
pages = {}
}

@conference{young2007random,
author={Young, S.J. and Scheinerman, E.R.},
title={Random dot product graph models for social networks},
booktitle={International Workshop on Algorithms and Models for the Web-Graph},
pages={138--149},
year={2007},
organization={Springer}
}

@article{JSSv096i05,
author={Knight, M. and Leeming, K. and Nason, G. and Nunes, M. },
 title={Generalized Network Autoregressive Processes and the GNAR Package},
 journal={Journal of Statistical Software},
 year={2020},
 volume={96},
 number={5},
 pages={1–36}
}

@article{armillotta2023,
author = {Armillotta,M. and Fokianos, K.},
title = {Count network autoregression},
journal = {Journal of Time Series Analysis},
year = {2023},
pages = {}
}

@article{betancourt2013hamiltonian,
author = {Betancourt, M. and Girolami, M.},
title = {Hamiltonian Monte Carlo for Hierarchical Models},
year = {2013},
pages = {}
}

@article{Konstantinos_2020,
author = {Fokianos, K. and St{\o}ve, B. and Tj{\o}stheim, D. and Doukhan, P.},
title = {{Multivariate count autoregression}},
journal = {Bernoulli},
publisher = {Bernoulli Society for Mathematical Statistics and Probability},
year = {2020},
volume = {26},
number = {1},
pages = {471 -- 499}
}

@article{FOKIANOS2021,
author = {Fokianos, K.},
title = {Multivariate Count Time Series Modelling},
journal = {Econometrics and Statistics},
year = {2021},
issn = {2452-3062}
}

@ARTICLE{Hoff20021090,
author={Hoff, P.D. and Raftery, A.E. and Handcock, M.S.},
title={Latent space approaches to social network analysis},
journal={Journal of the American Statistical Association},
year={2002},
volume={97},
number={460},
pages={1090-1098}
}

@article{Doukhan2017MultivariateCA,
author={Doukhan, P. and Fokianos, K. and St{\o}ve, B. and Tj{\o}stheim},
  title={Multivariate count autoregression},
  journal={Bernoulli},
  year={2017}
}

@book{varga2011geršgorin,
  author={Varga, R.S.},
  title={Ger{\v{s}}gorin and His Circles},
  isbn={9783540211006},
  lccn={2004104814},
  series={Springer Series in Computational Mathematics},
  year={2011},
  publisher={Springer Berlin Heidelberg}
}

@phdthesis{Andreassen_2013,
author = "Andreassen, C.M.",
title = "Models and inference for correlated count data",
year = "2013",
language = "English",
publisher = "Department of Mathematics, Aarhus University",
}

@ARTICLE{1987PhLB..195..216D,
       author = {{Duane}, S. and {Kennedy}, A.~D. and {Pendleton}, B.J. and {Roweth}, D.},
        title = "{Hybrid Monte Carlo}",
      journal = {Physics Letters B},
         year = 1987,
       volume = {195},
       number = {2},
        pages = {216-222}
}

@article{Neal2011MCMCUH,
author={Neal, R.M.},
title={MCMC Using Hamiltonian Dynamics},
journal = {Handbook of Markov Chain Monte Carlo},
year = {2012},
pages = {},
isbn = {9780429138508}
}

@ARTICLE{Sewell2016105,
author={Sewell, D.K. and Chen, Y.},
title={Latent space models for dynamic networks with weighted edges},
journal={Social Networks},
year={2016},
volume={44},
pages={105-116}
}

@ARTICLE{Sewell20151646,
author={Sewell, D.K. and Chen, Y.},
title={Latent Space Models for Dynamic Networks},
journal={Journal of the American Statistical Association},
year={2015},
volume={110},
number={512},
pages={1646-1657}
}

@article{latour_1997,
author={Latour, A.}, 
title={The Multivariate Ginar(p) Process}, 
journal={Advances in Applied Probability}, 
publisher={Cambridge University Press}, 
year={1997}, 
volume={29},
number={1}, 
pages={228–248}
}

@article{Ravishanker2014HierarchicalDM,
author={Ravishanker, N. and Serhiyenko, V. and Willig, M.R.},
  title={Hierarchical dynamic models for multivariate times series of counts},
  journal={Statistics and Its Interface},
  year={2014},
  volume={7},
  pages={559-570}
}

@article{Jung2008,
author = {Jung, R. and Liesenfeld, R. and Richard, J.F.},
title = {Dynamic Factor Models for Multivariate Count Data: An Application to Stock-Market Trading Activity},
journal = {Journal of Business \& Economic Statistics},
year = {2008},
volume = {29},
pages = {73-85}
}

@article{Jorgensen1996,
author = "J{\o}rgensen, B. and Christensen, S.L. and Song, P.X.-K. and Sun, L.",
title = "State-space models for multivariate longitudinal data of mixed types",
journal = "Canadian Journal of Statistics",
publisher = "John Wiley \& Sons, Inc.",
year = "1996",
language = "English",
volume = "24",
number = "3",
pages = "385--402",
issn = "0319-5724"
}

@article{Fokianos_Anders_Dag2008,
author = {Fokianos, K. and Rahbek, A. and Tjøstheim, D.},
title = {Poisson Autoregression},
journal = {Journal of the American Statistical Association},
year = {2008},
pages = {},
volume = {104}
}

@article{Davis2000,
author = {Davis, R. and Dunsmuir, W.T.M. and Wang, Y.},
title = {On autocorrelation in a Poisson regression model},
journal = {Biometrika},
year = {2000},
volume = {87},
pages = {491-505},
}

@article{Fokianos_Tjøstheim_2011,
author = {Fokianos, K. and Tjøstheim, D.},
title = {Log-linear Poisson autoregression},
journal = {Journal of Multivariate Analysis},
year = {2011},
pages = {563-578}
}

@article{Davis_Liu_2012,
author = {Davis, R. and Liu, H.},
title = {Theory and Inference for a Class of Observation-driven Models with
Application to Time Series of Counts},
journal = {Statistica Sinica},
year = {2012},
pages = {}
}

@article{Heinen_Erick_2007,
author = {Heinen, A. and Erick, R.},
title = {Multivariate autoregressive modeling of time series count data using copulas},
journal = {Journal of Empirical Finance},
year = {2007},
volume = {14},
pages = {564-583}
}

@article{Lee_Tjøstheim_2017,
author = {Lee, Y. and Lee, S. and Tjøstheim, D.},
title = {Asymptotic normality and parameter change test for bivariate Poisson INGARCH models},
journal = {TEST},
year = {2017},
volume = {27},
pages = {}
}

@article{Streett_2000,
author = {Streett, S.},
title = {Some observaton driven models for time series},
year = {2000},
pages = {}
}

@article{Ferland_Latour_Oraichi_2006,
author = {Ferland, R. and Latour, A. and Oraichi, D.},
title = {Integer-valued GARCH processes},
journal = {Journal of Time Series Analysis},
year = {2006},
volume = {27},
pages = {923-942}
}

@article{Davis_Fokianos_Holan_Joe_Livsey_Lund_Pipiras_Ravishanker_2021,
author = {Davis, R. and Fokianos, K. and Holan, S. and Joe, H. and Livsey, J. and Lund, R. and Pipiras, V. and Ravishanker, N.},
title = {Count Time Series: A Methodological Review},
journal = {Journal of the American Statistical Association},
year = {2021},
volume = {116},
pages = {1-50}
}

@article{Pedeli_Karlis_2013_03,
author = {Pedeli, X. and Karlis, D.},
title = {On composite likelihood estimation of a multivariate INAR(1) model},
journal = {Journal of Time Series Analysis},
year = {2013},
volume = {34},
pages = {206-220}
}

@article{Pedeli_Karlis_2013_11,
author = {Pedeli, X. and Karlis, D.},
title = {Some properties of multivariate INAR(1) processes},
journal = {Computational Statistics \& Data Analysis},
year = {2013},
volume = {67},
pages = {213-225}
}

@article{SCOTTO2014233,
author = {Scotto, M.G. and Weiß, C.H. and Silva, M.E. and Pereira, I.},
title = {Bivariate binomial autoregressive models},
journal = {Journal of Multivariate Analysis},
year = {2014},
volume = {125},
pages = {233-251},
issn = {0047-259X}
}

@article{Darolles_Fol_Lu_Sun2019,
  author={Darolles, S. and Fol, G.L. and Lu, Y. and Sun, R.},
  title={{Bivariate integer-autoregressive process with an application to mutual fund flows}},
  journal={Journal of Multivariate Analysis},
  year=2019,
  volume={173},
  number={C},
  pages={181-203},
  month={}
}

@article{Zeger1988ARM,
  author={Zeger, S.L.},
  title={A regression model for time series of counts},
  journal={Biometrika},
  year={1988},
  volume={75},
  pages={621-629}
}

@article{Harvey1989TimeSM,
author={Harvey, A. and Fernandes, C.A.C.},
  title={Time Series Models for Count or Qualitative Observations},
  journal={Journal of Business \& Economic Statistics},
  year={1989},
  volume={7},
  pages={407-417}
}

@article{Durbin_Koopman_2000,
author = {Durbin, J. and Koopman, S.J.},
title = {Time series analysis of non-Gaussian observations based on state space models from both classical and Bayesian perspectives},
journal = {Journal of the Royal Statistical Society: Series B (Statistical Methodology)},
year = {2000},
volume = {62},
number = {1},
pages = {3-56}
}

@article{Schnatter_Wagner_2006,
author = {Frühwirth-Schnatter, S. and Wagner, H.},
title = {Auxiliary Mixture Sampling for Parameter-Driven Models of Time Series of Small Counts with Applications to State Space Modelling},
journal = {Biometrika},
year = {2006},
volume = {93},
pages = {827-841}
}

@article{Ertekin_2015,
author = {Ertekin, Ş. and Rudin, C. and McCormick, T.},
title = {Reactive Point Processes: A New Approach to Predicting Power Failures in Underground Electrical Systems},
journal = {The Annals of Applied Statistics},
year = {2015},
volume = {9},
pages = {122-144}
}

@article{Brown_Partha_2004,
author = {Brown, E. and Kass, R. and Mitra, P.},
title = {Multiple neural spike train data analysis: state-of-the-art and future challenges},
journal = {Nature neuroscience},
year = {2004},
volume = {7},
pages = {456-61}
}

@article{Hall_Willett_2015,
author = {Hall, E. and Willett, R.},
title = {Online learning of neural network structure from spike trains},
journal = {International IEEE/EMBS Conference on Neural Engineering, NER},
year = {2015},
volume = {2015},
pages = {930-933}
}

@article{Livsey_Lund_Kechagias_Pipiras_2018,
author = {Livsey, J. and Lund, R. and Kechagias, S. and Pipiras, V.},
title = {Multivariate integer-valued time series with flexible autocovariances and their application to major hurricane counts},
journal = {The Annals of Applied Statistics},
year = {2018},
volume = {12},
pages = {408-431}
}

@book{book_Luetkepohl_2005,
author = {Luetkepohl, H.},
title = {The New Introduction to Multiple Time Series Analysis},
journal = {New Introduction to Multiple Time Series Analysis},
year = {2005},
pages = {},
isbn = {978-3-540-40172-8}
}

@book{box2015time,
  author={Box, G.E.P. and Jenkins, G.M. and Reinsel, G.C. and Ljung, G.M.},
  title={Time series analysis: forecasting and control},
  publisher={John Wiley \& Sons},
  year={2015}
}

@article{zhu2017network,
  author={Zhu, X. and Pan, R. and Li, G. and Liu, Y. and Wang, H.},
  title={Network vector autoregression},
  year={2017}
}

@article{zhu2020multivariate,
  author={Zhu, X. and Huang, D. and Pan, R. and Wang, H.},
  title={Multivariate spatial autoregressive model for large scale social networks},
  journal={Journal of Econometrics},
  year={2020},
  publisher={Elsevier},
  volume={215},
  number={2},
  pages={591--606}
}

@article{zhu2019network,
  author={Zhu, X. and Wang, W. and Wang, H. and H{\"a}rdle, W.K.},
  title={Network quantile autoregression},
  journal={Journal of econometrics},
  year={2019},
  publisher={Elsevier},
  volume={212},
  number={1},
  pages={345--358}
}

@article{knight2016modelling,
  author={Knight, M.I. and Nunes, M.A. and Nason, G.P.},
  title={Modelling, Detrending and Decorrelation of Network
  Time Series},
  journal={arXiv preprint arXiv:1603.03221},
  year={2016}
}

@article{kang2017dynamic,
  author={Kang, X. and Ganguly, A. and Kolaczyk, E.D.},
  title={Dynamic networks with multi-scale temporal structure},
  journal={arXiv preprint arXiv:1712.08586},
  year={2017}
}

@article{Chen2020CommunityNA,
  author={Chen, E.Y. and Fan, J. and Zhu, X.},
  title={Community network auto-regression for high-dimensional time series},
  journal={Journal of Econometrics},
  year={2020}
}

@article{ZHU2019145,
author = {Zhu, X. and Chang, X. and Li, R. and Wang, H.},
title = {Portal nodes screening for large scale social networks},
journal = {Journal of Econometrics},
year = {2019},
volume = {209},
number = {2},
pages = {145-157},
issn = {0304-4076}
}

@article{NetworkGarch_2020,
author = {Zhou, J. and Li, D. and Pan, R. and Wang, H.},
title = {Network Garch Model},
journal = {Statistica Sinica},
year = {2020},
volume = {30},
pages = {1-18}
}

@article{Zhu_Pan2020,
 author = {Zhu, X. and Pan, R.},
 title = {Grouped Network Vector Autoregression},
 journal = {Statistica Sinica},
 publisher = {Institute of Statistical Science, Academia Sinica},
 year = {2020},
 volume = {30},
 number = {3},
 pages = {1437--1462},
 urldate = {2023-06-15}
}

@article{bracher2022endemic,
  author={Bracher, J. and Held, L.},
  title={Endemic-epidemic models with discrete-time serial interval distributions for infectious disease prediction},
  journal={International Journal of Forecasting},
  publisher={Elsevier},
  year={2022},
  volume={38},
  number={3},
  pages={1221--1233}
}

@ARTICLE{Pandit_2020,
  author={Pandit, P. and Sahraee-Ardakan, M. and Amini, A.A. and Rangan, S. and Fletcher, A.K.},
  title={Generalized Autoregressive Linear Models for Discrete High-Dimensional Data}, 
  journal={IEEE Journal on Selected Areas in Information Theory}, 
  year={2020},
  volume={1},
  number={3},
  pages={884-896}
}

@ARTICLE{Salter-Townshend2012243,
author={Salter-Townshend, M. and White, A. and Gollini, I. and Murphy, T.B.},
title={Review of statistical network analysis: Models, algorithms, and software},
journal={Statistical Analysis and Data Mining},
year={2012},
volume={5},
number={4},
pages={243-264}
}

@ARTICLE{Handcock2007301,
author={Handcock, M.S. and Raftery, A.E. and Tantrum, J.M.},
title={Model-based clustering for social networks (with discussion)},
journal={Journal of the Royal Statistical Society. Series A: Statistics in Society},
year={2007},
volume={170},
number={2},
pages={301-354}
}

@ARTICLE{Krivitsky2008,
author={Krivitsky, P.N. and Handcock, M.S.},
title={Fitting position latent cluster models for social networks with latentnet},
journal={Journal of Statistical Software},
year={2008},
volume={24},
number={5}
}

@ARTICLE{Durante20171547,
author={Durante, D. and Dunson, D.B. and Vogelstein, J.T.},
title={Rejoinder: Nonparametric Bayes Modeling of Populations of Networks},
journal={Journal of the American Statistical Association},
year={2017},
volume={112},
number={520},
pages={1547-1552}
}

@ARTICLE{Durante201829,
author={Durante, D. and Dunson, D.B.},
title={Bayesian inference and testing of group differences in brain networks},
journal={Bayesian Analysis},
year={2018},
volume={13},
number={1},
pages={29-58}
}

@ARTICLE{angelo2019900,
author={{D'Angelo}, S. and Murphy, T.B. and {Alf\`{o}}, M.},
title={Latent space modelling of multidimensional networks with application to the exchange of votes in eurovision song contest},
journal={Annals of Applied Statistics},
year={2019},
volume={13},
number={2},
pages={900-930}
}

@ARTICLE{D_Angelo2020324,
author={{D'Angelo}, S. and {Alf\`{o}}, M. and Murphy, T.B.},
title={Modeling node heterogeneity in latent space models for multidimensional networks},
journal={Statistica Neerlandica},
year={2020},
volume={74},
number={3},
pages={324-341}
}

@article{hoff2005bilinear,
  author={Hoff, P.D.},
  title={Bilinear mixed-effects models for dyadic data},
  journal={Journal of the american Statistical association},
  publisher={Taylor \& Francis},
  year={2005},
  volume={100},
  number={469},
  pages={286--295}
}

@article{hoff2021additive,
  author={Hoff, P.D.},
  title={Additive and multiplicative effects network models},
  journal={Statistical Science},
  publisher={Institute of Mathematical Statistics},
  year={2021},
  volume={36},
  number={1},
  pages={34--50}
}

@ARTICLE{Gormley200790,
author={Gormley, I.C. and Murphy, T.B.},
title={A latent space model for rank data},
journal={Lecture Notes in Computer Science (including subseries Lecture Notes in Artificial Intelligence and Lecture Notes in Bioinformatics)},
year={2007},
volume={4503 LNCS},
pages={90-102}
}

@ARTICLE{Gollini2016246,
author={Gollini, I. and Murphy, T.B.},
title={Joint Modeling of Multiple Network Views},
journal={Journal of Computational and Graphical Statistics},
year={2016},
volume={25},
number={1},
pages={246-265}
}

@article{aliverti2019spatial,
  author={Aliverti, E. and Durante, D.},
  title={Spatial modeling of brain connectivity data via latent distance models with nodes clustering},
  journal={Statistical Analysis and Data Mining: The ASA Data Science Journal},
  publisher={Wiley Online Library},
  year={2019},
  volume={12},
  number={3},
  pages={185--196}
}

@article{hoff2007modeling,
  author={Hoff, P.D.},
  title={Modeling homophily and stochastic equivalence in symmetric relational data},
  booktitle = {Proceedings of the 20th International Conference on Neural Information Processing Systems},
publisher = {Curran Associates Inc.},
 year={2007},
pages = {657–664},
numpages = {8},
isbn = {9781605603520}
}

@article{hoff2011hierarchical,
  author={Hoff, P.D.},
  title={Hierarchical multilinear models for multiway data},
  journal={Computational Statistics \& Data Analysis},
  publisher={Elsevier},
  year={2011},
  volume={55},
  number={1},
  pages={530--543}
}

@article{barigozzi2019nets,
 author={Barigozzi, M. and Brownlees, C.},
 title={Nets: Network estimation for time series},
 journal={Journal of Applied Econometrics},
 publisher={Wiley Online Library},
 year={2019},
 volume={34},
 number={3},
 pages={347--364}  
}

@article{CLARK2021100493,
author = {Clark, N.J. and Dixon, P.M.},
title = {A class of spatially correlated self-exciting statistical models},
journal = {Spatial Statistics},
volume = {43},
pages = {100493},
year = {2021},
issn = {2211-6753}
}

@article{watanabe2010asymptotic,
  author={Watanabe, S. and Opper, M.},
  title={Asymptotic equivalence of Bayes cross validation and widely applicable information criterion in singular learning theory.},
  journal={Journal of machine learning research},
  year={2010},
  volume={11},
  number={12}
}

@article{vehtari2017practical,
  author={Vehtari, A. and Gelman, A. and Gabry, J.},
  title={Practical Bayesian model evaluation using leave-one-out cross-validation and WAIC},
  journal={Statistics and computing},
  publisher={Springer},
  year={2017},
  volume={27},
  pages={1413--1432}
}

@article{burkner2020approximate,
  author={B{\"u}rkner, P.C. and Gabry, J. and Vehtari, A.},
  title={Approximate leave-future-out cross-validation for Bayesian time series models},
  journal={Journal of Statistical Computation and Simulation},
  publisher={Taylor \& Francis},
  year={2020},
  volume={90},
  number={14},
  pages={2499--2523}
}

@book{BDA_Gelman2013,
author = {Gelman, A. and Carlin, J. and Stern, H. and Dunson, D. and Vehtari, A. and Rubin, D.},
title = {Bayesian Data Analysis},
year = {2013},
pages = {},
isbn = {9780429113079}
}

@book{franke1993multivariate,
  author={Franke, J. and Rao, T.S.},
  title={Multivariate First-Order Integer-Valued Autoregressions},
  series={Berichte der Arbeitsgruppe Technomathematik},
  publisher={Arbeitsgruppe Technomathematik, Univ.},
  year={1993}
}

@conference{mark2017network,
  author={Mark, B. and Raskutti, G. and Willett, R.},
  title={Network estimation via poisson autoregressive models},
  booktitle={2017 IEEE 7th International Workshop on Computational Advances in Multi-Sensor Adaptive Processing (CAMSAP)},
  organization={IEEE},
  year={2017},
  pages={1--5}
}

@article{COX1981,
author = {Cox, D. R. and Gudmundsson, G. and Lindgren, G. and Bondesson, L. and Harsaae, E. and Laake, P. and Juselius, K. and Lauritzen, S.L.},
title = {Statistical Analysis of Time Series: Some Recent Developments [with Discussion and Reply]},
 journal = {Scandinavian Journal of Statistics},
publisher = {[Board of the Foundation of the Scandinavian Journal of Statistics, Wiley]},
year = {1981},
volume = {8},
number = {2},
 pages = {93--115},
 ISSN = {03036898, 14679469}
}

@inbook{book_Francq,
author = {Francq, C. and Zakoian, J.-M.},
title = {Multivariate GARCH Processes},
booktitle = {GARCH Models},
publisher = {John Wiley \& Sons, Ltd},
year = {2019},
chapter = {10},
pages = {272-316},
isbn = {9781119313472}
}

@inbook{LPNM_2023,
author = {Kaur, H. and Rastelli, R. and Friel, N.},
title = {Latent Position Network Models},
booktitle = {The Sage Handbook of Social Network Analysis},
publisher = {SAGE Publications},
year = {2023},
chapter = {36},
pages = {526-541},
isbn = {9781529614671}
}

@inbook{amillotta2022generalized,
author = {Armillotta, M. and Fokianos, K. and Krikidis, I.},
title = {Generalized Linear Models Network Autoregression},
year = {2022},
pages = {112-125},
isbn = {978-3-030-97239-4}
}

@misc{betancourt2011geometry,
 author={Betancourt, M. and Stein, L.C.},
      title={The Geometry of Hamiltonian Monte Carlo}, 
      year={2011},
      eprint={1112.4118},
      archivePrefix={arXiv},
      primaryClass={stat.ME}
}

@article{ahelegbey2017bayesian,
author={Ahelegbey, D. and Carvalho, L. and Kolaczyk, E.},
  title={A Bayesian covariance graphical and latent position model for multivariate financial time series},
  journal={arXiv preprint arXiv:1712.06797},
  year={2017}
}

@article{Hoffman2019NeuTralizingBG,
 author={Hoffman, M.D. and Sountsov, P. and Dillon, J.V. and Langmore, I. and Tran, D. and Vasudevan, S.},
  title={NeuTra-lizing Bad Geometry in Hamiltonian Monte Carlo Using Neural Transport},
  journal={arXiv: Computation},
  year={2019},
month = {03},
pages = {}
}

@article{rastelli2023continuous,
  title={Continuous latent position models for instantaneous interactions},
  author={Rastelli, R. and Corneli, M.},
  journal={Network Science},
  volume={11},
  number={4},
  pages={560--588},
  year={2023},
  publisher={Cambridge University Press}
}

@article{Hardle_wang_yu2016,
title = {TENET: Tail-event driven network risk},
author = {Härdle, W.K. and Wang, W. and Yu, L.},
journal = {Journal of Econometrics},
volume = {192},
year = {2016},
pages = {499-513}
}

@article{DIEBOLD2014119,
title = {On the network topology of variance decompositions: Measuring the connectedness of financial firms},
author = {Diebold, F.X. and Yılmaz, K.},
journal = {Journal of Econometrics},
volume = {182},
number = {1},
pages = {119-134},
year = {2014},
issn = {0304-4076}
}

@Article{Billio_2021,
  author={Billio, M. and Getmansky, M. and Lo, A.W. and Pelizzon, L.},
  title={{Econometric measures of connectedness and systemic risk in the finance and insurance sectors}},
  journal={Journal of Financial Economics},
  year={2012},
  volume={104},
  number={3},
  pages={535-559}
}

@article{Dahlhaus_2000,
author = {Dahlhaus, R.},
year = {2000},
month = {08},
pages = {157-172},
title = {Graphical interaction models for multivariate time series1},
volume = {51},
journal = {Metrika}
}

@article{Shojaie_Michailidis_2010,
    author = {Shojaie, A. and Michailidis, G.},
    title = "{Discovering graphical Granger causality using the truncating lasso penalty}",
    journal = {Bioinformatics},
    volume = {26},
    number = {18},
    pages = {i517-i523},
    year = {2010},
    month = {09}
}

@article{Bermúdez_Karlis_2011,
author = {Bermúdez, L. and Karlis, D.},
year = {2011},
month = {07},
pages = {},
title = {Mixture of Bivariate Poisson Regression Models with an Application to Insurance},
volume = {56},
journal = {Computational Statistics \& Data Analysis}
}

@article{sewell2017latent,
  title={Latent space approaches to community detection in dynamic networks},
  author={Sewell, D.K. and Chen, Y.},
  journal={Bayesian analysis},
  volume={12},
  number={2},
  pages={351--377},
  year={2017},
  publisher={International Society for Bayesian Analysis}
}

@ARTICLE{Friel20166629,
author={Friel, N. and Rastelli, R. and Wyse, J. and Raftery, A.E.},
title={Interlocking directorates in Irish companies using a latent space model for bipartite networks},
journal={Proceedings of the National Academy of Sciences of the United States of America},
year={2016},
volume={113},
number={24},
pages={6629-6634}
}

@article{Gwee_2023,
   title={A Latent Shrinkage Position Model for Binary and Count Network Data},
   volume={1},
   ISSN={1936-0975},
   number={1},
   journal={Bayesian Analysis},
   publisher={Institute of Mathematical Statistics},
   author={Gwee, X.Y and Gormley, I.C. and Fop, M.},
   year={2023},
   month=jan }

@Book{NoceWrig06,
  author    = {Nocedal, J. and Wright, S.J.},
  publisher = {Springer},
  title     = {Numerical Optimization},
  year      = {2006},
  address   = {New York, NY, USA},
  edition   = {2e},
}

@Misc{STAN_SOFTWARE,
    title = {{RStan}: the {R} interface to {Stan}},
    author = {{Stan Development Team}},
    note = {R package version 2.32.6},
    year = {2024}
  }

@ARTICLE{sims_1980,
title = {Macroeconomics and Reality},
author = {Sim, C.},
year = {1980},
journal = {Econometrica},
volume = {48},
number = {1},
pages = {1-48}
}

@article{Lutkepohl_1999,
  author={Lütkepohl, H.},
  title={Vector autoregressions},
  year={1999},
  journal={Unpublished manuscript, Humboldt University of Berlin, Interdisciplinary Research Project 373: Quantification and Simulation of Economic Processes},
  type={SFB 373 Discussion Papers},
  number={1999,4}
}

@article{tafakori2022measuring,
  title={Measuring systemic risk and contagion in the European financial network},
  author={Tafakori, L. and Pourkhanali, A. and Rastelli, R.},
  journal={Empirical economics},
  volume={63},
  number={1},
  pages={345--389},
  year={2022},
  publisher={Springer}
}

@article{hledik2023dynamic,
  title={A dynamic network model to measure exposure concentration in the Austrian interbank market},
  author={Hledik, J. and Rastelli, R.},
  journal={Statistical Methods \& Applications},
  volume={32},
  number={5},
  pages={1695--1722},
  year={2023},
  publisher={Springer}
}

@article{johnson2008repeat,
  title={Repeat burglary victimisation: a tale of two theories},
  author={Johnson, S.D.},
  journal={Journal of Experimental Criminology},
  volume={4},
  pages={215--240},
  year={2008},
  publisher={Springer}
}

@article{yuan2019multivariate,
  title={Multivariate spatiotemporal hawkes processes and network reconstruction},
  author={Yuan, B. and Li, H. and Bertozzi, A.L. and Brantingham, P.J. and Porter, M.A.},
  journal={SIAM Journal on Mathematics of Data Science},
  volume={1},
  number={2},
  pages={356--382},
  year={2019},
  publisher={SIAM}
}

@article{mohler2011self,
  title={Self-exciting point process modeling of crime},
  author={Mohler, G.O. and Short, M.B. and Brantingham, P.J. and Schoenberg, F.P. and Tita, G.E.},
  journal={Journal of the american statistical association},
  volume={106},
  number={493},
  pages={100--108},
  year={2011},
  publisher={Taylor \& Francis}
}
\end{document}